\documentclass[journal,compsoc]{IEEEtran}    % eprint

\ifCLASSINFOpdf
\else
\fi

\usepackage{pdfpages}
\usepackage[bookmarksnumbered=true, bookmarksopen=true]{hyperref}
\usepackage{dsfont}
\usepackage[cmex10]{amsmath}
\usepackage{amssymb}
\usepackage{amsthm}
\usepackage{amsfonts}
\usepackage{cite}

\newtheorem{definition}{Definition}
\newtheorem{remark}[definition] {Remark}
\newtheorem{theorem}[definition]{Theorem}
\newtheorem{proposition}[definition]{Proposition}
\newtheorem{corollary}[definition]{Corollary}
\newtheorem{lemma}[definition]{Lemma}

\begin{document}
\title{A Framework For Fully-Simulatable $h$-Out-Of-$n$  Oblivious Transfer}

\author{ Zeng~Bing, Tang~Xueming, and~Chingfang~Hsu
\thanks{Zeng Bing is with the College of Computer Science and Technology,
        Huazhong University of Science and Technology, Wuhan City, Hubei 430074 China
(e-mail:zeng.bing.zb@gmail.com).} % <-this % stops a space

\thanks{Tang~Xueming is with the College of Computer Science and Technology,
        Huazhong University of Science and Technology, Wuhan City, Hubei 430074 China
(e-mail:tang.xueming.txm@gmail.com).} % <-this % stops a space
\thanks{Chingfang~Hsu is with the College of Computer Science and Technology,
        Huazhong University of Science and Technology, Wuhan City, Hubei 430074 China
        (e-mail:cherryjingfang@gmail.com).} % <-this % stops a space
}

\markboth{}%IEEE Transactions On Information Theory}%
{Zeng \MakeLowercase{\textit{et al.}}: A Framework For Fully-Simulatable
$h$-Out-Of-$n$  Oblivious Transfer}


\maketitle

\begin{abstract}
We present  a  framework for  fully-simulatable $h$-out-of-$n$  oblivious transfer ($OT^{n}_{h}$)  with security against non-adaptive malicious adversaries. The framework costs six communication rounds and costs at most $40n$ public-key operations in computational overhead.
Compared with the known protocols for fully-simulatable oblivious transfer that works in the plain mode (where there is no  trusted common reference string available) and proven to be secure under standard model (where there is no    random oracle available), the instantiation based on the decisional Diffie-Hellman assumption of  the framework  is the most efficient one, no matter seen from communication rounds or  computational overhead.

Our framework uses three abstract tools, i.e., perfectly binding commitment, perfectly hiding commitment and our new smooth projective hash.  This allows a simple and intuitive understanding of its security.

We instantiate the new smooth projective hash under the lattice assumption, the decisional Diffie-Hellman assumption, the decisional $N$-th residuosity assumption, the decisional quadratic residuosity
assumption. This indeed shows that the folklore that it is technically difficult to instantiate the projective hash framework under the lattice assumption is not true.  What's more, by using this lattice-based hash and lattice-based commitment scheme, we gain a  concrete protocol for $OT^{n}_{h}$  which is secure against quantum algorithms.
\end{abstract}



% \fi
\IEEEpeerreviewmaketitle

\section{Introduction}
\subsection{Oblivious transfer}
\IEEEPARstart{O}{blivious} transfer (OT), first introduced by \cite{rabin1981exchange} and later defined in another way with equivalent effect \cite{crpeau1987equivalence}  by \cite{even1985randomized}, is a fundamental primitive in cryptography and a concrete problem in the filed of secure multi-party computation.  Considerable  cryptographic protocols can be  built from it. Most remarkable,  \cite{Ishai2008fcot ,Kilian1988fcot,yao1986generate,goldreich1987play} proves that any secure multi-party computation can be based on a secure oblivious transfer protocol. In this paper, we concern a variant of OT,  $h$-out-of-$n$  oblivious transfer ($OT^{n}_{h}$).  $OT^{n}_{h}$ deals with the following scenario. A sender holds $n$ private messages $m_{1}, m_{2}, \ldots, m_{n}$. A receiver holds $h$ private positive integers $i_{1},  i_{2}, \ldots, i_{h}$, where $i_{1}<i_{2}<\ldots<i_{h} \leqslant n$. The receiver expects to get the messages $m_{i_{1}}, m_{i_{2}}, \ldots, m_{i_{h}}$ without leaking any information about his private input, i.e.,  the $h$ positive integers he holds. The sender expects all new knowledge learned by the receiver from their interaction is  at most $h$ messages. Obviously, the OT most literature refer to  is $OT^{2}_{1}$ and can be viewed as a special case of $OT^{n}_{h}$.

Considering a variety of attack we have to confront in real environment, a protocol for
$OT^{n}_{h}$ with security against malicious adversaries (a malicious adversary  may act in any arbitrary  malicious way to learn as much extra information as possible) is more desirable than the one with security against semi-honest adversaries (a semi-honest adversary,  on one side, honestly does everything told by a prescribed protocol; on one side, records the messages he sees to deduce extra information which is not supposed to be known to he).  Using Goldreich's compiler \cite{goldreich1987play ,gold2004found}, we can gain the former version  from the corresponding latter version. However, the  resulting protocol is prohibitive expensive for practical use, because it is embedded with so many invocations of zero-knowledge for NP. Thus, directly  constructing  the protocol  based on specific intractability assumptions seems more feasible.

The first step in this direction is independently made by \cite{naor2001efficient} and \cite{aiello2001priced} which respectively presents a two-round efficient protocol for $OT^{2}_{1}$  based on the decisional Diffie-Hellman (DDH) assumption. Starting from these works and using the tool smooth projective hashing,  \cite{kalai2005ot}  abstracts and generalizes the ideas of \cite{naor2001efficient,aiello2001priced}  to a framework for $OT^{2}_{1}$. Besides DDH assumption, the framework can be instantiated  under the decisional $N$-th residuosity (DNR) assumption and  decisional quadratic residuosity (DQR) assumption \cite{kalai2005ot}.

Unfortunately, these protocols (or frameworks) are only half-simulatable not fully-simulatable. By saying a protocol is fully-simulatable, we means that the protocol can be strictly proven its security under the real/ideal model simulation paradigm. The paradigm requires that for any adversary in the real world, there exists a corresponding adversary simulating him in the ideal world. Thus, the real adversary can not do more harm than the corresponding ideal adversary does. Therefore the security level of the protocol is guaranteed not to be lower than that of the ideal world. Undesirably, a half-simulatable protocol for $OT^{2}_{1}$  only provides a simulator in the case the  receiver is corrupted such as \cite{naor2001efficient,aiello2001priced} or in the case the  sender is corrupted such as \cite{kalai2005ot}.

Considering security, requiring a protocol to be fully-simulatable is necessary.  Specifically, a fully-simulatable  protocol provides security against all kinds of attacks, especially the future unknown attacks taken by any adversary whose computational resource  is fixed when constructing the protocol (generally, it is assumed that the adversaries run  arbitrary probabilistic polynomial-time) \cite{canetti2000security, gold2004found}, while a not fully-simulatable  protocol doesn't. For example, the protocols proposed by  \cite{naor2001efficient,aiello2001priced,kalai2005ot}
suffer the selective-failure attacks, in which a malicious sender can induce transfer failures that are dependent on the messages that the receiver requests \cite{naor2005computationally}.

Constructing   fully-simulatable protocols  for OT  with security against malicious adversaries naturally becomes the focus of
the research community.  ~\cite{camenisch2007simulatable} first presents such a fully-simulatable protocol. In detail, the OT is an adaptive $h$-out-$n$ oblivious transfer (denoted by $OT^{n}_{h\times 1}$ in related literature) and based on  $q$-Power Decisional Diffie-Hellman and $q$-Strong Diffie-Hellman assumptions. Unfortunately, these two assumptions are not standard assumptions used in cryptography  and seem significantly stronger than DDH, DQR and so on. Motivated by basing OT on weaker complexity assumption, ~\cite{green2007blind} presents a protocol for $OT^{n}_{h}$ using a
blind identity-based encryption which is  based on decisional bilinear Diffie-Hellman (DBDH) assumption. Using cut-choose technique,  \cite{lindell2008efficient} later presents two efficient  protocols for fully-simulatable $OT^{2}_{1}$ respectively based on DDH assumption and DNR assumption, where the DDH-based protocol is the most efficient one among these fully-simulatable works.

The protocols mentioned above are proved their securities  in the plain stand-alone model which not necessarily allows concurrent composition with other arbitrary malicious protocols. \cite{peikert2008ot} overcomes this weakness and further the research  by presenting a framework under common reference string (CRS) model for fully-simulatable, universally composable $OT^{2}_{1}$  and  instantiating the framework respectively under DDH, DQR and worst-case lattice assumption. It is notable that  conditioning on a trusted CRS is available, the DDH-based instantiation of the framework is the most efficient protocol for $OT^{2}_{1}$ no matter seen from the number of communication rounds or the computational overhead.  Recently, \cite{garay2009somewhat}, using a novel compiler and somewhat non-committing encryption they present, convert \cite{peikert2008ot}'s instantiations based on DDH, DQR to the corresponding protocols with higher security level. In more detail, the resulting protocols for $OT^{2}_{1}$ are secure against adaptive malicious adversaries, which corrupts the parties dynamically  based on his knowledge gathered so far.  Note that, the fully-simulatable protocols for $OT^{2}_{1}$ mentioned so far except the one presented by \cite{camenisch2007simulatable}   are only secure against non-adaptive malicious adversaries, which only corrupts the parties preset before the running of the protocol.

Though constructing  protocols for fully-simulatable $OT^{2}_{1}$ with security against malicious adversaries has been studied well, constructing  protocols for such $OT^{n}_{h}$ hasn't. We  note that there are some works aiming to extend known cryptographic protocols to $OT^{n}_{h}$.  \cite{naor1999oblivious} shows how to implementation $OT^{n}_{h}$ using $\log n$ invocation of $OT^{2}_{1}$ under half-simulation. A similar implementation for adaptive $OT^{n}_{h}$ can be seen in \cite{naor1999adaptive}.  What's more,
the same authors of \cite{naor1999oblivious,naor1999adaptive} propose a way to transform a singe-server private-information retrieval scheme (PIR) into an oblivious transfer scheme under half-simulation too \cite{naor2005computationally}. With the help of a random oracle,  \cite{ishai3extending} shows how to extend $k$ oblivious transfers (for some security parameter $k$) into many more, without much additional effort. However, the Random Oracle Model is risky.
First, \cite{Canetti2004ro} shows that a scheme is secure in the Random Oracle Model does not necessarily imply that a particular implementation of it (in the real world) is secure, or even that this scheme does not have  any "structural flaws". Second, \cite{Canetti2004ro} shows efficient implementing the random oracle is impossible. Later, \cite{Leurent2009Risky} finds that the random-oracle instantiations proposed by Bellare and Rogaway from 1993 and 1996, and the ones implicit in IEEE P1363 and PKCS standards are weaker than a random oracle. What is worse, \cite{Leurent2009Risky} shows that how the hash function defects deadly damages the securities of the cryptographic schemes presented in \cite{boneh2007space,bernstein2008proving}. Therefore, in this paper, we only consider the schemes which are fully-simulatable and without turning to a  random oracle.
To our best knowledge, only \cite{camenisch2007simulatable} and  \cite{green2007blind} respectively present such  fully-simulatable protocols for  $OT^{n}_{h}$. However, the assumptions the former uses are not standard and the latter uses is too expensive. Therefore, a well-motivated problem is to find a protocol or framework for efficient, fully-simulatable, secure against malicious adversaries  $OT^{n}_{h}$  under weaker complexity assumptions.

\subsection{Our Contribution}
In this paper, we present a framework for efficient, fully-simulatable, secure against
non-adaptive malicious adversaries $OT^{n}_{h}$ whose security is proven under stand model (i.e., without turning to a random oracle). To our best knowledge, this is the first  framework for such $OT^{n}_{h}$. The framework have the following features,
\begin{enumerate}
  \item  Fully-simulatable and secure against malicious adversaries without using a CRS.
         \cite{kalai2005ot}'s framework for $OT^{2}_{1}$ is half-simulatable. Thought
         \cite{peikert2008ot}'s framework for $OT^{2}_{1}$ is fully-simulatable, it doesn't work
         without a CRS.  What is more, how to provide a trusted
          CRS before the protocol run still is a unsolved  problem. The
          existing possible solutions, such as natural process suggested by
          \cite{peikert2008ot}, are only conjectures without formal proofs. The same problem
          remains in its adaptive version presented by   \cite{garay2009somewhat}.
           What is worse,  \cite{canetti2001uccom,canetti2006limitations} show that  even given a authenticated communication channel,  implementing  a universal composable protocol providing useful trusted CRS in the presence of malicious adversaries is impossible.
          Therefore, considering practical use, our framework are   better.
  \item Efficient. Compared with the existing  protocols for fully-simulatable $OT$ that without resorting to a CRS or a random oracle,  i.e., the protocols presented by \cite{camenisch2007simulatable,green2007blind,lindell2008efficient}, the DDH-based instantiation of  our  framework   costs the minimum number of   communication rounds and costs the minimum computational overhead.  Please see Section \ref{round} and Section \ref{overhead}  for the detailed comparisons.

      We  admit that, in the context of a trusted CRS is available and only
      $OT^{2}_{1}$ is needed, the  DDH-based instantiation of \cite{peikert2008ot}
      is the most efficient one.

  \item Abstract and modular. The framework is described using just three high-level
        cryptographic tools, i.e., perfectly binding commitment (PBC), perfectly hiding
        commitment (PHC) and our new smooth projective hash (denoted by $SPHDHC_{t,h}$ for simplicity).
        This allows a
        simple and intuitive understanding of its  security.
  \item Generally realizable. The high-level cryptographic tools PBC, PHC and  $SPHDHC_{t,h}$
        are  realizable from a variety of known specific assumptions, even future assumptions maybe.
        This makes our framework generally realizable. In particular,  we instantiate
        $SPHDHC_{t,h}$ from the DDH assumption, the DNR assumption, the DQR assumption and the lattice assumption. Instantiating PBC or PHC under
        specific assumptions is beyond the scope of this paper. Please  see
        \cite{goldreich1996zk,gold2001found} for such examples. Generally realizability is
        vital to make the framework live long, considering the future progress in breaking a
        specific  intractable problem. If this case happen, replacing the instantiation based on the broken    problem  with  that based on a unbroken problem suffices.
\end{enumerate}

What is more, we fix a folklore \cite{lindell2008efficient} that it appears technically difficult to instantiate the projective hash under lattice assumption by presenting  a lattice-based $SPHDHC_{t,h}$ instantiation. It is notable that we gain an $OT^{n}_{h}$ instantiation which is secure against quantum algorithms, using this lattice-based  $SPHDHC_{t,h}$ instantiation and  appropriate  lattice-based commitment schemes.  Considering that factoring integers and finding discrete logarithms are efficiently feasible for quantum algorithms \cite{shor1994algorithms,shor1999polynomial,shor1997polynomial}, this is  an example showing the benefits from  the generally realizability of the framework.

As an independent contribution, we present several propositions/lemmas related to the indistinguishability of
probability ensembles  defined by sampling polynomial instances. Such propositions/lemmas simplify
our security proof very much. We believe that they are as useful in security proof somewhere else  as in this paper.

\subsection{Our Approach}
We note that the smooth projective hash is a good abstract tool. Using this tool, \cite{kalai2005ot} in fact presents a framework for half-simulatable $OT^{2}_{1}$,  \cite{gennaro2006framework} present a framework for password-based authenticated key exchange
protocols. We also note that the cut-and-choose  is a good technique to make protocol fully-simulatable.  Using this tool,  \cite{lindell2008efficient} present several fully-simulatable protocol for $OT^{2}_{1}$,  \cite{lindell2007efficient} presents a general fully-simulatable protocol for two-party computation.  Indeed, we are inspired by such works. Our basic ideal is to use cut-and-choose technique and  smooth projective hash  to get a fully-simulatable
framework.

 Loosely speaking, a smooth projective hash (SPH) is a set of operations defined over two languages $\dot{L}$ and $\ddot{L}$, where $\dot{L} \cap \ddot{L} = \emptyset$. For any projective instance $\dot{x} \in \dot{L}$,  there are two ways to obtain its hash value,  i.e., the way using its hash key or the way using its projective key and its witness $\dot{w}$. For any smooth instance  $\ddot{x} \in \ddot{L}$, there is only one way to obtain its hash value,  i.e., the way using its hash key. The version of SPH presented by ~\cite{kalai2005ot} (denoted by $VSPHH$ for simplicity) holds a property called verifiable smoothness that can judge whether at least one of arbitrary two instances is smooth. Another property $VSPHH$ holds, called hard subset membership, makes sure $\ddot{x}$ and  $\dot{x}$ are computationally indistinguishable.

We observe that the $VSPHH$ indeed is easy to be extended to deal with $OT^{n}_{1}$, but seems difficult to be extended to deal with the  general $OT^{n}_{h}$. The reason is that, to hold verifiable smoothness, $\dot{x}$s and $\ddot{x}$s  have to be generated in a dependent way. This makes the verifiable smoothness  for multiple $\dot{x}$s and multiple $\ddot{x}$s (i.e., judge whether at least $n-h$ of arbitrary $n$ instances are smooth) difficult to hold without leaking information which is conductive to distinguish such $\dot{x}$s and $\ddot{x}$s. We also observe that, there is no way to construct a fully-simulatable framework using $VSPHH$, because there is no way to extract
the real input of the adversary in the case that the receiver is corrupted.

We define a new smooth projective hash called  $t$-smooth $h$-projective hash family that holds properties distinguishability,  hard subset membership, feasible cheating (denoted by $SPHDHC_{t,h}$ for simplicity).
The key solution in $SPHDHC_{t,h}$ to the mentioned problems is that  requiring each $\ddot{x}$ to hold a witness too. This  solution enables us to generate  $\dot{x}$s and $\ddot{x}$s  in a independent way. Correspondingly, the  verifiable smoothness  is not needed any more and replaced by a property called distinguishability, which provides a way to distinguish $\dot{x}$s and  $\ddot{x}$s if their witnesses are given.

Since  the receiver encodes his  input  as a permutation of $\dot{x}$s and $\ddot{x}$s, a  simulator can the extract the real input of the adversary in the case that the receiver is corrupted if their witnesses are available. Combining the  application of the technique cut-and-choose, a simulator can see such witnesses by rewinding the adversary. To extract  the real input of the adversary in the case that the sender is corrupted,  the property feasible cheating provides way to cheat out of the real input of the adversary. Naturally, all the properties and the correlated algorithm in $SPHDHC_{t,h}$ are   extended to deal with $n$ instances rather than only $2$ instances. Please see Section \ref{comparsph} for a detailed comparison this new hash with previous hash systems.

We show that  constructing $SPHDHC_{t,h}$ can be reduced  to constructing considerably simpler hash systems. Our lattice-based $SPHDHC_{t,h}$ instantiation is builded on the lattice-based cryptosystem presented by \cite{lindell2008efficient}. It is noticeable that it appears difficult to get lattice-based instantiation for $SPH$ \cite{lindell2008efficient}. Our solution is to let the instance $x$ ($x\in \dot{L} \cup \ddot{L}$) be available to the algorithm that is responsible for generating pair of the hash key and the projective key.
The other three intractability-assumption-based $SPHDHC_{t,h}$ instantiations can be ultimately  built from  known  SPH schemes  such as that presented by \cite{kalai2005ot} with necessary modifications.

Using $SPHDHC_{t,h}$ we construct the framework described with high-level as follows .
\begin{enumerate}
  \item  The receiver generates  hash parameters and appropriate many instance vectors,
         then sends them  to
         the  sender after disordering each vector.
  \item  The receiver and the sender  cooperate to toss coin to decide which vector
         to be opened.
  \item  The receiver opens the chosen instances, encodes his private input by
         reordering each unchosen vector and sends the resulting code, which in fact is a  sequence of permutations, to the sender.
  \item  The  sender checks that the chosen vectors are generated in the legal way
         which guarantees that the receiver learns at most $h$ message. If the check pass,
         the sender
         encrypts his private input (i.e., the $n$ messages he holds) using the hash values of the instances of the unchosen vectors in the way indicated by the code of receiver's private input, and sends the ciphertexts together with
         some auxiliary information  (i.e., the projective hash keys) that is conductive
         to decrypt some ciphertexts to the receiver.
  \item  The receiver decrypts the ciphertexts with the help of the auxiliary information and gains the messages he expects.
\end{enumerate}

Intuitively speaking, the receiver's security is implied by the property  hard subset membership of $SPHDHC_{t,h}$. This property  guarantees that the receiver can securely encode  his private input by   reordering each unchosen instance vector. The sender's security is implied by the cut-and-choose technique, which guarantees that
the probability that the adversaries controlling a corrupted  receiver learns extra new knowledge is negligible.

\subsection{Organization}
In Section \ref{Pre},  we describe the notations used in this  paper,  the security definition of $OT^{n}_{h}$, the definition of  commitment scheme.  In Section \ref{hashing}, we define our new hash system, i.e., $SPHDHC_{t,h}$.
In Section \ref{ot}, we construct our framework. In Section \ref{proof}, we prove the security of the framework. In Section \ref{reduceSPHDHC},  we reduce constructing $SPHDHC_{t,h}$ to  constructing considerably  simpler hash systems. In Section \ref{con_hash},  we  instantiate $SPHDHC_{t,h}$ under the lattice, DDH, DNR, DQR assumptions, respectively.

\section{Preliminaries} \label{Pre}
Most notations and concepts mentioned in this section originate from ~\cite{gold2001found,gold2004found,canetti2000security} which are basic literature in the
filed of secure multi-party computation (SMPC). We tailor them to the need of dealing with  $OT^{n}_{h}$.

\subsection{Basic Notations}     \label{Notion}
We denote an unspecified positive polynomial by $poly(.)$. We denote the set consists of all natural numbers  by $\mathbb{N}$.  For any $i\in \mathbb{N}$,  $[i] \stackrel{def}{=} \{1,2,\ldots,i \}$. We denote the set consists of all prime numbers  by $\mathbb{P}$.

%样本整体
We denote security parameter used to measure security and complexity by $k$. A function $\mu(.)$ is negligible in $k$, if there exists a positive constant integer $n_{0}$, for any $poly(.)$ and any $k$ which is greater than $n_{0}$ (for simplicity, we later call such $k$ sufficiently large $k$), it holds that $\mu(k)<1/poly(k)$. A probability ensemble $X \stackrel {def}{=} \{X(1^{k},a)\}_{k \in \mathbb{N},a \in \{0,1\}^{*}}$ is an infinite sequence of random variables indexed by $(k,a)$, where $a$ represents various types of inputs used to sample the instances according to the distribution of the random variable $X(1^{k},a)$. Probability ensemble $X$ is polynomial-time constructible, if there exists a probabilistic polynomial-time (PPT) sample algorithm $S_{X}(.)$ such that for any $a$, any $k$, the random variables $S_{X}(1^{k},a)$ and $X(1^{k},a)$ are identically distributed. We denote sampling an instance according to $X(1^{k},a)$ by $\alpha \leftarrow S_{X}(1^{k},a)$.

Let $X \stackrel {def}{=} \{X(1^{k},a)\}_{k \in \mathbb{N},a \in \{0,1\}^{*}}$ and $Y \stackrel {def}{=} \{Y(1^{k},a)\}_{k \in \mathbb{N},a \in \{0,1\}^{*}}$  be two probability ensembles.
They are computationally indistinguishable, denoted $X\stackrel{c}{=}Y$, if for any non-uniform PPT algorithm $D$ with an infinite auxiliary information sequence $z=(z_{k})_{k\in \mathds{N}}$ (where each $z_{k} \in \{0,1\}^{*} $), there exists a negligible function $\mu(.)$ such that for any sufficiently large $k$,
any $a$, it holds that
\begin{multline*}
    |Pr(D(1^{k},X(1^{k},a),a,z_{k})=1) - \\ Pr(D(1^{k},Y(1^{k},a),a,z_{k})=1)| \leqslant \mu(k)
\end{multline*}
They are  same, denoted $X=Y$, if for any sufficiently large $k$, any $a$, $X(1^{k},a)$ and $Y(1^{k},a)$ are defined in the same way. They are equal, denoted $X \equiv Y$, if for any sufficiently large $k$, any $a$,  the distributions of $X(1^{k},a)$ and $Y(1^{k},a)$ are identical. Obviously, if $X=Y$ then $X \equiv Y$; If $X \equiv Y$  then $X\stackrel{c}{=}Y$.

Let $\vec{x}$ be a vector (note that arbitrary binary string can be viewed as a vector). We denote
its $i$-th element by $\vec{x} \langle i \rangle$, denote its dimensionality by $\#\vec{x}$, denote
its length in bits by $|\vec{x}|$. For any positive integers set $I$, any vector $\vec{x}$,
$\vec{x} \langle I \rangle \stackrel{def}{=} (\vec{x} \langle i \rangle)_{i \in I, i\leq \#\vec{x}}$.

Let $M$ be a probabilistic (interactive) Turing machine. By $M_{r}(.)$  we denote $M$'s output generated at the end of an execution using randomness $r$.

Let $f:D \rightarrow R$. Let $D^{\prime} \subseteq \{0,1\}^{*}$. Then  $f(D^{\prime}) \stackrel{def}{=} \{ f(x)| x \in D^{\prime} \cap D \}$, $Range(f)\stackrel{def}{=} f(D)$.

Let $x \in_{\chi} Y$ denotes sampling an instance $x$ from domain $Y$ according to the distribution law (or probability density function
) $\chi$. Specifically, let $x \in_{U} Y$ denotes uniformly sampling an instance $x$ from domain $Y$.

\subsection{Security Definition Of A Protocol For $OT^{n}_{h}$} \label{secdef}
\subsubsection{Functionality Of $OT^{n}_{h}$}
$OT^{n}_{h}$ involves two parties, party $P_{1}$ (i.e., the sender) and party $P_{2}$
(i.e., the receiver). $OT^{n}_{h}$'s functionality is formally defined as follows
\begin{eqnarray*}
  f:\mathbb{N}\times \{0,1\}^{*}\times \{0,1\}^{*} & \rightarrow  & \{0,1\}^{*}\times \{0,1\}^{*} \\
  f(1^{k},\vec{m},H) &=&(\lambda, \vec{m}\langle H \rangle)
\end{eqnarray*}
where
\begin{itemize}
  \item $k$ is the public security parameter.
  \item $\vec{m} \in (\{0,1\}^{*})^{n}$ is $P_{1}$'s private input, and each $|\vec{m}\langle i \rangle|$ is the same.
  \item $H \in \Psi \stackrel{def}{=} \{ B| B \subseteq [n] \textrm{, } \# B=h \}$ is $P_{2}$'s private input.
  \item $\lambda$ denotes a empty string and is supposed to be got by $P_{1}$. That is,
        $P_{1}$ is supposed to get nothing.
  \item $\vec{m}\langle H \rangle$ is supposed to be got by $P_{2}$.
\end{itemize}
Note that, the length of all parties' private input have to be identical in SMPC (please see
~\cite{gold2004found} for the reason and related discussion). This means that $|\vec{m}|=|H|$ is
required. Without loss of generality, in this paper, we assume $|\vec{m}|=|H|$ always
holds, because padding can be easily used to meet such requirement.

Intuitively speaking, the security of $OT^{n}_{h}$ requires that $P_{1}$ can't learn any  new knowledge --- typically, $P_{2}$'s private input, from the interaction at all,  and $P_{2}$ can't learn more than $h$ messages held by $P_{1}$. To capture the security in a formal way, the concepts such as adversary, trusted third party, ideal world, real world were introduced. Note that the security target in this paper is to be secure against non-adaptive malicious adversaries, so only concepts related to this case is referred to in the following.

\subsubsection{Non-Adaptive Malicious Adversary}
Before running $OT^{n}_{h}$, the adversary $A$ has to corrupt all parties listed in $I\subseteq [2]$. In the case that $U \in \{P_{1}, P_{2}\}$ is not corrupted, $U$ will strictly follow the prescribed protocol as an honest party. In the case that party $U$ is corrupted, $U$ will be fully controlled by $A$ as a corrupted party. In this case, $U$ will have to pass all his knowledge to $A$ before the protocol runs and follows $A$'s instructions from then on --- so there is a probability that $U$  arbitrarily deviates from prescribed protocol. In fact, after $A$ finishes corrupting, $A$ and all corrupted parties have formed a coalition led by $A$ to learn as much extra knowledge, e.g. the honest parties' private inputs, as possible. From then on, they share knowledge with each other and coordinate their behavior. Without loss of generality, we can  view this coalition as follows. All corrupted parties are dummy. $A$ receives messages addressed to the members of the coalition and sends messages on behalf of the members.

Loosely speaking, we say $OT^{n}_{h}$ is secure, if and only if, for any malicious adversary $A$,
the knowledge $A$ learns in the real world is not more than that he learns in the ideal world. In other words, if and only if, for any malicious adversary $A$, what harm $A$ can do in real world is not more than what harm he can do in the ideal world. In the ideal world, there is an incorruptible trusted third party (TTP). All parties hand their private inputs to TTP. TTP computes $f$ and sends back $f(.)\langle i \rangle$ to $P_{i}$. In the real world, there is no TTP, and the computation of $f(.)$ is finished by $A$ and all parties' interactions.

\subsubsection{$OT^{n}_{h}$ In The Ideal World}
In the ideal world, an execution of $OT^{n}_{h}$  proceeds as follows.

Initial Inputs.
All entities know the public security parameter $k$. $P_{1}$ holds a private input
$\vec{m} \in (\{0,1\}^{*})^{n}$. Party $P_{2}$ holds a private input $H \in \Psi$. Adversary $A$ holds a name list $I\subseteq [2]$, a randomness $r_{A}\in \{0,1\}^{*}$ and an infinite auxiliary input sequence  $z=(z_{k})_{k\in \mathds{N}}$, where $z_{k}\in \{0,1\}^{*}$. Before proceeds to next stage, $A$ corrupts parties listed in $I$ and learns  $\vec{x} \langle I \rangle$, where
$\vec{x} \stackrel{def}{=} (\vec{m},H)$.

Submitting inputs to TTP.
Each honest party $P_{i}$ always submits its private input $\vec{x} \langle i \rangle$ unchanged to TTP. $A$ submits arbitrary string based on his knowledge to TTP for the corrupted parties.
The string TTP receives is a two-dimensional vector $\vec{y}$ which is formally described
as follows.
\begin{equation*}
    \vec{y}\langle i \rangle=
     \begin{cases}
           \vec{x} \langle i \rangle & \textrm{if $i \notin I$ },  \\
           \alpha  & \textrm{if } i \in  I
     \end{cases}
\end{equation*}
where $\alpha \in \{\vec{x} \langle i \rangle \} \cup  \{0,1\}^{|\vec{x} \langle i \rangle|} \cup  \{ abort_{i}\} $
and  $\alpha\leftarrow A(1^{k}, I, r_{A}, z_{k},\vec{x} \langle I \rangle)$.  Obviously, there is a probability that  $\vec{x}\neq \vec{y}$.

TTP computing $f$.
TTP checks that $\vec{y}$ is a valid input to $f$, i.e., no entry of $\vec{y}$ is of the form  $abort_{i}$. If $\vec{y}$ passes the check, then TTP computes $f$ and  sets $\vec{w}$ to be $f(1^{k},\vec{y})$.
Otherwise,  TTP sets $\vec{w}$ to be $(abort_{i},abort_{i})$. Finally, for each $i\in[n]$ TTP  hands $\vec{w}\langle i \rangle$ to each $P_{i}$  respectively and halts.

Outputs.
Each honest party $P_{i}$ always outputs the message $\vec{w}\langle i \rangle$ it obtains from the TTP. Each corrupted party $P_{i}$ outputs nothing (i.e., $\lambda$). The adversary outputs something generated by executing  arbitrary function of the information he gathers so far. Without loss of generality, this can be assumed to be  ($1^{k}, I, r_{A},
z_{k}, \vec{x} \langle I \rangle, \vec{w} \langle I \rangle$).

The output of the whole execution  in the ideal world, denoted by $Ideal_{f,I,A(z_{k})}(1^{k},\vec{m},H)$,  is defined by the outputs of all parties and that of the adversary as follows.
\begin{multline*}
    Ideal_{f, A(z),I}(1^{k}, \vec{x}, r_{A}) \langle i \rangle \\
     \stackrel{def}{=}
    \begin{cases}
          \begin{split}
             &  A\textrm{'s output, i.e., }(1^{k}, I, r_{A}, \\
              & \;\;\;  z_{k}, \vec{x} \langle I \rangle, \vec{w} \langle I \rangle ),
          \end{split}
            & i=0;\\
        \textrm{$P_{i}$'s output, i.e., } \lambda, & i \in I;\\
         \textrm{$P_{i}$'s output, i.e., }\vec{w} \langle i \rangle , & i \in [n] - I.\\
    \end{cases}
\end{multline*}
Obviously, $Ideal_{f, A(z),I}(1^{k}, \vec{x})$  is a random variable whose randomness is $r_{A}$.

\subsubsection{$OT^{n}_{h}$ In The Real World}
In the real world, there is no TTP. A  execution of $OT^{n}_{h}$ proceeds as follows.

Initial Inputs.
Initial input each entity holds in the real world is the same as in the ideal world but there are some difference as follows. A randomness $r_{i}$ is held by each party $P_{i}$. After finishes corrupting, in addition to the knowledge  $A$ learns in ideal world, the corrupted parties' randomness $\vec{r}\langle I \rangle$ is also learn by $A$, where $\vec{r} \stackrel{def}{=} (r_{1}, r_{2})$.

Computing $f$.
In the real world, computing $f$ is finished by all entities' interaction. Each honest party strictly follows the prescribed protocol (i.e., the concrete protocol, usually denoted $\pi$ , for $OT^{n}_{h}$). The corrupted parties have to follow $A$'s instructions and may arbitrarily deviate from prescribed protocol.

Outputs.
Each honest party $P_{i}$ always outputs what the prescribed protocol instructs. Each corrupted party $P_{i}$ outputs nothing. The adversary outputs something generated by executing  arbitrary function of the information he gathers so far. Without loss of generality, this can be assumed to be a string consisting of $1^{k}, I, r_{A}, \vec{r}\langle I \rangle, z_{k}, \vec{x} \langle I \rangle$ and messages addressed to the corrupted parties.

The output of the whole execution in the real world, denoted by $Real_{\pi,I,A(z_{k})}(1^{k},\vec{m},H, r_{A}, \vec{r})$, is defined by the outputs of all parties and that of the adversary as follows.
\begin{multline*}
  Real_{\pi,I,A(z_{k})}(1^{k},\vec{m},H, r_{A}, \vec{r}) \langle i \rangle \\
  \stackrel{def}{=}
    \begin{cases}
    \begin{split}
       &A's \textrm{ output, i.e., }  (1^{k}, I, r_{A},  \\
        & \;\;\;\; \vec{r}\langle I \rangle, z_{k}, \vec{x} \langle I \rangle, msg_{I}),
    \end{split}      & i=0;\\
     P_{i}'s \textrm{ output, i.e., } \lambda, & i \in I;\\
     \begin{split}
       &P_{i}'s \textrm{ output, i.e., what}\\
        & \;\;\;\;\textrm{instructed by } \pi ,
    \end{split}           & i \in [n] - I.\\
    \end{cases}
\end{multline*}
Obviously, $Real_{\pi,I,A(z_{k})}(1^{k},\vec{m},H)$ is a random variable whose randomnesses are $r_{A}$ and $\vec{r}$.

\subsubsection{Security Definition}
The security  of a protocol for $OT^{n}_{h}$ is formally captured by the following definition.

\begin{definition}[The security of a protocol for $OT^{n}_{h}$] \label{sec_def_ot}
Let $f$ denotes the functionality of  $OT^{n}_{h}$ and let $\pi$ be a  concrete protocol for $OT^{n}_{h}$. We say  $\pi$ securely computes $f$, if and only if for any non-uniform
probabilistic polynomial-time adversary $A$ with  an infinite sequence  $z=(z_{k})_{k\in \mathds{N}}$  in the real world, there exists a  non-uniform probabilistic expected polynomial-time adversary $A'$ with the same sequence in the ideal world  such that, for any $I\subseteq [2]$, it holds that

\begin{multline} \label{eq_ot_sec}
   \{Real_{\pi,I,A(z_{k})}(1^{k},\vec{m},H)\}_{\substack{k\in\mathds{N},
           \vec{m}\in (\{0,1\}^{*})^{n}\\  H \in \Psi,  z_{k} \in \{0,1\}^{*} } }
  \stackrel{c}{=}\\
  \{ Ideal_{f,I,A'(z_{k})}(1^{k},\vec{m},H)\} _{\substack{k\in\mathds{N},  \vec{m}\in
   (\{0,1\}^{*})^{n} \\  H \in \Psi,  z_{k}\in \{0,1\}^{*}} }
\end{multline}
where the parameters input  to the two probability ensembles are  same and each $\vec{m}\langle i\rangle$ is of the same length.  The adversary A' in the ideal world is called a simulator of the
adversary A in the real world.
\end{definition}

The concept, non-uniform probabilistic expected polynomial-time, mentioned in Definition \ref{sec_def_ot} is formulated in distinct way in distinct literature such as ~\cite{gold2001found,canetti2004adaptive}. We prefer to the following definition
\cite{katz2008handling}, because it is clearer in formulation  and more closely related to our issue.

\begin{definition}[$M_{1}$ runs in expected polynomial-time with respect to $M_{2}$] \label{def_exp_time}
Let $M_{1},M_{2}$ be two interactive Turing machines running a protocol. By $<M_{1}(x_{1}, r_{1}, z_{1}), M_{2}(x_{2}, r_{2}, z_{2})>(1^{k})$, we denote a running which starts with  $M_{i}$ holding a private input $x_{i}$, a randomness $r_{i}$, an auxiliary input $z_{i}$, the public security parameter $k$. By $IDN_{M_{1}}(<M_{1}(x_{1}, r_{1}, z_{1}), M_{2}(x_{2}, r_{2}, z_{2})>(1^{k}))$,
we denote the number of total direct deduction steps $M_{1}$ takes in the whole running. We say $M_{1}$ runs in expected polynomial-time with respect to $M_{2}$, if and only if there exists a polynomial $poly(.)$ such that for every $k\in\mathds{N}$, it holds that
\begin{multline*}
   \max(\{E_{R_{1},R_{2}}(IDN_{M_{1}}(<M_{1}(x_{1}, R_{1}, z_{1}),\\
    M_{2}(x_{2}, R_{2}, z_{2})>  (1^{k}))) | \\
     |x_{1}|=|x_{2}|=k,z_{1},z_{2}\in\{0,1\}^{*}\})    \leq poly(k)
\end{multline*}
where $ R_{1},R_{2}$ are random variables with uniform distribution over $\{0,1\}^{*}$.
\end{definition}

For Definition \ref{sec_def_ot}, it in fact requires that adversary $A$'s simulator $A'$ should run in expected polynomial-time with respect to TTP who computes $OT^{n}_{h}$'s functionality $f$.

We point out that the security definition presented in ~\cite{gold2001found,gold2004found,canetti2000security} requires the
simulator $A'$ to run in  strictly polynomial-time, but the one presented in
~\cite{canetti2004adaptive,lindell2007efficient,lindell2008efficient} allow $A'$ to run in
expected polynomial-time. Definition \ref{sec_def_ot} follows the latter. We argue that this is justified, since ~\cite{barak2004strict} shows that there is no (non-trivial) constant-round
zero-knowledge proof or argument having a strictly polynomial-time black-box simulator, which means allowing simulator to run in expected polynomial-time is essential for achieving constant-round protocols. See ~\cite{katz2008handling} for further discussion.

\subsection{Commitment Scheme}
In this section, we briefly introduce the cryptographic tool commitment scheme  which will be used in our framework. For the strict definition and the details, please see ~\cite{gold2001found} or ~\cite{goldreich1996zk}.

\begin{definition}[commitment scheme, non-strict description, ~\cite{gold2001found,goldreich1996zk}]
A  commitment scheme is a two-party protocol involving two phases.
\begin{itemize}
   \item Initial Inputs. At the beginning, all parties know the public security parameter $k$.
         The unbounded sender $P_{1}$ holds a randomness $r_{1}\in\{0,1\}_{}^{*}$, a
         value $m \in\{0,1\}_{}^{poly(k)}$ to be committed to, where the polynomial $poly(.)$
         is public. The PPT receiver $P_{2}$  holds a randomness $r_{2}\in
         \{0,1\}_{}^{*}$.
  \item  Commit Phase. $P_{1}$ computes a commitment, denoted $\alpha$, based on his knowledge, i.e.,   $\alpha \leftarrow P_{1}(1^{k}, m, r_{1})$, then $P_{1}$ send $\alpha$ to  $P_{2}$.
  \begin{itemize}
  \item The security for $P_{1}$ is implied by the property computationally hiding, which prevents
      $P_{2}$  from knowledge of the value committed by $P_{1}$. That is, for any  PPT  $P_{2}$, any $m_{1},m_{2} \in\{0,1\}_{}^{poly(k)}$,  it holds that
  \begin{multline*}
    \{ ViewC_{P_{2}}(<P_{1}(m), P_{2}>(1^{k})) \}_{k\in \mathds{N}} \\
    \stackrel{c}{=}
    \{ ViewC_{P_{2}}(<P_{1}(m'),P_{2}>(1^{k})) \}_{k\in \mathds{N}},
  \end{multline*}
where $ViewC_{P_{2}}(.)$ denotes $P_{2}$'s view  in  commit phase.
  \end{itemize}

   \item Reveal Phase. $P_{1}$ computes and sends a de-commitment, which typically consists of $m,r_{1}$, to $P_{2}$ to let $P_{2}$ know $m$. Receiving de-commitment, $P_{2}$ checks its validity. Typically $P_{2}$ checks that $\alpha = P_{1}(1^{k}, m, r_{1})$ holds. If de-commitment pass the check, $P_{2}$ knows and accepts $m$.
       \begin{itemize}
         \item  The security for $P_{2}$ is implied by the  property perfectly binding, which  guarantees that for any unbounded $P_{1}$, any $m_{1},m_{2} \in\{0,1\}_{}^{poly(k)}$ such that $m_{1}\neq m_{2}$, the probability that  $P_{2}$  accepts $m_{2}$ while $P_{1}$ commits to $m_{1}$ is zero, where  the probability is taken only over the randomness used by $P_{2}$.
       \end{itemize}
\end{itemize}
\end{definition}

The above definition defines perfectly binding commitment schemes (denoted by PBC). Relaxing the  property binding to allow the probability of  successful cheat of unbounded $P_{1}$ to be negligible, then the above definition defines  statically binding commitment schemes. Correspondingly, in the setting that $P_{1}$ is PPT and $P_{2}$ is unbounded, there exists
perfectly hiding commitment schemes (denoted by PHC) and statically hiding  commitment schemes, which provide perfectly hiding and statically hiding to $P_{1}$ respectively, and  only computationally binding to $P_{2}$.  If a property is secure against unbounded adversaries, we say this property is information-theoretically secure. We remark that there is no commitment scheme holding  both information-theoretically binding and information-theoretically hiding.

\section{A New  Smooth Projective Hash - $SPHDHC_{t,h}$} \label{hashing}
\subsection{The Definition Of $SPHDHC_{t,h}$}
In this section, we  define a new smooth projective hash --- $t$-smooth $h$-projective hash family that holds properties distinguishability,  hard subset membership, feasible cheating, denoted $SPHDHC_{t,h}$ for simplicity, which will be used to construct
our framework for $OT^{n}_{h}$. In section \ref{con_hash}, we instantiate $SPHDHC_{t,h}$ respectively under four distinct intractability assumptions.

Let us recall some related works before defining $SPHDHC_{t,h}$. ~\cite{carter1979universal,wegman1981new} present the classic notation of "universal hashing". Based on "universal hashing",  ~\cite{cramer2002universal} first introduces the concept of universal projective hashing, smooth projective hashing and hard subset membership problem in terms of languages and sets. In order to construct a framework for password-based authenticated key exchange, ~\cite{gennaro2006framework} modifies  such definition to some extent. That is, smoothness is defined over every instance of a language rather than a randomly chosen instance. ~\cite{kalai2005ot} refines the modified version in terms of the procedures used to implement it. What is more, a new requirement called verifiable smoothness is added to the hashing so as to construct a framework for $OT^{2}_{1}$. The resulting hashing is called verifiablely-smooth projective hash family that has hard subset membership property (denoted by $VSPHH$ for simplicity).   Note that, the framework presented by ~\cite{kalai2005ot} is not fully-simulatable.  The difference between $SPHDHC_{t,h}$ and the works mentioned above will be  under a  detailed  discussion after we define $SPHDHC_{t,h}$.

For clarity in presentation, we assume $n=h+t$ always holds and introduce additional notations.
Let $R=\{(x,w)|x,w \in \{0, 1\}^{*} \}$ be a relation, then $L_{R} \stackrel{def}{=}\{x|x \in \{0, 1\}^{*}, \exists w((x,w)\in R)\}$, $R(x) \stackrel{def}{=}\{w|(x,w) \in R\}$. $\Pi \stackrel{def}{=} \{ \pi| \pi:[n]\rightarrow [n], \pi\textrm{ is a permutation} \}$. Let $\pi \in \Pi$ (to comply with other literature, we also use $\pi$ somewhere to denote a protocol without bringing any confusion). Let $\vec{x}$ be an arbitrary vector.  By $\pi (\vec{x})$, we denote a vector resulted from applying  $\pi$ to $\vec{x}$. That is, $\vec{y}=\pi (\vec{x})$, if and only if $\forall i(i\in [d] \rightarrow \vec{x}\langle i \rangle = \vec{y} \langle \pi(i) \rangle) \wedge \forall i(i \notin [d] \rightarrow \vec{x}\langle i \rangle = \vec{y} \langle i \rangle)$ holds,  where $d \stackrel{def}{=} \min(\#\vec{x},n)$.

\begin{definition}[$t$-smooth $h$-projective hash family that holds properties distinguishability,  hard subset membership and feasible cheating]   \label{SPHDHC}
$\mathcal{H}=(PG, IS, DI, KG, Hash, pHash, Cheat)$ is an $t$-smooth $h$-projective hash family that holds properties distinguishability,  hard subset membership and feasible cheating ($SPHDHC_{t,h}$), if and only if $\mathcal{H}$ is specified as follows
\begin{itemize}
  \item The parameter-generator $PG$ is a  PPT algorithm that takes a security parameter $k$
        as input and outputs a family parameter $\Lambda$, i.e., $\Lambda \leftarrow PG(1^{k})$.
        $\Lambda$ will be used as a parameter to define three relations
        $R_{\Lambda}, \dot{R}_{\Lambda}\textrm{ and } \ddot{R}_{\Lambda}$, where $R_{\Lambda}=  \dot{R}_{\Lambda}\cup \ddot{R}_{\Lambda}$. Moreover,  $\dot{R}_{\Lambda} \cap \ddot{R}_{\Lambda}= \emptyset$  are supposed  to hold.

  \item The instance-sampler $IS$ is a PPT algorithm that takes a security parameter $k$,
        a family parameter $\Lambda$ as input and outputs a vector $\vec{a}$, i.e., $\vec{a}\leftarrow
        IS(1^{k},\Lambda)$.

        Let $\vec{a}=((\dot{x}_{1},\dot{w}_{1}), \ldots, (\dot{x}_{h},\dot{w}_{h}),
        (\ddot{x}_{h+1},\ddot{w}_{h+1}), \ldots, \\ (\ddot{x}_{n},\ddot{w}_{n}))^{T}$ be a vector generated by $IS$. We call
        each $\dot{x}_{i}$ or $\ddot{x}_{i}$  an instance of $L_{R_{\Lambda}}$. For each pair $(\dot{x}_{i},\dot{w}_{i})$ (resp.,   $(\ddot{x}_{i},\ddot{w}_{i})$), $\dot{w}_{i}$ (resp., $\ddot{w}_{i}$) is called a witness of $\dot{x}_{i}\in L_{\dot{R}_{\Lambda}}$
        (resp.,  $\ddot{x}_{i}\in L_{\ddot{R}_{\Lambda }})$.
        Note that, by this way we  indeed have defined the relationship $R_{\Lambda}, \dot{R}_{\Lambda}\textrm{ and } \ddot{R}_{\Lambda}$ here. The properties smoothness and projection we will mention later  make sure $\dot{R}_{\Lambda} \cap \ddot{R}_{\Lambda}= \emptyset$ holds.

        For simplicity in formulation later, we introduce some additional notations here.
        For $\vec{a}$ mentioned above,  $\vec{x}^{\vec{a}}\stackrel{def}{=} (\dot{x}_{1}, \ldots,
        \dot{x}_{h},\ddot{x}_{h+1}, \ldots, \ddot{x}_{n})^{T}$,
        $\vec{w}^{\vec{a}}\stackrel{def}{=} (\dot{w}_{1}, \ldots, \dot{w}_{h},\ddot{w}_{h+1},
        \ldots, \ddot{w}_{n})^{T}$. What is more, we abuse notation $\in$ to some extent. We write $\vec{x} \in Range( IS(1^{k},\Lambda))$ if and only if there exists a vector $\vec{x}^{\vec{a}}$ such that $ \vec{x}^{\vec{a}}= \vec{x} \textrm{ and } \vec{a} \in Range( IS(1^{k},\Lambda))$.  We write $x \in Range( IS(1^{k},\Lambda))$ if and only if there exists a vector $\vec{x}$ such that $ \vec{x} \in Range( IS(1^{k},\Lambda))$ and $x$ is an entry of $\vec{x}$.
  \item The distinguisher  DI is a PPT algorithm that takes a security parameter $k$, a family parameter $\Lambda$  and a pair strings $(x, w)$ as input and outputs an indicator bit $b$, i.e., $b \leftarrow DI(1^{k},\Lambda, x, w)$.
  \item The key generator $KG$ is a PPT algorithm that takes a security parameter $k$, a
        family
        parameter $\Lambda$ and an instance $x$ as input and outputs a hash key and a projection key, i.e., $(hk, pk) \leftarrow KG(1^{k}, \Lambda, x)$.
  \item The hash $Hash$ is a PPT algorithm that takes a security parameter $k$, a family
        parameter $\Lambda$, an instance $x$ and a hash key $hk$ as input and outputs a value $y$, i.e., $y \leftarrow Hash(1^{k}, \Lambda, x, hk)$.
  \item The projection $pHash$ is a PPT algorithm that takes a security parameter $k$, a family
        parameter $\Lambda$, an instance $x$, a witness $w$ of $x$  and a projection key $pk$ as input and outputs a value $y$, i.e., $y \leftarrow pHash(1^{k}, \Lambda, x, pk, w)$.
  \item The cheat $Cheat$ is a PPT algorithm  that  takes a security parameter $k$, a family
        parameter $\Lambda$ as input and outputs $n$ elements of $\dot{R}_{\Lambda}$, i.e., $((\dot{x}_{1},\dot{w}_{1}), \ldots (\dot{x}_{n},\dot{w}_{n}))\leftarrow Cheat(1^{k}, \Lambda)$.
\end{itemize}
and $\mathcal{H}$ has the following properties
\begin{enumerate}
  \item Projection. Intuitively speaking, it requires that for any instance $\dot{x} \in L_{\dot{R}_{\Lambda}}$, the hash value of $\dot{x}$ is obtainable with the help of its witness $\dot{w}$. That is, for any sufficiently large $k$, any $\Lambda \in Range(PG(1^{k}))$, any $(\dot{x},\dot{w})$ generated by $IS(1^{k},\Lambda)$, any $(hk,pk)\in Range(KG(1^{k},\Lambda, \dot{x}))$, it holds that
      \begin{equation*}
        Hash(1^{k}, \Lambda, \dot{x}, hk)=pHash(1^{k}, \Lambda, \dot{x},pk,\dot{w})
      \end{equation*}

  \item Smoothness. Intuitively speaking, it requires that for any instance vector $\vec{\ddot{x}} \in L_{\ddot{R}_{\Lambda}}^{t}$, the hash values of $\vec{\ddot{x}}$ are random and unobtainable unless their hash keys are known. That is, for any $\pi \in \Pi$, the two probability ensembles  $Sm_{1} \stackrel{def}{=} \{Sm_{1}(1^{k})\}_{k \in \mathds{N}}$ and $Sm_{2} \stackrel{def}{=} \{Sm_{2}(1^{k})\}_{k \in \mathds{N}}$, defined as follows, are computationally indistinguishable, i.e., $Sm_{1} \stackrel{c}{=} Sm_{2}$.

      $SmGen_{1}(1^{k})$: $\Lambda \leftarrow PG(1^{k})$, $\vec{a}\leftarrow IS(1^{k},\Lambda)$, $\vec{x} \leftarrow \vec{x}^{\vec{a}}$, for each $j \in [n]$ operates as follows: $(hk_{j},pk_{j})\leftarrow KG(1^{k},\Lambda, \vec{x}\langle j\rangle)$, $y_{j} \leftarrow Hash(1^{k}, \Lambda, \vec{x}\langle j\rangle, hk_{j})$, $\overrightarrow{xpky} \langle j \rangle \leftarrow (\vec{x}\langle j\rangle, pk_{j}, y_{j})$. Finally outputs $(\Lambda, \overrightarrow{xpky})$.

      $SmGen_{2}(1^{k})$: compared with $SmGen_{1}(1^{k})$, the only difference is that for each $j \in [n]-[h]$,  $y_{j} \in _{U} Range( Hash(1^{k}, \Lambda, \vec{x}\langle j\rangle, .))$.

      $Sm_{i}(1^{k})$: $(\Lambda, \overrightarrow{xpky}) \leftarrow SmGen_{i}(1^{k})$, $\widetilde{\overrightarrow{xpky}} \leftarrow \pi(\overrightarrow{xpky})$, finally outputs $(\Lambda, \widetilde{\overrightarrow{xpky}})$.
  \item Distinguishability.  Intuitively speaking, it requires that  the  DI can distinguish
  the projective instances and smooth instances with the help of their witnesses. That is,  it requires that  the  DI correctly computes the following function.
  \begin{gather*}
        \zeta :\mathds{N} \times (\{0,1\}^{*})^{3} \rightarrow \{0,1\} \\
          \zeta(1^{k},\Lambda,x, w)=  \begin{cases}
                                     0 & \textrm{ if }(x, w) \in \dot{R}_{\Lambda},\\
                                     1 & \textrm{ if }(x, w) \in \ddot{R}_{\Lambda},\\
                                     \textrm{undefined} & \textrm{ otherwise }.
                                   \end{cases}
  \end{gather*}
  \item Hard Subset Membership. Intuitively speaking, it requires that for any $\vec{x} \in Range( IS(1^{k},\Lambda))$,
         $\vec{x}$ can be disordered without being detected. That is,  for any $\pi \in \Pi$, the two probability ensembles $HSM_{1} \stackrel{def}{=} \{HSM_{1}(1^{k})\}_{k\in \mathds{N}}$ and $HSM_{2} \stackrel{def}{=} \{HSM_{2}(1^{k})\}_{k\in \mathds{N}}$,  specified as follows, are computationally indistinguishable, i.e., $HSM_{1} \stackrel{c}{=}HSM_{2}$.\\
            $HSM_{1}(1^{k})$: $\Lambda \leftarrow PG(1^{k})$, $\vec{a}\leftarrow IS(1^{k},\Lambda)$, finally outputs $(\Lambda, \vec{x}^{\vec{a}})$.\\
            $HSM_{2}(1^{k})$: Operates as same as $HSM_{1}(1^{k})$ with an exception that finally outputs $(\Lambda, \pi(\vec{x}^{\vec{a}}))$.

  \item  Feasible Cheating.  Intuitively speaking, it requires that there is a way to cheat to generate a $\vec{x}$ which is supposed to fall into $L_{\dot{R}_{\Lambda}}^{h}\times L_{\ddot{R}_{\Lambda}}^{t} $ but actually falls into  $L_{\dot{R}_{\Lambda}}^{n}$ without being caught.
  That is,   for any $\pi \in \Pi$, for any $\pi' \in \Pi$, the two probability
         ensembles $HSM_{2}$ and $HSM_{3}\stackrel{def}{=}\{HSM_{3}(1^{k})\}_{k\in \mathds{N}}$ are computationally indistinguishable, i.e., $HSM_{2} \stackrel{c}{=}HSM_{3}$, where $HSM_{2}$ is defined above and $HSM_{3}$ is defined as follows.

         $HSM_{3}(1^{k})$:$\Lambda \leftarrow PG(1^{k})$, $\vec{a}\leftarrow Cheat(1^{k})$, finally outputs $(\Lambda, \pi'(\vec{x}^{\vec{a}}))$.
\end{enumerate}
\end{definition}

\begin{remark}[The Witnesses Of The Instances]
The main use of the witnesses of an instance $\dot{x} \in L_{\dot{R}_{\Lambda}}$ is to project and gain the hash value of $x$. In contrast, with respect to an instance $\ddot{x} \in L_{\ddot{R}_{\Lambda}}$, it services  as a proof of $\ddot{x} \in L_{\ddot{R}_{\Lambda}}$. The property distinguishability guarantees that given the needed witness, the projective instances and the smooth instances are distinguishable.
For $OT^{n}_{h}$, this means that a receiver can use the witnesses of $\ddot{x}$ to persuade a sender to believe that the receiver is unable to gain the hash value of $\ddot{x}$.
\end{remark}

\begin{remark}[Hard Subset Membership] \label{rehsm}
The property  hard subset membership guarantees that for any $\vec{x} \in Range(IS(1^{k}, \Lambda))$, any $\pi \in \Pi$, any PPT adversary $A$, the advantage of $A$ identifying an entry of $\pi(\vec{x})$ falling into $L_{\dot{R}_{\Lambda}}$ (resp., $L_{\ddot{R}_{\Lambda}}$) with probability over prior knowledge
$h/n$ (resp.,  $t/n$) is negligible. That is, seen from $A$, every entry of $\pi(\vec{x})$ seems the same.

With respect to $OT^{n}_{h}$, this means that the receiver can encode his private input into a permutation of a vector $\vec{x} \in L_{R_{\Lambda}}^{n}$ without leaking any information. For example, if the receiver expects to gain $\vec{m}\langle H \rangle$, then he may generates a $\vec{x}$ and randomly chooses a permutation $\pi \in \Pi$ such that $\pi(\vec{x}) \langle i \rangle \in L_{\dot{R}_{\Lambda}}$ for each $i \in H$.   Any PPT adversary knows no new knowledge  about $H$ if only given $\pi(\vec{x})$.

However, if the witnesses of the instances of  $\vec{x}$ are available (the simulator can gain the witnesses by rewinding the adversary), then the receiver's input is known.
Therefore, there is way for the simulator to extract the real input of the adversary controlling the corrupted receiver.
\end{remark}

\begin{remark}[Feasible Cheating] \label{reCheat}
In our framework for $OT^{n}_{h}$, the sender uses the hash values of the instances generated by the receiver to encrypt its private inputs. The property feasible cheating makes cheating out of the sender's  all private inputs feasible. Note that, this is a key for the simulator to extract the real inputs of the adversary controlling the corrupted sender. Therefore, it is conductive to construct a fully-simulatable protocol for $OT^{n}_{h}$.
\end{remark}

\subsection{The Difference Between $SPHDHC_{t,h}$ And Related Hash Systems} \label{comparsph}
Now we discuss the difference between our $SPHDHC_{t,h}$ and related  hash systems previous works present or use. For simplicity, we only compare our $SPHDHC_{t,h}$ with the hash system  $VSPHH$  which is presented by ~\cite{kalai2005ot}. We argue that this is justified, on the one hand, the version of ~\cite{kalai2005ot} is the version holding most properties  among previous works. On the other hand, the aim  of ~\cite{kalai2005ot} is the closest to ours. They aim to construct a framework for $OT^{2}_{1}$ which actually is half-simulatable, while we aim to  establish  a fully-simulatable framework for  $OT^{n}_{h}$.

Loosely speaking, our $SPHDHC_{t,h}$ can be viewed as a generalized version of $VSPHH$. Indeed, $VSPHH$ resembles $SPHDHC_{1,1}$ very much and can be converted into $SPHDHC_{1,1}$ though some  modification is needed. The essential differences are listed as follows.
\begin{enumerate}
  \item The key difference is that, besides each projective instance $\dot{x}$  holding a witness $\dot{w}$,  $SPHDHC_{t,h}$ also requires each smooth instance $\ddot{x}$ to hold a witness  $\ddot{w}$.
  \item To deal with $OT^{n}_{h}$, $SPHDHC_{t,h}$ extends the $IS$ algorithm to generate $h$ $\dot{x}$s and $t$ $\ddot{x}$s in a invocation. As a natural result, $SPHDHC_{t,h}$ extends the property smoothness to hold with respect to  $t$ $\ddot{x}$s, and extends the property  hard subset membership to hold with respect to  $h$ $\dot{x}$s  and $t$ $\ddot{x}$s.

  \item In $VSPHH$ there exists a instance test $IT$ algorithm that takes  two instances as input and outputs a bit indicating whether at least one of the two instances is  smooth, i.e., $b\leftarrow IT(x_{1},x_{2})$.
      $SPHDHC_{t,h}$ discards this verifiability of smoothness and the correlated $IT$, and instead provides a distinguisher $DI$  algorithm which is conducive to apply the technique cut-and-choose.
  \item $SPHDHC_{t,h}$ requires a additional property feasible cheating and the necessary algorithm $Cheat$. This property  provides a simulator with  a way  to extract the real inputs of the adversary in the case that the sender is corrupted.
  \item $SPHDHC_{t,h}$ extends $KG$ algorithm such that the information of the instance is available to it. This makes constructing hash system  easier. In indeed, this makes lattice-based hash system come true which  is thought difficult by \cite{lindell2008efficient}.
\end{enumerate}

We observe that the $VSPHH$ indeed is easy to be extended to deal with $OT^{n}_{1}$, but seems difficult to be extended to deal with the  general $OT^{n}_{h}$. The reason is that, to hold verifiable smoothness, $\dot{x}$s and $\ddot{x}$s  have to be generated in a dependent way. This makes designing $IT$ dealing with $n$ instances  without leaking information which is conductive to distinguish such $\dot{x}$s and $\ddot{x}$s difficult. Therefore, even constructing a  framework for  $OT^{n}_{h}$  that is
half-simulatable as  \cite{kalai2005ot} seems impossible. We also observe that, there is no way to construct a fully-simulatable framework using $VSPHH$, because there is no way to extract
the real input of the adversary in the case that the receiver is corrupted.

The difficulties mentioned above can be overcame by requiring each $\ddot{x}$ to hold a  witness too. Since  the receiver encodes his  input  as a permutation of $\dot{x}$s and $\ddot{x}$s, a  simulator can the extract the real input of the adversary in the case that the receiver is corrupted if their witnesses are available. Combining the  application of the technique cut-and-choose, a simulator can see such witnesses by rewinding the adversary. What is more, the implementation of $DI$ is easier than that of its predecessor $IT$.  Because the operated object essentially is a pair of the form $(x,w)$ which is simpler than  $(x_{1},\ldots , x_{n})$  which is the general form of the objects operated by $IT$.

\section{Constructing A Framework For Fully-simulatable $OT^{n}_{h}$ }\label{ot}
In this section, we construct a framework for $OT^{n}_{h}$. In the framework, we will use
a PPT algorithm, denoted $\Gamma$ , that receiving  $B_{1}, B_{2} \in \Psi$, outputs a uniformly chosen  permutation $\pi \in_{U}  \Pi$ such that $\pi(B_{1}) = B_{2}$, i.e.,  $\pi \leftarrow  \Gamma(B_{1}, B_{2})$. We give an example implementation of $\Gamma$ as follows.

$\Gamma(B_{1}, B_{2})$: First, $E \leftarrow \emptyset$, $C \leftarrow [n] - B_{1}$. Second, for each $j \in B_{2}$, then $i \in_{U} B_{1}$, $B_{1} \leftarrow B_{1}-\{ i \}$, $E \leftarrow E \cup \{ j \rightleftharpoons i\}$. Third, $D \leftarrow [n] - B_{2}$, for each $j \in D$, then $i \in_{U} C$, $C \leftarrow C-\{ i \}$, $E \leftarrow E \cup \{ j \rightleftharpoons i\}$.
Fourth, define $\pi$ as $\pi(i)= j$ if and only if $j \rightleftharpoons i  \in  E$. Finally,  outputs $\pi$.

\subsection{The Framework For $OT^{n}_{h}$}
\begin{itemize}
  \item Common inputs: All entities know the public security parameter $k$, an positive polynomial
        $poly_{s}(.)$, a  $SPHDHC_{t,h}$ (where $n=h+t$) hash system $\mathcal{H}$,  a information-theoretically hiding commitment
        scheme (denoted by IHC), a information-theoretically binding commitment scheme (denoted by IBC).
  \item Private Inputs: Party $P_{1}$ (i.e., the sender) holds a private input $\vec{m}  \in (\{0,1\}^{*})^{n}$ and a randomness $r_{1}\in \{0,1\}^{*}$. Party $P_{2}$ ( i.e., the  receiver) holds a private input $H \in \Psi$ and a randomness $r_{2}\in \{0,1\}^{*}$. The adversary $A$ holds a name list $I\subseteq [2]$ and
        a  randomness $r_{A}\in\{0,1\}^{*}$.
  \item Auxiliary Inputs: The adversary $A$ holds an infinite auxiliary input sequence
        $z=(z_{k})_{k\in \mathds{N}},z_{k} \in \{0,1\}^{*}$.
\end{itemize}

The protocol works as follow. For clarity, we omit some trivial error-handlings such as $P_{1}$ refusing to send  $P_{2}$ something which is supposed to be sent. Handling such errors is easy.  $P_{2}$ halting and outputting $abort_{1}$ suffices.

\begin{itemize}
    \item Receiver's step (R1): \label{gen_para} $P_{2}$ generates hash parameters
         and samples instances.
         \begin{enumerate}
          \item $P_{2}$ samples $poly_{s}(k)$ instance vectors. Let $K \stackrel{def}{=} poly_{s}(k)$. $P_{2}$ does: $\Lambda  \leftarrow PG(1^k )$; for each $i \in [K]$, $\vec{a}_{i} \leftarrow IS(1^k,\Lambda)$. Without
             loss of generality, we assume
             $\vec{a}_{i}=((\dot{x}_{1},\dot{w}_{1}),\ldots,(\dot{x}_{h},\dot{w}_{h}),
             (\ddot{x}_{h+1},\ddot{w}_{h+1}),\ldots,\\(\ddot{x}_{n},\ddot{w}_{n}))^{T}$.
          \item $P_{2}$ disorders each instance vector.

           For each
             $i \in [K]$,
            $P_{2}$ uniformly chooses a permutation $\pi^{1}_{i} \in_{U} \Pi$,
            then $ \tilde{\vec{a}}_{i} \leftarrow \pi^{1}_{i}(\vec{a}_{i})$.
          \item $P_{2}$ sends the instances and the corresponding hash parameters, i.e., $(\Lambda,\tilde{\vec{x}}_{1},\tilde{\vec{x}}_{2},\ldots,\tilde{\vec{x}}_{K})$,
             to  $P_{1}$, where $\tilde{\vec{x}}_{i} \stackrel{def}{=}
             \vec{x}^{\tilde{\vec{a}}_{i} }$ (correspondingly,  $\tilde{\vec{w}}_{i}
             \stackrel{def}{=}  \vec{w}^{\tilde{\vec{a}}_{i} }$).
         \end{enumerate}
   \item Receiver's step (R2-R3)/Sender's step (S1-S2):
         \label{tossing coin} $P_{1}$ and $P_{2}$ cooperate to toss coin to choose
         instance vectors to open.
         \begin{enumerate}
         \item $P_{1}$: $s\in_{U} \{0,1\}^{K}$, sends $IHC(s)$ to $P_{2}$.
         \item \label{tossing coin_p1_f}
          $P_{2}$: $s' \in_{U} \{0,1\}^{K}$, sends $IBC(s')$ to $P_{1}$.
         \item $P_{1}$ and $P_{2}$ respectively sends each other the
          de-commitments
          to $IHC(s)$ or $IBC(s')$, and respectively checks the received de-commitments are valid. If the check fails, $P_{1}$ ($P_{2}$ respectively) halts  and outputs $abort_{2}$ ($abort_{1}$ respectively). If no check fails, then they proceed
          to next step.
          \item $P_{1}$ and $P_{2}$ share a common randomness $r=s \oplus s'$.
          The instance vectors  whose index fall into $CS
            \stackrel{def}{=} \{i|r\langle i \rangle =1, i\in[K]\}$ (correspondingly,
             $\overline{CS}\stackrel{def}{=}[K]-CS$)  are chosen to open.
          \end{enumerate}
  \item Receiver's step (R4):
        $P_{2}$ opens the chosen instances to $P_{1}$, encodes and sends his private input to $P_{1}$.
        \begin{enumerate}
         \item $P_{2}$ opens the chosen instances to prove that the instances he generates are legal.

             $P_{2}$ sends $((i,j,\tilde{\vec{w}}_{i}\langle j \rangle))_{i\in CS, j  \in J_{i}}$ to  $P_{1}$,  where
             $J_{i} \stackrel{def}{=}\{j| \tilde{\vec{x}}_{i}\langle j \rangle \in L_{\ddot{R}_{\Lambda}}, j\in [n] \}$.
         \item $P_{2}$ encodes his private input and sends the resulting code to
             $P_{1}$.

             Let $G_{i} \stackrel{def}{=}\{j|\tilde{\vec{x}}_{i}\langle
              j\rangle \in  L_{\dot{R}_{\Lambda}}, i\in \overline{CS} \}$. For each
              $i\in \overline{CS}$, $P_{2}$ does $\pi^{2}_{i} \leftarrow
             \Gamma(G_{i},H)$, sends $(\pi^{2}_{i})_{i
              \in\overline{CS}}$  to $P_{1}$. That is,
              $P_{2}$ encode his private input into sequences such as
              $\pi^{2}_{i}(\tilde{\vec{x}}_{i})$ where $i\in \overline{CS}$.
       \end{enumerate}
   Note that  $P_{2}$ can send $((i,j,\tilde{\vec{w}}_{i}\langle j
   \rangle))_{i\in CS, j\in J_{i}}$ and
   $(\pi^{2}_{i})_{i \in \overline{CS}}$ in one step.

   \item Sender's step (S3): $P_{1}$ checks the chosen instances, encrypts and sends his
         private input to $P_{2}$.
       \begin{enumerate}
        \item \label{pro_p1_check}
           $P_{1}$ verifies that each chosen instance vectors is legal, i.e., the number of the entries belonging
           to $L_{\dot{R}_{\Lambda}}$ is not more than $h$.

           $P_{1}$ checks
           that, for each $i\in CS$, $\#J_{i}\geq n-h$,  and for each $j\in
           J_{i}$, $VF(1^{k},\Lambda,\tilde{\vec{x}}_{i}\langle j\rangle,
           \tilde{\vec{w}}_{i}\langle j \rangle)$ is $1$. If the check fails, $P_{1}$
           halts and outputs $abort_{2}$, otherwise $P_{1}$ proceeds to next step.
        \item $P_{1}$ reorders the entries of each unchosen instance vector in
           the way told by $P_{2}$.

           For each $i\in \overline{CS}$, $P_{1}$ does
           $\tilde{\tilde{\vec{x}}}_{i}\leftarrow \pi^{2}_{i}
           (\tilde{\vec{x}}_{i})$.
         \item $P_{1}$ encrypts and sends his private input to $P_{2}$ together with some auxiliary messages.

             For each  $i\in
            \overline{CS}$, $j
          \in [n]$,  $P_{1}$ does: $(hk_{ij},pk_{ij} )
           \leftarrow  KG(1^k
           ,\Lambda, \tilde{\tilde{\vec{x}}}_{i}\langle j\rangle)$,
           $\beta_{ij}  \leftarrow
            Hash(1^k ,\Lambda,\tilde{\tilde{\vec{x}}}_{i}\langle j\rangle, hk_{ij})$,
          $\vec{ \beta_{i} } \stackrel{def}{=} (\beta _{i1},
          \beta _{i2},\dots,\beta_{in})^{T}$,$\vec{c}
          \leftarrow \vec{m}\oplus (\oplus_{i \in \overline{CS}} \vec {\beta
          _i})$, $\overrightarrow{pk}_{i}\stackrel{def}{=}
          (pk_{i1},pk_{i2},\ldots,pk_{in})^{T}$,
          sends $\vec{c}$ and $(\overrightarrow{pk}_{i})_{i \in \overline{CS}
           }$  to $P_{2}$.
        \end{enumerate}
   \item  Receiver's step (R5): $P_{2}$ decrypts the ciphertext $\vec{c}$ and gains the message he want.

   For each $i\in \overline{CS}$, $j \in H$, $P_{2}$ operates:
   $\beta'_{ij} \leftarrow pHash(1^k ,\Lambda ,\tilde{\tilde{ \vec{ x}}}_i
   \langle j\rangle, \overrightarrow{pk}_{i} \langle j\rangle,
    \tilde{\tilde{\vec{w}}}_i \langle j\rangle)$, $m'_{j} \leftarrow
   \vec{c}\langle
   j\rangle \oplus (\oplus_{i \in \overline {CS} } \beta'_{ij})$. Finally,
    $P_{2}$ gains   the messages $(m'_{j})_{j\in H}$.
  \end{itemize}

\subsection{The Correctness Of The Framework}
Now let us check the correctness of the framework, i.e., the framework works in the case that $P_{1}$ and $P_{2}$ are honest. For each $i\in \overline {CS}$, $j\in H$, we know
\begin{gather*}
    \vec{c}\langle j\rangle = \vec{m}\langle j\rangle \oplus (\oplus_{i \in
               \overline{CS}} \vec {\beta_i}\langle j\rangle) \\
   m'_{j} = \vec{c}\langle j\rangle \oplus (\oplus_{i \in \overline {CS} }
                     \beta'_{ij})
\end{gather*}
Because of the projection of $\mathcal{H}$, we know
\begin{equation*}
    \vec {\beta_i}\langle j\rangle = \beta'_{ij}
\end{equation*}
So we have
\begin{equation*}
     \vec{m}\langle j\rangle = m'_{j}
\end{equation*}
This means what $P_{2}$ gets is $\vec{m}\langle H \rangle$ that indeed is $P_{2}$ wants.

\subsection{The Security Of The Framework}
With respect to the security of the framework, we have the  following theorem.

\begin{theorem}[The protocol is secure against the malicious adversaries]\label{the_sec}
Assume that $\mathcal{H}$ is an $t$-smooth $h$-projective hash family that holds properties distinguishability,  hard subset membership and feasible cheating,  $IHC$ is a information-theoretically hiding commitment, $IBC$ is a information-theoretically  binding commitment. Then, the protocol securely computes the oblivious transfer
functionality in the presence of  non-adaptive malicious adversaries.
\end{theorem}

We defer the strick proof of Theorem \ref{the_sec} to section \ref{proof} and first give an intuitive analysis here as a warm-up. For the security of $P_{1}$, the framework should prevent $P_{2}$ from  gaining more than $h$ messages. Using cut and choose technique, $P_{1}$ makes sure with some probability that each instance vector contains no more than $h$ projective instance, which leads to  $P_{2}$ learning extra messages is difficult. The following theorem guarantees that this probability is overwhelming.

\begin{theorem}\label{p2cheat}
Assume that the commitment schemes  employed in the framework are a perfectly hiding commitment and a perfectly binding commitment. Then, in  case that  $P_{1}$ is honest and $P_{2}$ is corrupted, the probability that $P_{2}$ cheats to obtain more than $h$ messages is at most $1/2^{poly_{s}(k)}$.
\end{theorem}

\begin{IEEEproof}
According to the framework, there are two necessary conditions for $P_{2}$'s success in the cheating.
\begin{enumerate}
  \item $P_{2}$ has to generate at least one illegal  $\vec{x}_{i}$ which contains more
         than $h$ entries belonging to $L_{\dot{R}_{\Lambda}}$. If not, $P_{2}$ cann't correctly decrypt more than  $h$ entries of $\vec{c}$, because of
         the smoothness of $\mathcal{H}$. Without loss of generality, we assume the illegal
         instance vectors are $\vec{x}_{l_{1}},\vec{x}_{l_{2}},\ldots,\vec{x}_{l_{d}}$.
  \item   All illegal instance vectors are lucky not to be chosen and all the instance vectors unchosen just are  the illegal instance vectors,  i.e., $\overline{CS}
         = \{l_{1},l_{2},\ldots,l_{d}\}$. We prove this claim  in two case.
         \begin{enumerate}
          \item  In the case that  $\overline{CS} \neq \{l_{1},l_{2},\ldots,l_{d}\}$ and
            $\overline{CS} - \{l_{1},l_{2},\ldots,l_{d}\} = \emptyset $, there exists
            $j(j \in [d] \wedge l_j \in CS)$. So $P_{1}$ can detect $P_{2}$'s cheating and  $P_{2}$ will gain nothing.
         \item   In the case that $\overline{CS} \neq \{l_{1},l_{2},\ldots,l_{d}\}$ and
            $\overline{CS} - \{l_{1},l_{2},\ldots,l_{d}\} \neq \emptyset $, there exists  $j(j \in \overline{CS}\wedge \vec{x}_{j} \textrm{ is legal})$.  Because of the smoothness of $\mathcal{H}$, $P_{2}$ cannot correctly decrypt   more than $h$ entries of $\vec{c}$.
        \end{enumerate}
\end{enumerate}

Now, let us estimate the probability that the second necessary condition is met.
Note that, $IHC(s)$ is a perfectly hiding commitment, $IBC(s')$ is a perfectly binding commitment, and $P_{1}$ is honest, so the shared
randomness $r$ is uniformly distributed. We have
\begin{equation*}
    \begin{split}
       Pr(\overline{CS}= \{l_1 ,l_2 , \ldots ,l_d \})& = (1/2)^d (1/2)^{poly_{s}(k)- d}  \\
                                     &=1/2^{poly_{s}(k)}
     \end{split}
\end{equation*}
This means that the probability that $P_{2}$ cheats to obtain more than $h$ messages
is at most $1/2^{poly_{s}(k)}$.
\end{IEEEproof}
From the proof of Theorem \ref{p2cheat}, it is easy to see that if the commitment schemes  employed  are
the ones with statically properties,  the probability that $P_{2}$ cheats to obtain more than $h$ messages  is negligible too, since the upper-bound of this probability deviates $1/2^{poly_{s}(k)}$ at most negligible distance.

For the security of $P_{2}$, the framework first should prevent $P_{1}$ from learning $P_{2}$'s private input. There is a potential risk in Step R4 where $P_{2}$ encodes
his private input. From Remark \ref{rehsm}, we know that hard subset membership guarantees that for any PPT  malicious $P_{1}$, without being given $\pi^{1}_{i}$, the probability that  $P_{1}$ learns any new knowledge is negligible. Thus $P_{2}$'s encoding is safe. Besides cheating $P_{2}$ of private input, it seems there is another obvious attack
that malicious $P_{1}$ sends invalid messages, e.g. $pk_{ij}$ which  $(hk_{ij},pk_{ij})\notin Range(KG(1^{k},\Lambda,x_{ij}))$, to $P_{2}$. This attack in fact doesn't matter. Its effect is equal to that of $P_{1}$'s altering his real input, which is allowed in the ideal world too.

\subsection{The Communication Rounds}\label{round}
Step R1 and Step R2 can  be taken in one round. Step R5 is  taken without communication. Each of other steps is taken in one round. Therefore, the total number of the communication rounds is six.

Compared with existing  fully-simulatable protocols for oblivious transfer that without resorting to a random oracle or a trusted common reference string (CRS), our protocol is the most efficient one. On counting the total communication rounds of a protocol, we  count that of the modified version. In the modified version,  the  consecutive  communications of the same direction are   combined into one round. The protocol for $OT^{n}_{h\times 1}$ of ~\cite{camenisch2007simulatable} costs one, two zero-knowledge proofs of knowledge respectively in initialization and in transfer a message, where each zero-knowledge proofs of knowledge is performed in four rounds.   The whole protocol costs at least ten rounds. The protocol for $OT^{n}_{h}$ of ~\cite{green2007blind} costs one zero-knowledge proof of knowledge in
initialization which is performed in  three rounds at least, one protocol to extract a secret key corresponding to the identity of a message which is performed in four rounds, one zero-knowledge proof of knowledge in transfer a message which is performed in  three rounds at least. We point out  that the interactive proof of knowledge of a  discrete logarithm modulo a prime, presented by ~\cite{Schnorr1991signature}  and taken as a zero-knowledge proof of knowledge protocol  in ~\cite{green2007blind}, to our best knowledge,  is not known to be zero-knowledge. However, turning to  the techniques of $\Sigma$-protocol, \cite{cramer2000efficient} make it zero-knowledge at cost of increment
of three rounds in communication,  which in turn induces the increment in communication rounds  of the protocol of  ~\cite{green2007blind}. Taking all into consideration, this  protocol costs at least ten rounds.
The protocol for $OT^{2}_{1}$ of ~\cite{lindell2008efficient} costs six rounds.

\subsection{The Computational Overhead}\label{overhead}
We measure the computational overhead of the framework in terms of the number of public key operations (i.e., operations based on trapdoor functions, or similar operations)
, because the overhead of public key operations, which depends on the length of their inputs, is greater than that of symmetric key operations (i.e., operations based on one-way functions) by orders of magnitude. Please see \cite{lu2009secloss} to know which cryptographic operation is public key  operation or private key operation.

As to the framework, the public key operations are $Hash(.)$ and  $pHash(.)$, and the symmetric key operations are $IHC(.)$ and  $IBC(.)$.  In Step S3, $P_{1}$ takes $n\cdot \#\overline{CS}$ invocations of $Hash(.)$ to encrypt his private input. In Step R5, $P_{2}$ takes $h\cdot \#\overline{CS}$ invocations of $pHash(.)$ to decrypt the messages he want. The value of $\#\overline{CS}$ is $poly_{s}(k)$, $poly_{s}(k)/2$, respectively, in the worst case and  in the average case.
Thus, fixing the problem we tackle (i.e., fixing the values of $n$ and $h$), the efficiency only depends on the value of $poly_{s}(k)$. In Section \ref{proof} where we strictly prove the security of the framework, we'll see that in the case that only $P_{2}$ is corrupted,  our simulator doesn't consider a situation in the real world that arises with probability at most $1/2^{poly_{s}(k)}$. Therefore, setting $poly_{s}(k)=40$ is secure enough to use our framework in practice. In the worst case the computational overhead mainly   consists of $40n$ invocations of $Hash()$ taken by $P_{1}$ and $40h$ invocations of $pHash()$ taken by $P_{2}$; in the average case the computational overhead mainly   consists of $20n$ invocations of $Hash()$ taken by $P_{1}$ and $20h$ invocations of $pHash()$ taken by $P_{2}$.

We point out that, our simulator also may fail (with negligible probability) in the case that $P_{1}$ is corrupted, but the probability of this event arising
depends on the computational hiding of IBC and on the computational binding of IHC rather than
the value of $poly_{s}(k)$ and has no influence on computational overhead. So we don't need to  take this case into consideration here.

Compared with  existing fully-simulatable  protocols for oblivious transfer that without resorting to a random oracle or a trusted CRS, our DDH-based instantiation that will be presented in Section \ref{hash_ddh} is the most efficient one in computational overhead.
The operations of the protocol in \cite{camenisch2007simulatable} are based on the non-standard assumptions, i.e., $q$-Power Decisional Diffie-Hellman and $q$-Strong Diffie-Hellman (q-SDH) assumptions,  which both are
associated with bilinear groups. \cite{cheon2006security} indicates that  q-SDH-based operations are more expensive that standard-assumption-based operations. The operations of the protocol in \cite{green2007blind} are based on  Decisional Bilinear Diffie-Hellman (DBDH) assumption.  Since bilinear curves are considerably more expensive than regular Elliptic curves \cite{galbraith2008pairings} and DDH is obtainable from Elliptic curves, the  operations in \cite{camenisch2007simulatable,green2007blind}  are  considerably more expensive than that DDH-based operations.  Therefore,  our DDH-based instantiation  are more  efficient than the protocols presented by \cite{camenisch2007simulatable,green2007blind}.  The  DDH-based protocol  for $OT^{2}_{1}$ presented by \cite{lindell2008efficient} also are very efficient. However,  it can be viewed as a  specific case of our framework, thought some modification of the protocol is needed.

We have to admit that, in the context of a trusted CRS is available and only
$OT^{2}_{1}$ is needed, \cite{peikert2008ot}'s  DDH-based instantiation,  which is two-round efficient and of two public key encryption operations and one public key decryption operation, is the most efficient one,  no matter seen from the number of communication rounds or the computational overhead.

\section{A Security Proof Of The Framework} \label{proof}
We prove Theorem \ref{the_sec} holds in this section.  For notational clarity, we denote the entities,
the parties and the adversary in the real world by $P_{1}$, $P_{2}$, $A$, and denote the corresponding entities in the ideal world by $P_{1}'$, $P_{2}'$, $A'$. In the light of the parties being corrupted, there are four cases to be considered and we prove Theorem \ref{the_sec} holds in each case. For simplicity,  we assume that the commitment schemes employed are a perfectly binding commitment scheme and a perfectly hiding
commitment scheme. If the statically ones are employed, the proof can be done in the same way with a slight modification.

We don't know how to construct a strictly polynomial-time simulator for the adversary in the real world, in the case that only $P_{1}$  or $P_{2}$ is corrupted. Instead, expected polynomial-time simulators are constructed (see section \ref{secdef} for the justification), which results in a failure  of standard black-box reduction technique. Fortunately, the problem and its derived problems can be solved  using the technique given by \cite{goldreich1996zk}.

\subsection{In the case that $P_{1}$ Is Corrupted}
In the case that $P_{1}$ is corrupted, $A$ takes the full control of $P_{1}$ in the real world. Correspondingly, $A$'s simulator,  $A'$, takes the full control of $P_{1}'$ in the ideal world, where $A'$ is constructed as follow.

\begin{itemize}
  \item Initial input: $A'$ holds   the same $k$, $I \stackrel{def}{=}\{1\}$, $z=(z_{k})_{k\in \mathds{N}}$, as   $A$.  What is more, $A'$ holds a uniform distributed randomness $r_{A'}\in\{0,1\}^{*}$.   The parties $P_{1}'$ and $P_{1}$, whom $A'$ and $A$ respectively is to corrupt, hold the same $\vec{m}$.
  \item $A'$ works as follows.
  \begin{itemize}
    \item Step Sm1:
          $A'$ corrupts  $P_{1}'$ and learns $P_{1}'$'s private input $\vec{m}$.
          Let $\bar{A}$ be a copy of $A$, i.e., $\bar{A}=A$. $A'$ use $\bar{A}$ as
          a subroutine. $A'$ fixes the initial inputs of $\bar{A}$ to be identical to
          his except that fixes the randomness of $\bar{A}$ to be a uniformly distributed value. $A'$ activates $\bar{A}$,  and  supplies $\bar{A}$  with $\vec{m}$ before $\bar{A}$ engages in the protocol for $OT^{n}_{h}$.

          In the following steps, $A'$  builds an environment for $\bar{A}$ which simulates the real world. That is,  $A'$ disguises himself as  $P_{1}$ and $P_{2}$ at the same time to interact with  $\bar{A}$.

   \item  Step Sm2:
          $A'$ uniformly chooses a randomness $r\in_{U} \{0,1\}^{K}$ ($K \stackrel{def}{=}poly_{s}(k)$) as the shared randomness. Let $CS$ and $\overline{CS}$ be the sets decided by $r$.  For each $i \in CS$, $A'$ honestly generates the hash parameters and
          instance vectors. For each $i \in \overline{CS}$, $A'$ calls $Cheat(1^{k})$  to generate such  parameters and vectors. $A'$ sends these  hash parameters and instance vectors to $\bar{A}$.

\begin{remark}
From the remark \ref{rehsm}, we know that each entry of the instance vector generated by $Cheat(1^{k})$ is projective. If such instance vectors are not chosen to be open, then
the probability of $\bar{A}$ detecting this fact is negligible, and $A'$ can extract the real input of $\bar{A}$, which is we want.
\end{remark}

   \item Step Sm3:
         $A'$  plays the role of $P_{2}$ and executes Step R2-R3 of the framework to cooperate with $\bar{A}$ to toss coin.  When tossing coin is completed successfully, $A'$ learns and records the value $s$ $\bar{A}$ commits to.

\begin{remark}
The aim of doing this tossing coin  is to know the randomness $s$
$\bar{A}$ choses. What $A'$ will do next is to take $IBC(r \oplus s)$ as his commitment to redo tossing coin.
\end{remark}

   \item  Step Sm4:  $A'$ repeats the following procedure, denoted $\Upsilon$, until $\bar{A}$ correctly reveals the recorded value $s$.

          $\Upsilon$: $A'$ rewinds  $\bar{A}$ to the end of Step S1 of the framework. Then, taking $IBC_{\gamma}(r \oplus s)$ as his commitment, $A'$ executes Step R2 and R3 of the framework, where $\gamma$ is a fresh randomness uniformly chosen.

\item  Step Sm5: Now $A'$ and $\bar{A}$ shares the common randomness $r$.
        $A'$ executes Step R4 of the framework as the honest $P_{2}$ do.
        On receiving $\vec{c}$ and $(\overrightarrow{pk}_{i})_{i \in
        \overline{CS}}$,  $A'$ correctly decrypts all entries of
        $\vec{c}$ and gains  $\bar{A}$'s full real private input $\vec{m}$.
        Then $A'$  sends $\vec{m}$ to the $TTP$.
\item Step Sim6:
    When $\bar{A}$ halts, $A'$ halts with outputting what $\bar{A}$ outputs.
\end{itemize}
\end{itemize}

Without considering Step Sim4,  $A'$ is polynomial-time. However, taking  Step Sim4 into consideration, this is not true any more. Let $q(\alpha)$, $p(\alpha)$ respectively denotes the probability that $\bar{A}$ correctly reveals his commitment in Step Sim3 and in Procedure $\Upsilon$, where $\alpha \stackrel{def}{=}(1^{k},z_{k},I,\vec{m},r_{\bar{A}})$.  Then, the expected times of repeating $\Upsilon$  in Step Sim4 is $q(\alpha)/p(\alpha)$. Since the view $\bar{A}$
holds before revealing his commitment in Step Sim3 is different from that in procedure $\Upsilon$,
$q(\alpha)$, $p(\alpha)$ are distinct. What the computational secrecy of $IBC$  guarantees  and only guarantees is $|q(\alpha)-p(\alpha)|= \mu(.)$. However, there is a risk that $q(\alpha)/p(\alpha)$ is not bound by a polynomial.  For example, $q(\alpha)=1/2^{k}, p(\alpha)=1/2^{2k}$, which result in  $q(\alpha)/p(\alpha)= 2^{k}$. This is a big problem and gives rise to many other difficulties we will encounter later.

Fortunately, \cite{goldreich1996zk} encounters and solves the same problem and its derived problem  as ours. In a little more details, \cite{goldreich1996zk} presents a protocol, in which $P_{1}$, $P_{2}$ respectively sends a perfectly hiding commitment, a perfectly binding commitment, and the corresponding de-commitments  to each other as the situation of tossing coin of our framework. To
prove the security in the case that $P_{1}$ is corrupted, \cite{goldreich1996zk} constructs a simulator in the same way as ours and encounters the same problem as ours.

Using the idea of \cite{goldreich1996zk}, we can overcome such problem too. Specifically, an expected polynomial-time  simulator can be obtained by replacing Step Sim4 with Step $Sim4.1$, $Sim4.2$ given as follow.

\begin{itemize}
  \item Step $Sim4.1$:  $A'$  estimates the value of $q(\alpha)$. $A'$ repeats the
        following procedure, denoted $\Phi$,  until the number of the time of $\bar{A}$ correctly revealing his commitment is  up to $poly(k)$, where $poly(.)$ is a big enough polynomial.

        $\Phi$: $A'$ rewinds  $\bar{A}$ to the end of Step S1 of the framework and $A'$ honestly executes Step R2 and R3 of the framework to interact with it.

        Denote the number of times that $\Phi$ is repeated by $d$, then $q(\alpha)$ is estimated as  $\tilde{q}(\alpha) \stackrel{def}{=}poly(k)/d$.

  \item Step $Sim4.2$: $A'$ repeats the procedure $\Upsilon$. In case
     $\bar{A}$ correctly reveals the recorded value $s$, $A'$ proceeds to the next step. In case $\bar{A}$ correctly reveals a value which is different from $s$, $A'$ outputs $ambiguity_{1}$ and halts. In case the number of the time of repeating $\Upsilon$ exceeds
     the value of $poly(k)/\tilde{q}(\alpha)$, $A'$ outputs $timeout$ and halts.
\end{itemize}

\begin{proposition}
The simulator $A'$ is expected polynomial-time.
\end{proposition}

\begin{IEEEproof}
Conditioning on  Step $Sim4.1$ is executed, the expected value of $d$ is $poly(k)/q(\alpha)$. Choosing a  big enough $poly(.)$, $\tilde{q}(\alpha)$ is within a constant factor of $q(\alpha)$ with probability $1-2^{poly(k)}$.  Therefore, the expected running time of $A'$,
\begin{equation*}
\begin{split}
ExpTime_{A'} & \leq  Time_{Sim1}+ Time_{Sim2} + Time_{Sim3}  \\
                          &+ q(\alpha) \cdot (Time_{\Phi}\cdot poly(k)/q(\alpha)+\\
                          &Time_{\Upsilon}\cdot poly(k)/\tilde{q}(\alpha))\\
                         & + Time_{Sim5} + Time_{Sim6}
\end{split}
\end{equation*}
, is bounded by a polynomial.
\end{IEEEproof}

What is more, we have
\begin{enumerate}
  \item The probability that $A'$ outputs $timeout$ is negligible.
  \item The probability that $A'$ outputs $ambiguity_{1}$ is negligible.
  \item The output of $A'$ in the ideal world and the output of $A$ in the real world are computationally indistinguishable, i.e.,
\begin{multline}\label{X2Y}
    \{Ideal_{f, \{1\},A'(z_{k})}(1^{k},\vec{m},H)\langle 1\rangle \} _{ \substack{k\in\mathds{N},\vec{m}\in (\{0,1\}^{*})^{n} \\ H \in \Psi,z_{k}\in \{0,1\}^{*}} }
      \stackrel{c}{=} \\
    \{ Real_{\pi, \{1\},A(z_{k})}(1^{k},\vec{m},H) \langle 1\rangle \} _{\substack{k\in\mathds{N},\vec{m}\in (\{0,1\}^{*})^{n} \\ H \in \Psi,z_{k}\in \{0,1\}^{*}}}
\end{multline}
\end{enumerate}

Since the  propositions above can be proven in the same way as \cite{goldreich1996zk}, we don't iterate such details here.

\begin{proposition}
In the case that $P_{1}$ was corrupted, i.e., $I=\{1\}$, the equation \eqref{eq_ot_sec} required by Definition  \ref{sec_def_ot} holds.
\end{proposition}

\begin{IEEEproof}
First let us focus on the real world.  $A$'s real input can be formulated as $\gamma \leftarrow A(1^{k},\vec{m},z_{k},r_{A}, r_{1})$.
Note that in this case, $P_{2}$'s output is a determinate function of $A$'s
real input. Since $A$'s real input is in its view, without loss of generality, we assume $A$'s output, denoted $\alpha$, constains  its real input. Therefore, $P_{2}$'s output is a determinate function of $A$'s output, where the function is
\begin{gather*}
    g(\alpha)=\begin{cases}
                 abort_{1}     & \textrm{if } \gamma=abort_{1},\\
                 \gamma \langle H \rangle  & otherwise.
              \end{cases}
\end{gather*}
Let $h(\alpha)\stackrel{def}{=}(\alpha, \lambda, g(\alpha))$. Then we have
\begin{multline*}
Real_{\pi,\{1\},A(z_{k})}(1^{k},\vec{m},H) \equiv\\
  h(Real_{\pi,\{1\},A(z_{k})}(1^{k},\vec{m},H)\langle 0 \rangle)
\end{multline*}

Similarly, in the ideal world, we have
\begin{multline*}
Ideal_{f, \{1\},A'(z_{k})}(1^{k},\vec{m},H) \stackrel{c}{=}\\
  h(Ideal_{f, \{1\},A'(z_{k})}(1^{k},\vec{m},H)\langle 0 \rangle)
\end{multline*}

We use $\stackrel{c}{=}$  not $\equiv$ here because there is a negligible probability that
$A'$ outputs $timeout$ or $ambiguity_{1}$, which makes $h(.)$ undefined.

Let $X(1^{k},\vec{m},H,z_{k}, \{1\}) \stackrel{def}{=}
Real_{\pi,\{1\},A(z_{k})}(1^{k},\vec{m},H)\langle 0  \rangle$,
$Y(1^{k},\vec{m},H,z_{k}, \{1\}) \stackrel{def}{=}
Ideal_{f, \{1\},A'(z_{k})}(1^{k},\vec{m},H)\langle 0 \rangle$. Following equation \eqref{X2Y}, $X\stackrel{c}{=}Y$. Let $F \stackrel{def}{=}(h)_{k \in \mathds{N}}$.
What is more, assume that $A'$ runs in a strictly polynomial-time.
According to Proposition \ref{fuind} we will present in Section \ref{con_hash},
the proposition holds.

In fact, $A'$  doesn't run in strictly polynomial-time, which results in
a failure of above standard reduction.  Fortunately, this difficulty can be overcome by truncating the rare executions of $A'$ which are too long, then applying standard reduction techniques.
Since the details is the same as \cite{goldreich1996zk}, we don't give them here and please see \cite{goldreich1996zk} for them.
\end{IEEEproof}

\subsection{In the case that $P_{2}$ Is Corrupted}
In the case that $P_{2}$ is corrupted, $A$ takes the full control of $P_{2}$ in the real world.
Correspondingly, $A'$ takes the full control of $P_{2}'$ in the ideal world.
We construct $A'$ as follows.

\begin{itemize}
  \item Initial input:
        $A'$ holds the same $k$, $I \stackrel{def}{=}\{2\}$, $z=(z_{k})_{k\in \mathds{N}}$ as $A$, and holds a uniformly distributed  randomness $r_{A'}\in\{0,1\}^{*}$.
        The parties $P_{2}'$ and $P_{2}$ hold  the same private input $H$.

  \item $A'$ works as follows.

  \begin{itemize}
    \item Step Sim1:  $A'$ corrupts  $P_{2}'$ and learns $P_{2}'$'s private input $H$. $A'$ takes $A$'s copy  $\bar{A}$ as  a subroutine, fixes $\bar{A}$'s initial input,  activates $\bar{A}$, supplies $\bar{A}$  with $H$, builds an environment for $\bar{A}$ in the same way as $A'$ does in the case that $P_{1}$ is corrupted.
    \item Step Sim2: Playing the role of $P_{1}$,  $A'$ honestly executes the sender's
          steps until reaches Step S3.3.  If Step S3.3 is reached, $A'$ records the shared randomness $r$ and the messages, denoted $msg$, which he sends to $\bar{A}$.
          Then $A'$  proceeds to next step. Otherwise, $A'$ sends $abort_{2}$ to TTP, outputs what
          $\bar{A}$  outputs and halts.
    \item Step Sim3:
          $A'$ repeats the following procedure, denoted $\Xi$, until the hash parameters and the instance vectors $\bar{A}$ sends in Step R1 passes the check. $A'$ records the shared randomness $\tilde{r}$, the messages $\bar{A}$ sends to open the chosen instance vectors.

          $\Xi$: $A'$ rewinds $\bar{A}$ to the beginning of Step R2, and honestly follows
           sender's steps until reaches Step S3.3 to interact with  $\bar{A}$.

           Note that, in each repeating $\Xi$,  the value $A'$ commits to and the randomness used to generate the commitment
           in Step S1 are fresh and uniformly chosen.
    \item Step Sim4:
          \begin{enumerate}
            \item In case $r=\tilde{r}$, $A'$ outputs $failure$ and halts;
            \item In case $r \neq \tilde{r} \wedge  \forall i( r\langle i \rangle  \neq \tilde{r} \langle i \rangle \rightarrow r\langle i \rangle =1 \wedge   \tilde{r}\langle i \rangle =0)$, $A'$ runs from scratch;
            \item Otherwise, i.e., in case $r \neq \tilde{r} \wedge \exists i( r\langle i\rangle =0 \wedge  \tilde{r}\langle i \rangle =1)$,
                $A'$ records the first one, denoted $e$, of these $i$s and proceeds to next step.
          \end{enumerate}

\begin{remark} \label{re_extra}
The aim of Step Sim3 and Sim4 is to prepare to extract the real input of $\bar{A}$. If the third case happens, then $A'$ knows each entry of  $\tilde{\vec{x}}_{e}$ he sees
in Step Sim2 belong to $L_{\dot{R}_{\Lambda_{e}}}$ or $L_{\ddot{R}_{\Lambda_{e}}}$. What is more, $\tilde{\vec{x}}_{e}$ is indeed a legal instance vector. This is because $\tilde{\vec{x}}_{e}$ passes the check executed by $A'$ in Step Sim3.
Combing $\pi^{2}_{e}$ received in Step Sim2, $A'$ knows the real input of $\bar{A}$.

Note that, $\bar{A}$'s initial input is fixed by $A'$ in Step Sim1. So receiving the same messages, $\bar{A}$ responds in the same way. Therefore, rewinding $\bar{A}$ to the beginning of Step R2, sending the message sent in Step Sim2, $A'$ can reproduce the same scenario as he meets in Step Sim2.
\end{remark}

\item Step Sim5:
          $A'$ rewinds $\bar{A}$ to the beginning of Step R2 of the framework, and sends $msg$  previously recorded to $\bar{A}$ in order. According to the analysis  of Remark \ref{re_extra}, $A'$ can extract $\bar{A}$'s real
          input $H'$. $A'$ does so and sends $H'$ to TTP and receives message $\vec{m}\langle H' \rangle$.
\item Step Sim6:
        $A'$ constructs $\vec{m}'$ as follows. For each  $i \in H'$, $\vec{m}' \langle i \rangle \leftarrow \vec{m}\langle i \rangle$. For each  $i \notin H'$, $\vec{m}' \langle i \rangle \in_{U} \{0,1\}^{*}$. Playing the role of $P_{1}$ and taking $\vec{m}'$  as his real input, $A'$ follows Step S3.3 to complete the interaction with $\bar{A}$.
\item Step Sim6:
        When $\bar{A}$ halts, $A'$ halts with outputting what $\bar{A}$ outputs..
\end{itemize}
\end{itemize}

Note that $S$ doesn't simulate a situation  in the real world that $A$ cheats $P_{1}$ of more than $h$ message. Fortunately, Theorem \ref{p2cheat} guarantees
that this situation arises with probability  at most $1/2^{poly_{s}(k)}$ and so can be ignored.

\begin{proposition} \label{pro_time}
The simulator $A'$ is expected polynomial-time.
\end{proposition}

\begin{IEEEproof}
First, let us focus on  Step Sim3. In each repetition of $\Xi$, because of the perfectly hiding of $IHC(.)$, and the uniform distribution of the value $A'$ commits to, the chosen instance vectors are uniformly distributed. This lead to the probability that $\bar{A}$ passes the check in each repetition is the same. Denote this probability by $p$. The expected time of Step Sim3 is
\begin{equation*}
   ExpTime_{Sim3} = (1/p) \cdot Time_{\Xi}
\end{equation*}

Under the same analysis, the probability that $\bar{A}$ passes the check  in Step Sim2 is $p$ too.
Then, the expected time that $A'$ runs once  from Step Sim1 to the beginning of Step Sim4  is
\begin{equation*}
\begin{split}
    OncExpTime_{Sim1 \rightarrow Sim4}& \leq  Time_{Sim1} + Time_{Sim2} \\
       &+ p \cdot ExpTime_{Sim3} \\
      &= Time_{Sim1} + Time_{Sim2}\\
      &+ Time_{\Xi}
  \end{split}
\end{equation*}

Second, let us focus on Step Sim4,  especially the case that $A'$ needs to run from scratch.
Note that  the initial inputs  $A'$ holds is the same in each  trial. Thus the probability that $A'$ runs from scratch in each trial is the same. We denote this probability by $1-q$.
Then the expected time that $A'$ runs from Step Sim1 to the beginning of Step Sim5 is
\begin{equation*}
\begin{split}
ExpTime_{Sim1 \rightarrow Sim5} & \leq  (1+ 1/q ) \\
                                &\cdot (OncExpTime_{Sim1 \rightarrow Sim4}\\
                                &+  Time_{ Sim4}) \\
                                &= (1+ 1/q ) \cdot (Time_{Sim1} + \\
                                &Time_{Sim2}+ Time_{\Xi}+Time_{ Sim4})
\end{split}
\end{equation*}
The reason there is $1$ here is that $A'$ has to run from scratch at least one time in any case.

The expected running time of $A'$ in a whole execution is
\begin{equation} \label{eq_exptm_a'}
\begin{split}
   ExpTime_{A'}& \leq  ExpTime_{Sim1 \rightarrow Sim5}+  Time_{Sim5}\\
         &+ Time_{Sim6} \\
    &=(1+ 1/q ) \cdot (Time_{Sim1} + Time_{Sim2} \\
    &+ Time_{\Xi}+Time_{ Sim4})\\
    &+Time_{Sim5}+ Time_{Sim6}
\end{split}
\end{equation}

Third, let us estimate the value of $q$, which is the probability that $A'$ does not run from scratch in a trial. We  denote this event by $C$. It's easy to see that event $C$ happens, if and only if one of the following events happens.
\begin{enumerate}
  \item  Event $B$ happens, where $B$ denotes the even that $A'$ halts before reaching Step Sim3.
  \item  Event $\bar{B}$ happens and $R=\tilde{R}$, where $R$ and $\tilde{R}$ respectively denotes the random variable which is defined as the shared randomness $A'$ gets in Step Sim2 and Step Sim3.
  \item  Event $\bar{B}$ happens and there exists $i$ such that $R \langle i
          \rangle=0 \wedge  \tilde{R}\langle i \rangle =1$  .
\end{enumerate}
So
\begin{equation}\label{eq_q}
   \begin{split}
     q  =& Pr(C) \\
        =&Pr(B) + Pr (\bar B \cap R = \tilde R) \\
        &+ Pr (\bar B \cap \exists
          i(R\langle i \rangle  = 0 \wedge \tilde R\langle i \rangle  = 1))\\
        =& Pr(B) + Pr(\bar B)\cdot( Pr (R = \tilde R|\bar B) \\
        &+ Pr (\exists i(R\langle i \rangle  = 0 \wedge
             \tilde R\langle i \rangle  = 1)|\bar B))
   \end{split}
\end{equation}

Let $S_{1} \stackrel{def}{=} \{ (r,\tilde r)|(r,\tilde r) \in (\{ 0,1\} ^K )^2
, r = \tilde r\}$, $S_{2} \stackrel{def}{=} \{ (r,\tilde r)|(r,\tilde r) \in (\{ 0,1\} ^K )^2  , r \ne \tilde r , \forall i(r\langle i \rangle  \ne \tilde r\langle i \rangle  \to r\langle i \rangle  = 1 \wedge \tilde r\langle i \rangle  = 0)\}$, $S_{3} \stackrel{def}{=} \{ (r,\tilde r)|(r,\tilde r) \in (\{ 0,1\} ^K )^2 , r \ne \tilde r , \exists i(i \in [K] \wedge r\langle i \rangle  = 0 \wedge \tilde r\langle i \rangle  = 1)\}$. It is easy to see that
 $S_{1}$, $S_{2}$, $S_{3}$ constitute a complete partition of $(\{0,1\}^{K})^{2}$ and
$\#S_{1} = 2^{K}$, $\#S_{2} = \#S_{3}= (2^{K} \cdot 2^{K}- 2^{K})/2$.

Because of the perfectly hiding of $IHC(.)$, and the uniform distribution of the value $A'$ commits to, $R$ and $\tilde{R}$ are all uniformly distributed. We have
\begin{equation}\label{eq_e1}
      Pr(R=\tilde R|\bar B)= \# S_{1} /\# (\{ 0,1\} ^K )^2
                          =1/2^K
\end{equation}
and
\begin{equation}\label{eq_e2}
\begin{split}
 Pr(\exists i(R\langle i \rangle  = 0 \wedge \tilde R\langle i \rangle  = 1)|\bar B)
    &= \# S_3 /\# (\{ 0,1\} ^K )^2 \\
    & = 1/2 - 1/2^{K + 1}
\end{split}
\end{equation}

Combining equation \eqref{eq_q}, \eqref{eq_e1} and \eqref{eq_e2}, we have
\begin{equation} \label{eq_qq}
   \begin{split}
      q  & = Pr(B) + Pr(\bar B)( 1/2 + 1/2^{K + 1}) \\
          & = 1/2+ 1/2^{K + 1} + (1/2 - 1/2^{K + 1}) Pr(B)\\
          & > 1/2
    \end{split}
\end{equation}

Combining equation \eqref{eq_exptm_a'} and  \eqref{eq_qq}, we have
\begin{equation*}
\begin{split}
ExpTime_{A'} & < 3(Time_{Sim1} + Time_{Sim2} \\
                &+ Time_{\Xi}+Time_{ Sim4})\\
                & +Time_{Sim5}+ Time_{Sim6}
\end{split}
\end{equation*}
which means the expected running time of $A'$  is bound by a polynomial.
\end{IEEEproof}

\begin{lemma} \label{pro_pro_fail}
The probability that $A'$ outputs $failure$ is less than  $1/2^{K-1}$.
\end{lemma}

\begin{IEEEproof}
Let $X$ be a random variable defined as the number of the trials in a whole execution.
From the proof of Proposition \ref{pro_time}, we know two facts. First,  $Pr(X=i)=(1-q)^{i-1}q< 1/2^{i-1}$. Second, in each trial the event $A'$ outputs $failure$ is the combined event of $\bar{B}$ and $R=\tilde{R}$,  where the combined event happens with the following probability.
\begin{equation*}
        Pr(\bar B \cap R=\tilde{R})
           =Pr(\bar B) Pr(R=\tilde{R} | \bar B)
          \leq  Pr(R=\tilde{R} | \bar B)
\end{equation*}
Combining equation \eqref{eq_e1}, this probability is not more than $1/2^{K}$. Therefore, the probability that $A'$ outputs $failure$  in a whole execution is
\begin{equation*}
\begin{split}
 \sum^{\infty}_{i=1}Pr(X=i)Pr(\bar B \cap R=\tilde{R})
   &< (1/2^{K})\cdot \sum^{\infty}_{i=1}1/2^{i-1} \\
    & =1/2^{K-1}
\end{split}
\end{equation*}
\end{IEEEproof}

\begin{lemma} \label{pro_av_a'v_ind2}
The output of the adversary $A$  in the real world  and that of the simulator $A'$ in the ideal world are computationally indistinguishable, i.e.,
\begin{gather*}
  \{ Real_{\pi,\{2\},A(z_{k})}(1^{k},\vec{m},H)\langle 0 \rangle \}_{\substack{ k\in\mathds{N},\vec{m}\in
      (\{0,1\}^{*})^{n}\\ H \in \Psi, z_{k} \in \{0,1\}^{*}}}
      \stackrel{c}{=} \\
    \{ Ideal_{f, \{2\},A'(z_{k})}(1^{k},\vec{m},H)\langle 0 \rangle \}_{\substack{ k\in\mathds{N},\vec{m}\in  (\{0,1\}^{*})^{n}\\ H \in \Psi, z_{k} \in \{0,1\}^{*}}}
\end{gather*}
\end{lemma}

\begin{IEEEproof}
First, we claim that the outputs of $A'$ and $\bar{A}$ are computationally indistinguishable.
The only point that the output of  $A'$  is different from that of $\bar{A}$ is  $A'$ may
outputs $failure$. Since the probability that this point arises is negligible, our claim holds.

Second, we claim that the outputs of $A$ and $\bar{A}$ are computationally indistinguishable.
The only point that the view of $\bar{A}$ is different from that of $A$ is that the ciphertext $\bar{A}$ receives is generated by encrypting $\vec{m}'$ not $\vec{m}$. Fortunately, $SPHDHC_{t,h}$'s property smoothness guarantees that the ciphertext  generated in the two way are computationally indistinguishable. Therefore, our claim holds.

Combining the two claims, the proposition holds.
\end{IEEEproof}

\begin{proposition}
In the case that $P_{2}$ was corrupted, i.e., $I=\{2\}$, the equation \eqref{eq_ot_sec} required by Definition  \ref{sec_def_ot} holds.
\end{proposition}

\begin{IEEEproof}
Note that the honest parties $P_{1}$ and $P'_{1}$ end up with outputting nothing. Thus, the fact that the outputs of $A'$ and $A$ are computationally indistinguishable, which is supported by Lemma \ref{pro_av_a'v_ind2}, directly prove this proposition holds.
\end{IEEEproof}

\subsection{Other  Cases}
In the case that both $P_{1}$ and $P_{2}$ are corrupted, $A$ takes the full control of the two
corrupted parties. In the ideal world, a similar situation also holds with respect to $A'$,
$P'_{1}$ and $P'_{2}$. Liking in previous cases, $A'$ uses $A$'s copy,  $\bar{A}$, as a subroutine and builds a simulated environment for $\bar{A}$. $A'$ provids  $\bar{A}$ with $P'_{1}$ and $P'_{2}$'s initial inputs before $\bar{A}$ engages in the protocol. When $\bar{A}$ halts, $A'$ halts with outputting what $\bar{A}$ outputs.  Obviously, $A'$ runs in strictly polynomial-time and the equation \eqref{eq_ot_sec} required by Definition  \ref{sec_def_ot} holds in this case.

In the case that none of $P_{1}$ and $P_{2}$ is  corrupted.  The simulator $A'$ is constructed as follows.
$A'$ uses  $\bar{A}$, $\bar{P}_{1}$, $\bar{P}_{2}$ as subroutines, where $\bar{A}$, $\bar{P}_{1}$, $\bar{P}_{2}$, respectively, is the copy of $A$, $P_{1}$ and $P_{2}$. $A'$ fixes  $\bar{A}$'s initial inputs in the same way  as in previous cases. $A'$  chooses an  arbitrary $\bar{\vec{m}} \in (\{0,1\}^{*})^{n}$ and  a  uniformly distributed randomness  $\bar{r}_{1}$ as $\bar{P}_{1}$'s initial inputs.
$A'$  chooses an  arbitrary $\bar{H} \in \Psi$ and  a  uniformly distributed randomness  $\bar{r}_{2}$ as $\bar{P}_{2}$'s initial inputs.  $A'$ actives these  subroutines and make the communication between $\bar{P}_{1}$ and $\bar{P}_{2}$  available to $\bar{A}$. Note that, in the case that none of $P_{1}$ and $P_{2}$ is  corrupted, what  adversaries can see in real life only is the communication between honest parties. When $\bar{A}$ halts, $A'$ halts with outputting what $\bar{A}$ outputs.  Obviously, $A'$ runs in strictly polynomial-time and the equation \eqref{eq_ot_sec} required by Definition  \ref{sec_def_ot} holds in this case.

\section{How To  Construct $SPHDHC_{t,h}$ Easily}\label{reduceSPHDHC}
$SPHDHC_{t,h}$ holds so many properties that constructing it  from scratch is not always easy. In this section, we reduce  constructing $SPHDHC_{t,h}$ to  constructing seemingly  simpler hash systems. A idea naturally arising is that  generating the instances independently in essence  to obtain the required properties. We  keep this idea in mind  to proceed to construct $SPHDHC_{t,h}$.

\subsection{Smoothness} \label{howto_sm}
In this section, we describe how to obtain smoothness for a hash family. First, we introduce a lemma from ~\cite{gold2004found}.
\begin{lemma}[~\cite{gold2004found}] \label{polyind}
Let $X \stackrel{def}{=}\{X(1^{k},a)\}_{k\in \mathds{N}, a \in \{0,1\}^{*}}$ and $Y \stackrel{def}{=}\{Y(1^{k},a)\}_{k\in \mathds{N}, a \in \{0,1\}^{*}}$ be two polynomial-time constructible probability ensembles, and $X \stackrel{c}{=}Y$, then
\begin{equation*}
    \vec{X} \stackrel{c}{=} \vec{Y}
\end{equation*}
where  $\vec{X}\stackrel{def}{=}\{\vec{X}(1^{k},a)\}_{\substack{ k \in \mathbb{N} \\ a \in \{0,1\}^{*}} }$,
$\vec{X}(1^{k},a) \stackrel{def}{=}(X_{i}(1^{k},a))_{i \in [poly(k)]}$,
each $ X_{i}(1^{k},a)=X(1^{k},a)$,
$\vec{Y}\stackrel{def}{=}\{\vec{Y}(1^{k},a)\}_{\substack{ k \in \mathbb{N} \\ a \in \{0,1\}^{*}}}$,
$\vec{Y}(1^{k},a) \stackrel{def}{=} (Y_{i}(1^{k},a))_{i \in [poly(k)]}$,
each $Y_{i}(1^{k},a)=Y(1^{k},a)$,
and all $X_{i}(1^{k},a)$, $Y_{i}(1^{k},a)$ are independent.
\end{lemma}

\begin{proposition} \label{fuind}
Let $X \stackrel{def}{=}\{X(1^{k},a)\}_{k\in \mathds{N}, a \in \{0,1\}^{*}}$ and $Y \stackrel{def}{=}\{Y(1^{k},a)\}_{k\in \mathds{N}, a \in \{0,1\}^{*}}$ be two polynomial-time constructible probability ensembles, $X \stackrel{c}{=}Y$, $F \stackrel{def}{=}(f_{k})_{k\in \mathds{N}}$, $f_{k}:\{0,1\}^{*} \rightarrow \{0,1\}^{*}$ is polynomial-time computable, then
\begin{equation*}
    F(X) \stackrel{c}{=} F(Y)
\end{equation*}
where $F(X) \stackrel{def}{=}\{f_{k}(X(1^{k},a))\}_{k\in \mathds{N},a \in \{0,1\}^{*}}$,$F(Y) \stackrel{def}{=}\{f_{k}(Y(1^{k},a))\}_{k\in \mathds{N},a \in \{0,1\}^{*}}$.
\end{proposition}

\begin{IEEEproof}
Assume the proposition is false, then there exists a non-uniform PPT distinguisher $D$ with an infinite sequence $z=(z_{k})_{k\in \mathds{N}}$, a polynomial $poly(.)$, an infinite positive integer set $G\subseteq \mathds{N}$ such that, for each $k\in G$, it holds that
\begin{multline*}
     |Pr(D(1^{k},z_{k}, a, f_{k}(X(1^{k},a)))=1) -\\
     Pr(D(1^{k},z_{k}, a, f_{k}(Y(1^{k},a)))=1)|
    \geq 1/poly(k)
\end{multline*}

We construct a distinguisher $D'$ with an infinite sequence $z=(z_{k})_{k\in \mathds{N}}$ for the ensembles $X$ and $Y$ as follows.

$D'(1^{k},z_{k},a,\gamma)$: $\delta \leftarrow f_{k}(\gamma)$, finally outputs
$D(1^{k},z_{k},a, \delta)$.

Obviously, $D'(1^{k},z_{k},a, X(1^{k},a))=D(1^{k},z_{k},a, f_{k}(X(1^{k},a))$, $D'(1^{k},z_{k},a, Y(1^{k},a))=D(1^{k},z_{k},a, f_{k}(Y(1^{k},a))$. So we have
\begin{multline*}
      |Pr(D'(1^{k},z_{k},a, X(1^{k},a))=1) -\\
    Pr(D'(1^{k},z_{k}, a,Y(1^{k},a))=1)|
    \geq 1/poly(k)
\end{multline*}

This contradicts the fact $X \stackrel{c}{=}Y$.
\end{IEEEproof}

\begin{lemma}  \label{ind_smoo}
Let $\mathcal{H}=(PG, IS, DI, KG, Hash, pHash, Cheat)$ be a Hash Family. $n\stackrel{def}{=}h+t$.
For each $i \in [2]$ and $j \in [n]$, $Sm^{j}_{i} \stackrel{def}{=}\{Sm^{j}_{i}(1^{k}) \}_{k \in \mathds{N}}
\stackrel{def}{=}\{(SmGen_{i}(1^{k})\langle 1 \rangle,SmGen_{i}(1^{k})\langle 2 \rangle \langle j \rangle)\}_{k \in \mathds{N}}$, where $SmGen_{i}(1^{k})$ is defined in Definition \ref{SPHDHC}. If $\mathcal{H}$ meets the following three conditions
\begin{enumerate}
  \item All random variables $SmGen_{i}(1^{k})\langle 2 \rangle \langle j \rangle$ are
        independent, where $i \in [2]$,$j \in [n]-[h]$.
  \item $Sm^{h+1}_{1} = \ldots  =  Sm^{n}_{1}$, and $Sm^{h+1}_{2} =
        \ldots =  Sm^{n}_{2}$.
  \item $Sm^{h+1}_{1} \stackrel{c}{=} Sm^{h+1}_{2}$.
\end{enumerate}
then $\mathcal{H}$ has property smoothness.
\end{lemma}

\begin{IEEEproof}
Following Lemma  \ref{polyind},
\begin{multline*}
  \{(Sm^{h+1}_{1}(1^{k}),\ldots, Sm^{n}_{1}(1^{k}))\}_{k \in \mathds{N}}\\ \stackrel{c}{=} \{(Sm^{h+1}_{2}(1^{k}),\ldots, Sm^{n}_{2}(1^{k}))\}_{k \in \mathds{N}}
\end{multline*}
holds. Let $\vec{X} \stackrel{def}{=} \{(Sm^{1}_{1}(1^{k}),\ldots, Sm^{n}_{1}(1^{k}))\}_{k \in \mathds{N}}$, and $\vec{Y} \stackrel{def}{=} \{(Sm^{1}_{2}(1^{k}),\ldots, Sm^{n}_{2}(1^{k}))\}_{k \in \mathds{N}}$. From the definition of $SmGen_{i}(1^{k})$, we notice that, for each $j\in [h]$  $Sm^{j}_{1}(1^{k})=Sm^{j}_{2}(1^{k})$. So it holds that
\begin{equation*}
  \vec{X}  \stackrel{c}{=} \vec{Y}
\end{equation*}
Since each $Sm^{j}_{i}(1^{k})$ is polynomial-time constructible, thus both$\vec{X}$ and $\vec{Y}$ are polynomial-time constructible. Let $F \stackrel{def}{=}(\pi)_{k \in \mathds{N}}$, where $\pi \in \Pi$. Following Proposition \ref{fuind}, we have $F(\vec{X}) \stackrel{c}{=} F(\vec{Y})$, i.e.,
\begin{multline*}
  \{ \pi(Sm^{1}_{1}(1^{k}),\ldots, Sm^{n}_{1}(1^{k}))\}_{k \in \mathds{N}}
\\ \stackrel{c}{=} \{ \pi(Sm^{1}_{2}(1^{k}),\ldots, Sm^{n}_{2}(1^{k}))\}_{k \in \mathds{N}}
\end{multline*}

Notice that $SmGen_{1}(1^{k})\langle 1 \rangle = SmGen_{2}(1^{k})\langle 1 \rangle$, we have
\begin{multline*}
   \{(SmGen_{1}(1^{k})\langle 1 \rangle, \pi(SmGen_{1}(1^{k})\langle 2 \rangle ))\}_{k \in \mathds{N}} \\ \stackrel{c}{=}
\{(SmGen_{2}(1^{k})\langle 1 \rangle, \pi(SmGen_{2}(1^{k})\langle 2 \rangle ))\}_{k \in \mathds{N}}
\end{multline*}
That is
\begin{equation*}
  Sm_{1} \stackrel{c}{=} Sm_{2}
\end{equation*}
, which meets the requirement of the smoothness.
\end{IEEEproof}

Loosely speaking, following Lemma \ref{ind_smoo}, given a hash family $\mathcal{H}$, if each $\ddot{x}$ was sampled in an independent way and its projective key is useless to obtain the
value of $Hash(1^{k}, \Lambda,\ddot{x},.)$,  then $\mathcal{H}$ is smooth.

\subsection{Hard Subset Membership} \label{howto_hsm}
In this section, we deal with how to obtain hard subset membership for a hash family.

\begin{proposition} \label{xy_f(xy)_ind}
Let $X \stackrel{def}{=} \{X(1^{k},a)\}_{k\in \mathds{N}, a\in \{0,1\}^{*}}$ and
$Y \stackrel{def}{=} \{Y(1^{k},a)\}_{k\in \mathds{N}, a\in \{0,1\}^{*}}$ be two polynomial-time constructible probability ensembles, and $X \stackrel{c}{=} Y$. Then
$$\overrightarrow{XY} \stackrel{c}{=} \Phi(\overrightarrow{\widetilde{XY}})$$
where $\overrightarrow{XY}$ and $\Phi(\overrightarrow{\widetilde{XY}})$ are two probability
ensembles defined as follows.
\begin{itemize}
  \item $\overrightarrow{XY}\stackrel{def}{=}\{\overrightarrow{XY}(1^{k},a)\}_{k\in
        \mathds{N}, a\in \{0,1\}^{*}}$, $\overrightarrow{XY}(1^{k},a) \stackrel{def}{=}(X_{1}(1^{k},a), \ldots,
        X_{poly_{1}(k)}(1^{k},a), Y_{poly_{1}(k)+1}(1^{k},a), \ldots, \\ Y_{poly(k)}(1^{k},a))$, each $X_{i}(1^{k},a)=X(1^{k},a)$, each $Y_{i}(1^{k},a)=Y(1^{k},a)$, $poly_{1}(.) \leq poly(.)$, all $X_{i}(1^{k},a)$ and
        $Y_{i}(1^{k},a)$ are independent;
  \item $\Phi(\overrightarrow{\widetilde{XY}})\stackrel{def}{=}
        \{\Phi_{k}(\overrightarrow{\widetilde{XY}}(1^{k},a))\}_{k\in \mathds{N}, a\in \{0,1\}^{*}}$, $\overrightarrow{\widetilde{XY}}(1^{k},a)=\overrightarrow{XY}(1^{k},a)$, $\Phi \stackrel{def}{=} (\Phi_{k})_{k \in \mathds{N}}$, each $\Phi_{k}$ is a permutation
        over $[poly(k)]$.
\end{itemize}
\end{proposition}

\begin{IEEEproof}
In case $\Phi_{k}([poly_{1}(k)]) \subseteq [poly_{1}(k)]$, it obviously holds. We proceed to prove it also holds in case $\Phi_{k}([poly_{1}(k)])\nsubseteq [poly_{1}(k)]$. Assume it does not hold in this case, then there exists a non-uniform PPT distinguisher $D$ with an infinite sequence $z=(z_{k})_{k\in \mathds{N}}$, a polynomial $poly_{2}(.)$, a infinite positive integer set $G\subseteq \mathds{N}$ such that, for each $k\in G$,
\begin{multline}\label{pr1}
   |Pr(D(1^{k},z_{k},a, \overrightarrow{XY}(1^{k},a))=1)   \\
   - Pr(D(1^{k},z_{k},a,\Phi_{k}(\overrightarrow{\widetilde{XY}}(1^{k},a))=1)|\\
    \geq 1/poly_{2}(k)
\end{multline}
$V \stackrel{def}{=}\{i|i\in [poly_{1}(k)],\Phi_{k }(i)\in [poly(k)]-[poly_{1} (k)]\}$.
We list the elements of $V$ in order as $i_{1 } < ... < i_{j} ... < i_{\# V}$. Let $V_{j} \stackrel{def}{=} \{ i_{1} , \ldots ,i_{j} \}$. We define the following permutations over $[poly(k)]$.
\begin{gather*}
\Phi_{k}^{0'}(i) = i \quad i\in [poly(k)] \\
\Phi_{k}^{0}(i) = \begin{cases}
                      i &  i\in V \cup \Phi_{k}(V)  \\
                      \Phi_{k}(i) &  i\in [poly(k)]-V- \Phi_{k}(V) \\
                   \end{cases} \\
\end{gather*}
For $j\in[\#V]$,
\begin{equation*}
\Phi_{k}^{j}(i) = \begin{cases}
                      i &  i\in (V-V_{j}) \cup \Phi_{k}(V-V_{j}),\\
                      \Phi_{k}(i) &  i\in [poly(k)]-(V-V_{j}) -\Phi_{k}(V-V_{j}).
                   \end{cases}
\end{equation*}
It is easy to see that $\overrightarrow{\widetilde{XY}}(1^{k},a) = \Phi_{k}^{0'}(\overrightarrow{\widetilde{XY}}(1^{k},a)) \equiv \Phi_{k}^{0}(\overrightarrow{\widetilde{XY}}(1^{k},a))$, and $\Phi_{k}=\Phi_{k}^{\#V}$. Since $\overrightarrow{XY}(1^{k},a)= \overrightarrow{\widetilde{XY}}(1^{k},a)$,
then $\overrightarrow{XY}(1^{k},a)
\stackrel{c}{=}\Phi_{k}^{0}(\overrightarrow{\widetilde{XY}}(1^{k},a))$.
So we have
\begin{multline}\label{pr2}
    |Pr(D(1^{k},z_{k},a, \overrightarrow{XY}(1^{k},a))=1) - \\ Pr(D(1^{k},z_{k},a,\Phi_{k}(\overrightarrow{\widetilde{XY}}(1^{k},a)))=1)| \\
= |Pr(D(1^{k},z_{k},a, \Phi_{k}^{0}(\overrightarrow{\widetilde{XY}}(1^{k},a)))=1) - \\ Pr(D(1^{k},z_{k},a,\Phi_{k}^{\#V}(\overrightarrow{\widetilde{XY}}(1^{k},a)))=1)|
\end{multline}
Following triangle inequality, we have
\begin{multline}\label{pr3}
   |Pr(D(1^{k},z_{k},a, \Phi_{k}^{0}(\overrightarrow{\widetilde{XY}}(1^{k},a)))=1) -\\
   Pr(D(1^{k},z_{k},a,\Phi_{k}^{\#V}(\overrightarrow{\widetilde{XY}}(1^{k},a)))=1)| \leq \\ \sum^{\#V}_{j=1}
   |Pr(D(1^{k},z_{k},a, \Phi_{k}^{j-1}(\overrightarrow{\widetilde{XY}}(1^{k},a)))=1) -\\
   Pr(D(1^{k},z_{k},a,\Phi_{k}^{j}(\overrightarrow{\widetilde{XY}}(1^{k},a)))=1)|
\end{multline}
Combining equation \eqref{pr1} \eqref{pr2} \eqref{pr3}, we have
\begin{multline*}
    \sum^{\#V}_{j=1}
    |Pr(D(1^{k},z_{k},a,\Phi_{k}^{j-1}(\overrightarrow{\widetilde{XY}}(1^{k},a)))=1)-\\
    Pr(D(1^{k},z_{k},a,\Phi_{k}^{j}(\overrightarrow{\widetilde{XY}}(1^{k},a)))=1)|
   \geq  1/poly_{2}(k)
\end{multline*}
So there exists $j\in [\#V]$ such that
\begin{multline}\label{pr4}
    |Pr(D(1^{k},z_{k},a,\Phi_{k}^{j-1}(\overrightarrow{\widetilde{XY}}(1^{k},a)))=1) -\\
    Pr(D(1^{k},z_{k},a,\Phi_{k}^{j}(\overrightarrow{\widetilde{XY}}(1^{k},a)))=1)|\\
   \geq  1/(\#V \cdot poly_{2}(k))
\end{multline}
According to the definition of $\Phi_{k}^{j-1},\Phi_{k}^{j}$, the differences between them are the values of points $i_{j},\Phi_{k}(i_{j})$. Similarly, the only differences between $\Phi_{k}^{j-1}(\overrightarrow{\widetilde{XY}}(1^{k},a))$ and $\Phi_{k}^{j}(\overrightarrow{\widetilde{XY}}(1^{k},a))$ are the $i_{j}$-th and $\Phi_{k}(i_{j})$-th entries, i.e., $\Phi_{k}^{j-1}(\overrightarrow{\widetilde{XY}}(1^{k},a))\langle i_{j} \rangle = X(1^{k},a)$, $\Phi_{k}^{j-1}(\overrightarrow{\widetilde{XY}}(1^{k},a))\langle \Phi_{k}(i_{j})  \rangle = Y(1^{k},a)$, $\Phi_{k}^{j}(\overrightarrow{\widetilde{XY}}(1^{k},a))\langle i_{j} \rangle = Y(1^{k},a)$, $\Phi_{k}^{j}(\overrightarrow{\widetilde{XY}}(1^{k},a))\langle \Phi_{k}(i_{j})  \rangle = X(1^{k},a)$.\\
Let $\overrightarrow{MXY} \stackrel{def}{=} \{\overrightarrow{MXY}(1^{k},a)\}_{k\in \mathds{N}, a \in\{0,1\}^{*}}$, where $\overrightarrow{MXY}(1^{k},a)$ is defined as follows. For each $d \in [poly(k)]$,
\begin{equation*}
\overrightarrow{MXY}(1^{k},a)\langle d \rangle=
   \begin{cases}
      \Phi_{k}^{j-1}(\overrightarrow{\widetilde{XY}}(1^{k},a))\langle d \rangle
                    & d \neq \Phi_{k}(i_{j})\\
      X(1^{k},a)    & d = \Phi_{k}(i_{j})
   \end{cases}
\end{equation*}
The difference between $\overrightarrow{MXY}(1^{k},a)$ and
$\Phi_{k}^{j-1}(\overrightarrow{\widetilde{XY}}(1^{k},a))$ is
that $\overrightarrow{MXY}(1^{k},a)\langle \Phi_{k}(i_{j}) \rangle =X(1^{k},a)$,
$\Phi_{k}^{j-1}(\overrightarrow{\widetilde{XY}}(1^{k},a))\langle \Phi_{k}(i_{j}) \rangle
=Y(1^{k},a)$.
The difference between $\overrightarrow{MXY}(1^{k},a)$ and
$\Phi_{k}^{j}(\overrightarrow{\widetilde{XY}}(1^{k},a))$ is
that $\overrightarrow{MXY}(1^{k},a)\langle i_{j} \rangle =X(1^{k},a)$,
$\Phi_{k}^{j}(\overrightarrow{\widetilde{XY}}(1^{k},a))\langle i_{j} \rangle=Y(1^{k},a)$.
Following triangle inequality, we have
\begin{multline}\label{pr5}
   |Pr(D(1^{k},z_{k},a,\Phi_{k}^{j-1}(\overrightarrow{\widetilde{XY}}(1^{k},a)))=1) -\\
    Pr(D(1^{k},z_{k},a, \overrightarrow{MXY}(1^{k},a))=1)| + \\
   |Pr(D(1^{k},z_{k},a, \overrightarrow{MXY}(1^{k},a))=1) -\\
    Pr(D(1^{k},z_{k},a, \Phi_{k}^{j}(\overrightarrow{\widetilde{XY}}(1^{k},a)))=1)| \\
    \geq
  |Pr(D(1^{k},z_{k},a,\Phi_{k}^{j-1}(\overrightarrow{\widetilde{XY}}(1^{k},a)))=1) -\\
   Pr(D(1^{k},z_{k},a,\Phi_{k}^{j}(\overrightarrow{\widetilde{XY}}(1^{k},a)))=1)|
\end{multline}
Combining \eqref{pr4} \eqref{pr5}, we know that
\begin{multline}\label{pr6}
   |Pr(D(1^{k},z_{k},a,\Phi_{k}^{j-1}(\overrightarrow{\widetilde{XY}}(1^{k},a)))=1) -\\
    Pr(D(1^{k},z_{k},a, \overrightarrow{MXY}(1^{k},a))=1)|  \\
    \geq
   1/( 2\#V \cdot poly_{2}(k))
\end{multline}
or
\begin{multline}\label{pr7}
   |Pr(D(1^{k},z_{k},a, \overrightarrow{MXY}(1^{k},a))=1) -\\
    Pr(D(1^{k},z_{k},a, \Phi_{k}^{j}(\overrightarrow{\widetilde{XY}}(1^{k},a)))=1)|  \\
    \geq
   1/( 2\#V \cdot poly_{2}(k))
\end{multline}
holds.  Without loss of generality, we assume equation \eqref{pr6} holds (in case equation \eqref{pr7} holds, the proof can be done in  similar way).
We can construct a distinguisher $D'$ with an infinite sequence $z=(z_{k})_{k\in \mathds{N}}$ for the probability ensembles $X$ and $Y$ as follows.

$D'(1^{k},z_{k},a,\gamma)$: $\overrightarrow{xy}\langle \Phi_{k}^{j-1}(i)\rangle  \leftarrow
S_{X}(1^{k},a) \;  \forall i \in [poly_{1} (k)]$,
$\overrightarrow{xy} \langle \Phi_{k}^{j-1}(i)\rangle  \leftarrow S_{Y}(1^{k},a) \;  \forall i \in [poly(k)]-[poly_{1}(k)]- \{\Phi_{k}(i_{j})\}$,
$\overrightarrow{xy} \langle \Phi_{k}(i_{j}) \rangle  \leftarrow \gamma$,
finally outputs   $D(1^{k},z_{k},a,\overrightarrow{xy})$.

Obviously, if $\gamma$ is sampled from $Y(1^{k}, a)$, then $\overrightarrow{xy}$ is an instance of $\Phi_{k}^{j-1}(\overrightarrow{\widetilde{XY}}(1^{k},a))$; if $\gamma$ is sampled from $X(1^{k}, a)$, then $\overrightarrow{xy}$ is an instance of $\overrightarrow{MXY}(1^{k},a)$.  So we have
\begin{multline}\label{pr8}
   |Pr(D'(1^{k},z_{k},a, X(1^{k},a))=1) -\\
    Pr(D'(1^{k},z_{k}, a,Y(1^{k},a))=1)|= \\
    | Pr(D(1^{k},z_{k},a, \overrightarrow{MXY}(1^{k},a))=1)-\\
     Pr(D(1^{k},z_{k},a,\Phi_{k}^{j-1}(\overrightarrow{\widetilde{XY}}(1^{k},a)))=1)|
\end{multline}
Combining \eqref{pr6} \eqref{pr8}, we have
\begin{multline*}
   |Pr(D'(1^{k},z_{k},a, X(1^{k},a))=1) -\\
    Pr(D'(1^{k},z_{k}, a,Y(1^{k},a))=1)|
     \geq   1/( 2\#V \cdot poly_{2}(k))
\end{multline*}
This contradicts the fact $X \stackrel{c}{=}Y$. Therefore, the proposition also holds in case $\Phi_{k}([poly_{1}(k)])\nsubseteq [poly_{1}(k)]$ too.
\end{IEEEproof}

\begin{lemma} \label{hsm}
Let $\mathcal{H}=(PG, IS, DI, KG, Hash, pHash, Cheat)$ be a hash family. Let $n\stackrel{def}{=}h+t$.
For each $i \in [n]$, $HSM^{i} \stackrel{def}{=} \{ HSM^{i}(1^{k})\}_{k \in \mathds{N}}$,
$HSM^{i}(1^{k}) \stackrel{def}{=}(HSM_{1}(1^{k}) \langle 1 \rangle,HSM_{1}(1^{k}) \langle i + 1 \rangle)$, where $HSM_{1}(1^{k})$ is defined in Definition \ref{SPHDHC}. If $\mathcal{H}$ meets the following three conditions,
\begin{enumerate}
  \item All variables $HSM_{1}(1^{k}) \langle i+1 \rangle$ are independent,
        where $i\in [n]$.
  \item $HSM^{1} = \ldots =  HSM^{h}$, $HSM^{h+1} = \ldots =  HSM^{n}$.
  \item $HSM^{1} \stackrel{c}{=}  HSM^{h+1}$.
\end{enumerate}
then $\mathcal{H}$  has property hard subset membership.
\end{lemma}

\begin{IEEEproof}
Let $\pi \in \Pi $,  $X \stackrel{def}{=} HSM^{1}$, $Y \stackrel{def}{=} HSM^{h+1}$,
$\Phi  = (\pi)_{k\in \mathds{N}}$, $poly_{1}(.) \stackrel{def}{=}h$, $poly(.)
\stackrel{def}{=}n$. Following Proposition \ref{xy_f(xy)_ind}, we know
\begin{equation*}
\overrightarrow{XY} \stackrel{c}{=} \Phi(\overrightarrow{\widetilde{XY}})
\end{equation*}
That is
\begin{multline*}
((HSM_{1}(1^{k}) \langle 1 \rangle, HSM_{1}(1^{k}) \langle 2 \rangle), \ldots  \\
(HSM_{1}(1^{k}) \langle 1 \rangle, HSM_{1}(1^{k}) \langle n + 1 \rangle)) \stackrel{c}{=}\\
(HSM_{2}(1^{k}) \langle 1 \rangle, HSM_{2}(1^{k}) \langle 2 \rangle),\ldots \\
(HSM_{2}(1^{k}) \langle 1 \rangle, HSM_{2}(1^{k}) \langle n + 1 \rangle))
\end{multline*}
where $HSM_{1}(1^{k})$, $HSM_{2}(1^{k})$ are taken from  Definition \ref{SPHDHC}.  Note that $HSM_{1}(1^{k}) \langle 1 \rangle = HSM_{2}(1^{k}) \langle 1 \rangle$, so
\begin{multline*}
(HSM_{1}(1^{k}) \langle 1 \rangle, HSM_{1}(1^{k}) \langle 2 \rangle, \ldots, HSM_{1}(1^{k}) \langle n + 1 \rangle)
\stackrel{c}{=}\\
(HSM_{2}(1^{k}) \langle 1 \rangle, HSM_{2}(1^{k}) \langle 2 \rangle, \ldots, HSM_{2}(1^{k}) \langle n + 1 \rangle)
\end{multline*}
i.e.,
\begin{equation*}
 HSM_{1}\stackrel{c}{=} HSM_{2}
\end{equation*}
, which meets the requirement of the property hard subset membership.
\end{IEEEproof}

Loosely speaking, Lemma \ref{hsm} shows that, given a hash family $\mathcal{H}$, if random variables  $IS(1^{k},\Lambda) \langle 1\rangle,\ldots,IS(1^{k},\Lambda) \langle n\rangle$  are independent,  $IS(1^{k},\Lambda) \langle 1\rangle,\ldots, IS(1^{k},\Lambda) \langle h\rangle$  sample $\dot{x}$ from $L_{\dot{R}_{\Lambda}}$ in the same way , $IS(1^{k},\Lambda) \langle h+1\rangle,\ldots,IS(1^{k},\Lambda) \langle n\rangle$ sample $\ddot{x}$ from $L_{\ddot{R}_{\Lambda}}$  in the same way,  $L_{\dot{R}_{\Lambda}}$ and $L_{\ddot{R}_{\Lambda}}$ are computationally indistinguishable,  then $\mathcal{H}$ has hard subset membership.

\subsection{Reducing To Constructing Considerably Simpler Hash}
In this section, we reduce  constructing $SPHDHC_{t,h}$ to constructing considerably simpler hash.

\begin{definition}[smooth projective hash family that holds properties distinguishability and hard subset membership]
\label{SPHDH}
$\mathcal{H}=(PG, IS, DI, KG, Hash, pHash)$ is a  smooth projective hash family that holds properties distinguishability and hard subset membership (SPHDH), if and only if $\mathcal{H}$ is specified as follows
\begin{itemize}
  \item The algorithms $PG$,  $DI$, $KG$, $Hash$, and $pHash$ are specified as same as in $SPHDHC_{t,h}$'s definition, i.e.,   Definition \ref{SPHDHC}.

  \item The instance-sampler $IS$ is a PPT algorithm that takes a security parameter $k$,
        a family parameter $\Lambda$, a work mode $\delta\in \{0,1\}$ as input and outputs a instance along with its witness $(x,w)$, i.e., $(x,w) \leftarrow IS(1^{k},\Lambda, \delta)$.

         Correspondingly, we define relations $R_{\Lambda},\dot{R}_{\Lambda},\ddot{R}_{\Lambda}$
         as follows.
         $\dot{R}_{\Lambda} \stackrel{def}{=} \cup_{k\in \mathds{N}} Rang(IS(1^{k},\Lambda, 0))$,   $\ddot{R}_{\Lambda} \stackrel{def}{=} \cup_{k\in \mathds{N}} Rang(IS(1^{k},\Lambda, 1))$,
         $R_{\Lambda}\stackrel{def}{=} \dot{R}_{\Lambda} \cup \ddot{R}_{\Lambda}$.

\end{itemize}
and $\mathcal{H}$ has the following properties
\begin{enumerate}
  \item The properties projection and  distinguishability are specified as same as in $SPHDHC_{t,h}$'s definition, i.e.,   Definition \ref{SPHDHC}.

  \item Smoothness. Intuitively speaking, it requires that for any instance  $\ddot{x} \in L_{\ddot{R}_{\Lambda}}$, the hash value of $\ddot{x}$  is unobtainable unless its hash key is known. That is, the two probability ensembles  $Sm_{1} \stackrel{def}{=} \{Sm_{1}(1^{k})\}_{k \in \mathds{N}}$ and $Sm_{2} \stackrel{def}{=} \{Sm_{2}(1^{k})\}_{k \in \mathds{N}}$ defined as follows, are computationally indistinguishable, i.e., $Sm_{1} \stackrel{c}{=} Sm_{2}$.

      $Sm_{1}(1^{k})$: $\Lambda \leftarrow PG(1^{k})$, $(\ddot{x},\ddot{w})\leftarrow IS(1^{k},\Lambda,1)$,  $(hk, pk)\leftarrow KG(1^{k},\Lambda, \ddot{x})$, $y \leftarrow Hash(1^{k}, \Lambda, \ddot{x}, hk)$. Finally outputs $(\Lambda, \ddot{x}, pk, y )$.

      $Sm_{2}(1^{k})$: compared with $Sm_{1}(1^{k})$, the only difference is that $y \in _{U} Range( Hash(1^{k}, \Lambda, \ddot{x}, .))$.

  \item Hard Subset Membership. Intuitively speaking, it requires that the instances of
        $L_{\dot{R}_{\Lambda}}$ and that of $L_{\ddot{R}_{\Lambda}}$ are computationally indistinguishable. That is, the two probability ensembles  $Hm_{1} \stackrel{def}{=} \{Hm_{1}(1^{k})\}_{k \in \mathds{N}}$ and $Hm_{2} \stackrel{def}{=} \{Hm_{2}(1^{k})\}_{k \in \mathds{N}}$ defined as follows, are computationally indistinguishable, i.e., $Hm_{1} \stackrel{c}{=} Hm_{2}$.

        $Hm_{1}(1^{k})$: $\Lambda \leftarrow PG(1^{k})$, $(\dot{x}, \dot{w}) \leftarrow IS(1^{k},\Lambda,0)$, finally outputs $(\Lambda,\dot{x} )$.\\
        $Hm_{2}(1^{k})$: $\Lambda \leftarrow PG(1^{k})$, $(\ddot{x}, \ddot{w}) \leftarrow IS(1^{k},\Lambda,1)$, finally outputs $(\Lambda,\ddot{x} )$.
\end{enumerate}
\end{definition}

It is easy to see that the projection and smoothness are two contradictory properties. That is, for any instance $x$, it holds at most one of the two. Therefore, $\dot{R}_{\Lambda} \cap \ddot{R}_{\Lambda} = \emptyset$.

\begin{theorem}[reduce constructing $SPHDHC_{t,h}$  to constructing SPHDH]\label{SPHHCW2SPHDH}
Given a SPHDH $\mathcal{H}$, then we can efficiently gain a  $SPHDHC_{t,h}$ $\mathcal{\overline{H}}$.
\end{theorem}
\begin{IEEEproof}
Let $\mathcal{H}=(PG, IS, DI, KG, Hash, pHash)$. First,  we  construct a new hash system $\mathcal{\overline{H}}=(\overline{PG}, \overline{IS}, \overline{DI}, \overline{KG}, \overline{Hash}, \overline{pHash}, \overline{Cheat})$ as follows.
\begin{itemize}
  \item The procedures $\overline{PG}$,  $\overline{DI}$, $\overline{KG}$, $\overline{Hash}$, $\overline{pHash}$ directly take the corresponding procedures from $\mathcal{H}$.
  \item $\overline{IS}(1^{k}, \Lambda)$: For each $i\in [h]$, $\vec{a}\langle i \rangle \leftarrow IS(1^{k}, \Lambda, 0)$; for each $i\in[n]- [h]$, $\vec{a}\langle i \rangle \leftarrow IS(1^{k}, \Lambda, 1)$; finally outputs $\vec{a}$.
  \item $\overline{Cheat}(1^{k}, \Lambda)$: For each $i\in [n]$, $\vec{a}\langle i \rangle \leftarrow IS(1^{k}, \Lambda, 0)$; finally outputs $\vec{a}$.
\end{itemize}

Second, we prove $\mathcal{\overline{H}}$ is a $SPHDHC_{t,h}$. From the construction, we know that it remains to prove that $\mathcal{\overline{H}}$ holds properties smoothness, hard subset membership and  feasible cheating. However, this fact directly follows Lemma \ref{ind_smoo}, Lemma \ref{hsm} and  Lemma \ref{feasbcheat}. Therefore, $\mathcal{\overline{H}}$ is a $SPHDHC_{t,h}$.
\end{IEEEproof}

Sometimes it is not easy to gain smoothness for a hash family. In this case we have to  construct a hash family, defined as follows, as the first step to our goal.

\begin{definition}[$\epsilon$-universal projective hash family that holds properties distinguishability and hard subset membership]
\label{UPHDH}
$\mathcal{H}=(PG, IS, DI, KG, Hash, pHash)$ is a  $\epsilon$-universal projective hash family that holds properties distinguishability and hard subset membership ($\epsilon$-UPHDH), if and only if $\mathcal{H}$ is specified as follows.
\begin{itemize}
  \item All algorithms  are specified as same as in $SPHDH$'s definition, i.e.,   Definition \ref{SPHDH}.
\end{itemize}
and $\mathcal{H}$ has the following properties
\begin{enumerate}
  \item The properties projection, distinguishability and hard subset membership are specified as same as in   Definition \ref{SPHDH}.
  \item $\epsilon$-universality.  Intuitively speaking, it requires   the probability of guessing the hash value of $\ddot{x}$  is at most $\epsilon$. That is, for any sufficiently large $k$, any $\Lambda \in Range(PG(1^{k}))$, any $\ddot{x} \in Range(IS(1^{k},\Lambda,1))$, any $pk \in Range(KG(1^{k}, \Lambda, \ddot{x})\langle 2 \rangle)$, any $y \in Range( Hash(1^{k}, \Lambda, \ddot{x}, .))$, it holds that
$$ Pr( Hash(1^{k}, \Lambda, \ddot{x}, HK)=y |PK =pk) \leq \epsilon $$
where $(HK,PK) \leftarrow KG (1^{k}, \Lambda, \ddot{x})$, the probability is taken over the randomness of  $KG$.
\end{enumerate}
\end{definition}

Compared with SPHDH,  $\epsilon$-UPHDH relaxes the upper bound of the probability of guessing the hash value  of $\ddot{x}$ to a higher value. Assume $\epsilon<1$, as ~\cite{cramer2002universal,kalai2005ot}, we can efficiently gain a  SPHDH from a $\epsilon$-UPHDH.

\begin{theorem}\label{SPHDH2UPHDH}
Given a $\epsilon$-UPHDH $\mathcal{\widetilde{H}}$, where $\epsilon<1$, then we can efficiently gain a  SPHDH $\mathcal{H}$.
\end{theorem}

The way to prove this theorem is to construct a required algorithm, which can be gained  by a simply application of the Leftover Hash Lemma (please see \cite{luby1996pseudorandomness} for this lemma). The detailed construction essentially is the same as  \cite{cramer2002universal}. Considering the space, we don't iterate it here.

Combining Theorem \ref{SPHHCW2SPHDH} and Theorem \ref{SPHDH2UPHDH}, we have the following corollary.

\begin{corollary}[reduce constructing $SPHDHC_{t,h}$  to constructing $\epsilon$-UPHDH]\label{SPHHCW2UPHDH}
Given a $\epsilon$-UPHDH $\mathcal{\widetilde{H}}$, then we can efficiently gain a  $SPHDHC_{t,h}$ $\mathcal{\overline{H}}$.
\end{corollary}

\section{Constructing $SPHDHC_{t,h}$}\label{con_hash}
In this section, we construct  $SPHDHC_{t,h}$ respectively under the lattice assumption, the decisional Diffie-Hellman assumption, the decisional $N$-th residuosity assumption and the decisional quadratic residuosity assumption. Theorem \ref{SPHHCW2SPHDH} and Corollary \ref{SPHHCW2UPHDH} show that,  to construct a $SPHDHC_{t,h}$, what we need to do is  to construct a SPHDH or  construct a  $\epsilon$-UPHDH ($\epsilon<1$).

\subsection{A Construction Under The Decisional Diffie-Hellman Assumption } \label{hash_ddh}
\subsubsection{Background}
Let $Gen(1^{k})$ be an algorithm such that randomly chooses a cyclic group and outputs the group's  description $G=<g,q,*>$, where $g$, $q$, $*$ respectively is the generator, the order, the operation  of the group.

The DDH  problem is how to construct an algorithm to distinguish the two probability ensembles $DDH_{1} \stackrel{def}{=}\{ DDH_{1}(1^{k}) \}_{k\in\mathds{N}}$ and $DDH_{2}\stackrel{def}{=}\{ DDH_{2}(1^{k}) \}_{k\in\mathds{N}}$ which are formulate  as follows.
\begin{itemize}
  \item $DDH_{1}(1^{k})$: $<g,q,*>\leftarrow Gen(1^{k})$, $a \in_{U} Z_{q}$, $b \in_{U} Z_{q}$, $c \leftarrow ab$, finally outputs $(<g,q,*>,g^{a},g^{b},g^{c})$.
  \item $DDH_{2}(1^{k})$: Basically operates in the same way as $DDH_{1}(1^{k})$ except that
         $c \in_{U} Z_{q}$.
\end{itemize}

At present, there is no efficient algorithm  solving the problem. Therefore, it is assumed that $DDH_{1}\stackrel{c}{=}DDH_{2}$.

\subsubsection{Detailed Construction}
We now present our DDH-based instantiation  of  SPHDH   as follows. For simplicity,  we  assume the groups generated by $Gen(1^{k})$ is of prime order.
\begin{itemize}
  \item $PG(1^{k})$: $\Lambda \leftarrow Gen(1^{k}) $, finally outputs $\Lambda$.
  \item  $IS(1^{k},\Lambda, \delta)$: $(g,q,*) \leftarrow\Lambda$, $a\in_{U} Z_{q}$, $b \in_{U} Z_{q}$, $\dot{x} \leftarrow(g^{a},g^{b}, g^{ab})$, $\dot{w}\leftarrow(a,b)$,
        $c \in_{U}Z_{q}$, $\ddot{x} \leftarrow(g^{a},g^{b}, g^{c})$,
        $\ddot{w} \leftarrow(a,b)$,
        finally outputs  $(\dot{x},\dot{w})$ if $\delta=0$,  $(\ddot{x},\ddot{w})$ if $\delta=1$.
  \item $DI(1^{k},\Lambda,x,w)$: $(g,q,*) \leftarrow\Lambda$,
        $(\alpha, \beta, \gamma)\leftarrow x$,
        $(a,b) \leftarrow w$, if $(\alpha, \beta, \gamma)=(g^{a},g^{b}, g^{ab})$ holds,
        then outputs $0$;
        if $(\alpha, \beta)=(g^{a},g^{b})$ and
        $\gamma \neq g^{ab}$ holds, then outputs $1$.
  \item $KG(1^{k},\Lambda,x)$: $(g,q,*) \leftarrow\Lambda$, $(\alpha, \beta,
        \gamma)\leftarrow x$, $u \in _U Z_q$, $v \in _U Z_q$,
        $pk \leftarrow \alpha^{u}g^{v}$, $hk \leftarrow \gamma^{u}\beta^{v}$,
        finally outputs $(hk,pk)$.
  \item $Hash(1^k,\Lambda,x,hk)$: $y\leftarrow hk$, outputs $y$.
  \item $pHash(1^k ,\Lambda,x,pk,w)$: $(a,b) \leftarrow w$, $y\leftarrow pk^{b}$,
        finally outputs $y$.
\end{itemize}

\begin{lemma}
The hash system holds the property projection.
\end{lemma}

\begin{IEEEproof}
Let $(\dot{x}, \dot{w}) \in Range(IS(1^{k}, \Lambda, 0))$. Let $(hk,pk) \in Range(KG(1^{k},\Lambda,\dot{x}))$. Then,
\begin{equation*}
    \begin{split}
    Hash(1^k,\Lambda,\dot{x},hk)&= Hash(1^k,\Lambda,(g^{a},g^{b},g^{ab}),(g^{abu}g^{bv})) \\
                                &=g^{abu}g^{bv}
   \end{split}
\end{equation*}
\begin{equation*}
    \begin{split}
    pHash(1^k,\Lambda,\dot{x},hk, \dot{w})&= pHash(1^k,\Lambda,(g^{a},g^{b},g^{ab}),\\
                                   &\;\;\;\; (g^{au}g^{v}), (a,b)) \\
                                   &=g^{abu}g^{bv}
   \end{split}
\end{equation*}
That is,
\begin{equation*}
Hash(1^k,\Lambda,\dot{x},hk) =pHash(1^k ,\Lambda,\dot{x},pk,\dot{w})
\end{equation*}
\end{IEEEproof}

\begin{lemma}
Assuming $DDH$ is a hard problem, the hash system holds the property smoothness.
\end{lemma}

\begin{IEEEproof}
For this system, the probability ensembles $Sm_{1}$, $Sm_{2}$ mentioned in the definition of SPHDH  can be described as follows.
\begin{itemize}
  \item $Sm_{1}(1^{k})$: $\Lambda\leftarrow PG(1^{k})$, $(g,q,*) \leftarrow\Lambda$, $a\in_{U} Z_{q}$, $b \in_{U} Z_{q}$, $c \in_{U} Z_{q}$, $\ddot{x} \leftarrow(g^{a},g^{b}, g^{c})$,  $u \in_U Z_q$, $v \in_U Z_q$, $pk \leftarrow g^{au+v}$, $hk \leftarrow g^{cu+bv}$, $y\leftarrow hk$.
 Finally outputs $(\Lambda,\ddot{x},pk,y)$.
  \item $Sm_{2}(1^{k})$: Operates as same as $Sm_{1}(1^{k})$  with an exception that $y$ is  generated as follows. $d \in_{U} Z_{q}$, $y \leftarrow g^{d}$.
\end{itemize}
Because $b,c,u,v$ are chosen uniformly and $q$ is prime, both $cu$ and $bv$ are uniformly distributed over $Z_{q}$. Thus $cu+bv$ is uniformly distributed over $Z_{q}$ too. Therefore, $Sm_{1}  \equiv Sm_{2}$.
\end{IEEEproof}

\begin{lemma}
The hash system holds the property distinguishability.
\end{lemma}

The proof of this lemma is trivial, so we omit it.

\begin{lemma}
Assuming $DDH$ is a hard problem, the hash system holds the property hard subset membership.
\end{lemma}

\begin{IEEEproof}
For this system, the probability ensembles $Hm_{1}$, $Hm_{2}$  mentioned in the definition of SPHDH  can be described as follows.
\begin{itemize}
  \item $Hm_{1}(1^{k})$: $\Lambda\leftarrow PG(1^{k})$, $(g,q,*) \leftarrow\Lambda$, $a\in_{U} Z_{q}$, $b \in_{U} Z_{q}$, $\dot{x} \leftarrow(g^{a},g^{b}, g^{ab})$.
Finally outputs $(\Lambda,\dot{x})$.
  \item $Hm_{2}(1^{k})$: $\Lambda\leftarrow PG(1^{k})$,
$(g,q,*) \leftarrow\Lambda$, $a\in_{U} Z_{q}$, $b \in_{U} Z_{q}$, $c \in_{U}Z_{q}$, $\ddot{x} \leftarrow(g^{a},g^{b}, g^{c})$. Finally outputs $(\Lambda,\ddot{x})$.
\end{itemize}
Obviously, $Hm_{1}\stackrel{c}{=} Hm_{2}$.
\end{IEEEproof}

Combining all lemmas above, we have the following theorem.
\begin{theorem}  \label{theo_lattice_SPWH}
Assuming $DDH$ is a hard problem,  the hash system is a SPHDH.
\end{theorem}

\subsubsection{A Concrete Protocol For $OT^{n}_{h}$ Based On DDH}
It's known that the  encryption scheme presented by \cite{elgamal1985public} can be used as
a perfectly binding commitment scheme.  The  encryption scheme is   directly based on the problem of discrete log. Since the task of solving the problem DDH can be reduced to that of solving the problem discrete log, the encryption scheme is based on DDH essentially. The DDH-based commitment scheme presented by \cite{pedersen1991non} is a  perfectly hiding one. Therefore, using those two commitment schemes and our DDH-based $SPHDHC_{t,h}$, we gain  a  concrete protocol for $OT^{n}_{h}$ based only on DDH.  To reach the best efficiency, we should use the DDH of the group which is on elliptic curves. See Section \ref{overhead}
for further discussion.

\subsection{A Construction Under Lattice}\label{hash_lwe}
\subsubsection{Background}
Learning with errors (LWE) is an average-case problem. \cite{regev2009lattices} shows that its
hardness is implied by the worst-case hardness of standard lattice problem for quantum algorithms.

In lattice, the modulo operation is defined as $x \mod  y \stackrel{def}{=} x - \llcorner x/y \lrcorner y $. Then we know $x \mod  1 \stackrel{def}{=} x - \llcorner x \lrcorner$.
Let $\beta $ be an arbitrary positive real number. Let $\Psi_{\beta}$ be
a probability density function whose  distribution is over $[0,1)$ and obtained by  sampling from a normal variable with mean $0$ and standard deviation $\beta/\sqrt{2\pi}$ and reducing the result modulo 1, more specifically
\begin{align*}
  \Psi_{\beta}&:[0,1) \rightarrow R^{+} \\
    \Psi_{\beta}(r)& \stackrel{def}{=} \sum_{k=-\infty}^{\infty} \frac{1}{\beta} \exp (-\pi(\frac{r-k}{\beta})^{2 })
\end{align*}

Given  an arbitrary integer $q \geq 2$, an arbitrary probability destiny function $\phi:[0,1) \rightarrow R^{+}$,  the discretization of $\phi$ over $Z_{q}$ is defined as
\begin{align*}
  \bar{\phi} &: Z_{q} \rightarrow R^{+} \\
    \bar{\phi}(i) & \stackrel{def}{=}  \int^{(i+1/2)/q}_{(i-1/2)/q} \phi(x)dx
\end{align*}

$LWE$ can be formulated as follows.

\begin{definition}[Learning With Errors]
Learning with errors problem ($LWE_{q,\chi}$) is how to construct an efficient algorithm that
receiving $q,g,m,\chi, (\vec{a}_{i}, b_{i})_{i\in [m]}$, outputs $\vec{s}$ with nonnegligible probability. The input and the output is specified in the following way.

$q\leftarrow q(1^{k})$, $g\leftarrow g(1^{k})$, $m\leftarrow poly(1^{k})$, $\chi \leftarrow \chi(1^{k})$, $\vec{s} \in_{U} (Z_{q})^{k}$. For each $i\in [m]$,  $\vec{a}_{i} \in_{U} (Z_{q})^{k}$, $e_{i} \in_{\chi} Z_{q}$, $b_{i} \leftarrow \vec{s}^{T}\cdot \vec{a}_{i}+ e_{i} \mod  q$.

where $q,g$ are positive integers, $\chi: Z_{q} \rightarrow R^{+}$ is a probability density function.
\end{definition}

With respect to the hardness of $LWE$, \cite{regev2009lattices} proves that setting appropriate parameters, we can reduce  two worst-case standard lattice problems to $LWE$, which means $LWE$ is
a very hard problem.

\begin{lemma}[\cite{regev2009lattices}] \label{lwe2sivp}
Setting security parameter $k$ to be a value such that $q$ is a prime, $\beta \leftarrow \beta(1^{k})$, $\beta \in(0,1)$, and $\beta \cdot q > 2 \sqrt{k}$. Then the lattice problems $SIVP$ and $GapSVP$
can be reduced to $LWE_{q,\bar{\Psi}_{\beta}}$. More specifically, if there exists an efficient
(possibly quantum) algorithm that solves $LWE_{q,\bar{\Psi}_{\beta}}$, then there exists an efficient quantum algorithm solving the following worst-case lattice problems in the $l_{2}$ norm.
\begin{itemize}
  \item SIVP: In any lattice $\Lambda$ of dimension $k$, find a set of $k$ linearly independent
 lattice vectors of length within at most $\tilde{O}(k/\beta)$ of optimal.
  \item GapSVP: In any lattice $\Lambda$ of dimension $m$, approximate the length of a shortest nonzero lattice vector to within a $\tilde{O}(k/\beta)$ factor.
\end{itemize}
\end{lemma}

We emphasize the fact that the reduction of Lemma \ref{lwe2sivp} is quantum,  which implies
that any algorithm  breaking any cryptographic schemes which only based on $LWE$ is an
algorithm solving at least one of the problems SIVP and GapSVP.

%Let $LWE_{1} \stackrel{def}{=}\{ LWE_{1}(1^{k})\}_{k \in \mathds{N}}$ and $LWE_{2} \stackrel{def}{=} \{ LWE_{2}(1^{k})\}_{k \in \mathds{N}}$ be two probability ensembles defined as follows.
%\begin{itemize}
%  \item $LWE_{1}(1^{k})$: $q\leftarrow q(1^{k})$, $g\leftarrow g(1^{k})$, $m\leftarrow poly(1^{k})$, $\phi \leftarrow \phi(1^{k})$ and $\phi:[0,1) \rightarrow R^{+}$ is a probability density function, $\chi \leftarrow \bar{\phi}$, $\vec{s} \in_{U} (Z_{q})^{k}$, $\vec{a}_{i} \in_{U} (Z_{q})^{k}$, $e_{i} \in_{\chi} Z_{q}$, $b_{i} \leftarrow \vec{s}^{T}\cdot \vec{a}_{i}+ e_{i} \mod  q $. Finally outputs$(
%      (q,g,m,\phi),(\vec{a}_{i}, b_{i})_{i \in [m]})$.
%  \item $LWE_{2}(1^{k})$: Operates in the same way as $LWE_{1}(1^{k})$, except that
%        $b_{i} \in_{U} Z_{q}$.
%\end{itemize}
%
%\cite{regev2009lattices} shows that setting appropriate parameters, $LWE_{1}$ and $LWE_{2}$ are computationally indistinguishable.
%
%\begin{lemma}[\cite{regev2009lattices}] \label{lwe2ind}
%If $k \geq 1$ is an integer and $2 \leq q \leq poly_{2}(k)$ is a prime. Then solving
%$LWE_{q,\phi}$ can be reduced to distinguishing  $LWE_{1}$ and $LWE_{2}$. More specifically, if there exists a distinguisher for $LWE_{1}$ and $LWE_{2}$, then, there exists an efficient algorithm that solves $LWE_{q,\phi}$.
%\end{lemma}

How  to precisely set the parameters as values to gain a concrete $LWE$,  which is
as hard as required in Lemma \ref{lwe2sivp} is beyond the scope of this
paper. To  see such examples and more details, we recommend \cite{regev2009lattices} and \cite{peikert2008ot}.

The instantiation of \emph{SPHDH}, which we will present soon, needs to use a $LWE$-based public key cryptosystem  presented by \cite{Gentry2008Trapdoors}, which is a slight variant of \cite{regev2009lattices}'s cryptosystem.  This cryptosystem  is described as follow.
\begin{itemize}
  \item Message space: $\{0,1\}$.
  \item $Setup(1^{k})$: $q\leftarrow q(1^{k})\wedge
q \in \mathds{P}\wedge q\in[k^{2}, 2k^{2}]$, $m\leftarrow (1+\varepsilon)(k+1) \log q$ ( where $\varepsilon>0$ is an arbitrary constant), $\chi \leftarrow \bar{\Psi}_{\alpha(k)} \wedge \alpha(k) = o(1/(\sqrt{k} \log k))$  (e.g., $\alpha(k)=\frac{1}{\sqrt{k} (\log k)^{2}}$).
    $para \leftarrow (q,m,\chi)$, finally outputs $para$.
  \item $KeyGen(1^{k}, para)$: $A \in_{U}(Z_{q})^{m \times k}$,
        $\vec{s}\in_{U}(Z_{q})^{k}$, $\vec{e} \in_{\chi} (Z_{q})^{m}$ (which means each entry of $\vec{e}$ is independently drawn from  $Z_{q}$ according to $\chi$),
        $\vec{b}\leftarrow A \vec{s}+ \vec{e} \mod  q$,
        $pubk\leftarrow (A, \vec{b})$, $sk\leftarrow \vec{s}$,
        finally outputs a public-private key pair $(pubk, sk)$.
  \item $Enc(.)$, $Dec(.)$: Since $Enc(.)$, $Dec(.)$ are immaterial to understand this paper, we omit their detailed procedure here.
\end{itemize}

\cite{Gentry2008Trapdoors} shows that if $LWE_{q,\bar{\Psi}_{\alpha}}$ is hard,  choosing appropriate parameters, this cryptosystem holds the following properties.
\begin{enumerate}
  \item It provides security against chosen plaintext attack, though we only need semantic security here.
  \item For each  $A \in (Z_{q})^{m \times k}$,  we have
  \begin{equation*}
    Pr(\vec{b} \textrm{ is messy} | \vec{b}\in_{U}  (Z_{q})^{m} ) \geq 1-2/q^{k},
  \end{equation*}
where $\vec{b}$  is said to be messy if and only if,  $\forall m_{0},  m_{1} \in \{0,1\}$, the statistical distance between the  distribution of $Enc_{A,\vec{b}}(m_{0})$ and that of $Enc_{A,\vec{b}}(m_{1})$ is negligible. In other word, $\vec{b}$  is said to be messy if and only if, $Enc_{A,\vec{b}}(. )$ loses messages and so its ciphertext can't be decrypted using any private key $\vec{s}\in(Z_{q})^{k}$.
\item Given  $A \in (Z_{q})^{m \times k}$ and its trapdoor $T$, then there exists an efficient decision  algorithm $IsMessy$ holds the following two property. First, $Pr(IsMessy(A,T, \vec{b})=0 | \vec{b}\in_{U}  (Z_{q})^{m})$ is negligible. Second, $\vec{b}$ is indeed messy  if $IsMessy(A,T, \vec{b})=1$.
\end{enumerate}

\subsubsection{Detailed Construction}
We now present our LWE-based instantiation  of  \emph{SPHDH} as follows.
\begin{itemize}
  \item $PG(1^{k})$: $\Lambda \leftarrow Setup(1^{k })$, finally outputs $\Lambda$.
 \item  $IS(1^{k},\Lambda, b)$: $(q,m,\chi)\leftarrow \Lambda$, $A \in_{U}(Z_{q})^{m \times k}$ along with its trapdoor $T$, $\vec{s}\in_{U}(Z_{q})^{k}$,  $\vec{e} \in_{\chi} (Z_{q})^{m}$,
        $\dot{x}\leftarrow (A, A \vec{s}+ \vec{e} \mod q)$, $\dot{w}\leftarrow (0,\vec{s})$,
        uniformly chooses  $\vec{b}\in (Z_{q})^{m}$ such that $IsMessy(A,T, \vec{b})=1$ (recall that only negligible fraction of $\vec{b}$ are not messy, therefore such $\vec{b}$ can be efficiently chosen) ,
        $\ddot{x} \leftarrow (A, \vec{b})$, $\ddot{w}\leftarrow (1, T)$,
        finally outputs  $(\dot{x},\dot{w})$ if $b=0$,  $(\ddot{x},\ddot{w})$ if $b=1$.
  \item $DI(1^{k},\Lambda,x,w)$: $(q,m,\chi)\leftarrow \Lambda$, $(A, \vec{b})\leftarrow x$,
        $(i,\varrho) \leftarrow w $, if $i=1$ and $IsMessy(A,\varrho, \vec{b})=1$ holds, then outputs $1$; otherwise outputs $0$.
  \item $KG(1^{k},\Lambda,x)$: $(q,m,\chi) \leftarrow \Lambda$, $(A, \vec{b})\leftarrow x$,
        $a \in_U \{0,1\}$, $\alpha  \leftarrow Enc_{A,\vec{b}}(a)$, $hk \leftarrow a$,
       $pk \leftarrow \alpha $, finally outputs $(hk,pk)$.
  \item $Hash(1^k,\Lambda,x,hk)$: $(q,m,\chi)\leftarrow \Lambda$, $a \leftarrow hk$,
        finally outputs $a$.
  \item $pHash(1^k ,\Lambda,x,pk,w)$: $(m,q,\chi) \leftarrow \Lambda$, $\alpha \leftarrow pk$,
  $(i, \varrho) \leftarrow w$, $a \leftarrow Dec_{\varrho} (\alpha )$,  finally outputs $a$.
\end{itemize}

In the above construction of SPHDH, each instance holds a matrix $A$, which seems  expensive. However,
in the corresponding construction of $SPHDHC_{t,h}$, this overhead can be reduced by each instance vector sharing a matrix $A$. We point out that it's not secure that all instance vectors share a matrix $A$. The reason is that in this case,  seeing matrix $A$'s  trapdoor $T$ in Step S2 of the framework,  $P_{1}$ can distinguish  smooth instances and projective instances of the unchosen instance vectors, which leads to $P_{1}$ deducing $P_{2}$'s private input.

\begin{lemma}
Assuming $LWE$ is a hard problem, the hash system holds the property projection.
\end{lemma}

\begin{IEEEproof}
Let $\dot{x}=(A,\vec{b}) \in Range(IS(1^{k}, \Lambda, 0))$, $\dot{w}=(0,\vec{s})$. Obviously, $((A,\vec{b}),\vec{s})$ is a correct public-private key pair. Then, we have
\begin{equation*}
    Hash(1^k,\Lambda,\dot{x},hk)= a,
\end{equation*}
\begin{equation*}
\begin{split}
   pHash(1^k ,\Lambda,\dot{x},pk,\dot{w})&=Dec_{\vec{s}} (\alpha ) \\
    &= Dec_{\vec{s}} (Enc_{A,\vec{b}}(a) )\\
    & =a,
\end{split}
\end{equation*}
This means that for any $(\dot{x},\dot{w},\Lambda)$ generated by the hash system, it holds that
\begin{equation*}
Hash(1^k,\Lambda,\dot{x},hk) =pHash(1^k ,\Lambda,\dot{x},pk,\dot{w}).
\end{equation*}
\end{IEEEproof}

\begin{lemma}
The hash system holds the property smoothness.
\end{lemma}

\begin{IEEEproof}
For this system, the probability ensembles $Sm_{1}$, $Sm_{2}$ mentioned in the definition of SPHDH  can be described as follows.
\begin{itemize}
  \item $Sm_{1}(1^{k})$: $\Lambda\leftarrow PG(1^{k})$, $(q,m,\chi)\leftarrow \Lambda$, $A \in_{U}(Z_{q})^{m \times k}$ along with its trapdoor $T$, uniformly chooses  $\vec{b}\in (Z_{q})^{m}$ such that $IsMessy(A,T, \vec{b})=1$,
        $\ddot{x} \leftarrow (A, \vec{b})$, $\ddot{w}\leftarrow (1, T)$, $a \in_U \{0,1\}$, $\alpha  \leftarrow Enc_{A,\vec{b}}(a)$, $pk \leftarrow \alpha $, $y\leftarrow a$, finally outputs $(\Lambda,\ddot{x},pk,y)$.
  \item $Sm_{2}(1^{k})$: Operates as same as $Sm_{1}(1^{k})$  with an exception that $y \in_{U} \{0,1\}$.
\end{itemize}
Obviously, $Sm_{1}(1^{k}) \equiv Sm_{2}(1^{k})$.
\end{IEEEproof}

\begin{lemma} \label{DI_LWE}
Assuming $LWE$ is a hard problem, the hash system holds the property distinguishability.
\end{lemma}

\begin{IEEEproof}
Recalling the property of $IsMessy$, we know if ($A,\vec{b}$) isn't messy,  $IsMessy(A,T, \vec{b})=0$; if ($A,\vec{b}$) is messy,  $IsMessy(A,T, \vec{b})=1$  with a probability close to $1$. Thus, if  $(x,w)\in \ddot{R}_{\Lambda}$, $DI$ outputs $1$; if  $(x,w)\in \dot{R}_{\Lambda}$, $DI$ outputs $0$. $DI$ correctly
computes $\zeta$.
\end{IEEEproof}

\begin{lemma} \label{hsm_LWE}
Assuming $LWE$ is a hard problem, the hash system holds the property hard subset membership.
\end{lemma}

\begin{IEEEproof}
For this system, the probability ensembles $Hm_{1}$, $Hm_{2}$  mentioned in the definition of SPHDH  can be described as follows.
\begin{itemize}
  \item $Hm_{1}(1^{k})$: $\Lambda\leftarrow PG(1^{k})$, $(q,m,\chi)\leftarrow \Lambda$, $A \in_{U}(Z_{q})^{m \times k}$ along with its trapdoor $T$, $\vec{s}\in_{U}(Z_{q})^{k}$,  $\vec{e} \in_{\chi} (Z_{q})^{m}$,
        $\dot{x}\leftarrow (A, A \vec{s}+ \vec{e} \mod q)$, finally outputs $(\Lambda,\dot{x})$.
  \item $Hm_{2}(1^{k})$: $\Lambda\leftarrow PG(1^{k})$, $(q,m,\chi)\leftarrow \Lambda$, $A \in_{U}(Z_{q})^{m \times k}$ along with its trapdoor $T$, uniformly chooses  $\vec{b}\in (Z_{q})^{m}$ such that $IsMessy(A,T, \vec{b})=1$,
        $\ddot{x} \leftarrow (A, \vec{b})$,  finally outputs $(\Lambda,\ddot{x})$.
\end{itemize}
Obviously,  $Hm_{1}\stackrel{c}{=} Hm_{2}$.
\end{IEEEproof}

Combining  Lemma \ref{lwe2sivp}  and  above lemmas, we have the following theorem.
\begin{theorem}  \label{theo_lattice_SPWH}
If $SIVP$ or $GapSVP$ is a hard problem, the hash system is a SPHDH.
\end{theorem}

\subsubsection{A Concrete Protocol For $OT^{n}_{h}$ With Security  Against Quantum Algorithms}
The security proof of the framework guarantees that, any adversary
breaking the framework is an algorithm breaking  at least  one of cryptographic tools used in the
framework. Moreover, it's generally believed that lattice-based cryptography resists quantum attacks \cite{Micciancio2009Lattice}. Therefore, to gain an instantiation of our framework with security against quantum algorithms, it suffices to adopt lattice-based instantiations of the cryptographic tools. Thus, it remains to find a $IHC$ and a $IBC$ with such security level.
Though there exists  general methods  to construct perfectly binding commitments and perfectly hiding commitments from one-way functions (or one-way  permutations) (see \cite{gold2001found} Chapter 4), the resulting  lattice-based commitments seem too expensive. Thus, other approach is needed.

First, based on the result of \cite{Gentry2008Trapdoors},  we can get  a relatively efficient  statically binding commitment.

\begin{lemma}[\cite{Gentry2008Trapdoors}]\label{uniquesecret}
There exists an efficient algorithm for the lattice-based cryptosystem mentioned early  such that, for all but at most negligible fraction of public key generated by $KeyGen$, given a trapdoor for the  matrix $A$, and a  public key $(A, A^{T}\vec{s}+\vec{e})$, it can efficiently extract the unique  secret key $\vec{s}$.
\end{lemma}

The lattice-based cryptosystem can be used as a statically binding commitment in the following way. In commit phase, $P_{1}$ sends a public key $(A,\vec{b})$ along with $E_{A,\vec{b}}(m)$ to $P_{2}$.  The computationally hiding directly follows the security level of the cryptosystem. In reveal phase, $P_{1}$ sends the trapdoor of $A$, the value $m$, and the randomness used in commit phase to $P_{2}$. Following Lemma \ref{uniquesecret}, almost all legitimate public keys, respectively, correspond to  a unique private key. This guarantees that an encryption relative to legitimate public keys  have a unique decryption. Therefore, it holds statically binding.

Second, combining the works of \cite{Halevi1996Commitment,ajtai1996generating,goldreich1996collision,micciancio2008worst}, we can get a relatively  efficient  statically hiding commitment. \cite{Halevi1996Commitment} presents a efficient way to construct  statically hiding commitments from any collision-free hash. Assuming one of the lattice problems $SIVP$ and  $SVP$ is hard,  \cite{goldreich1996collision} shows that \cite{ajtai1996generating}'s lattice-based hash of suitably chosen parameters  is  collision-free. Under the assumption that $GapSVP^{2}_{\tilde{O}(n)}$ is hard in the worst case,  \cite{micciancio2008worst} later also shows that the lattice-based hash is collision-free.  Therefore, applying \cite{Halevi1996Commitment}'s method to \cite{ajtai1996generating}'s  hash, we get a lattice-based statically hiding commitment.

Now we can gain a concrete   protocol for $OT^{n}_{h}$ with security against quantum algorithms, this is summarized by the following theorem.

\begin{theorem} \label{ot_quantum}
Assuming that one of the lattice problems $SIVP$ and $GapSVP$ is hard for quantum algorithms,
instantiating the $OT^{n}_{h}$ framework with the lattice-based $SPHDHC_{t,h}$, and the lattice-based  commitment schemes  (no matter the ones that are got by applying the general method or the ones we suggests above), the resulting concrete protocol for $OT^{n}_{h}$ is secure against quantum algorithms.
\end{theorem}

\subsection{A Construction Under The Decisional N-th Residuosity  Assumption}\label{hash-nr}
\subsubsection{Verifiable-$\epsilon$-universal Projective Hash Family}
In this section, we  will  build a instantiation of $\epsilon$-UPHDH ($\epsilon<1$) from
a instantiation of a hash system called  verifiable-$\epsilon$-universal projective hash family by \cite{kalai2005ot}. Therefore, it is necessary to introduce the definition of this hash system.

\begin{definition}[verifiable-$\epsilon$-universal projective hash family, \cite{kalai2005ot}]\label{VUPH}
$\mathcal{H}=(PG, IS, IT, KG, Hash, pHash)$ is a  $\epsilon$-universal projective hash family ($\epsilon$-VUPH), if and only if $\mathcal{H}$ is specified as follows.
\begin{itemize}
  \item The algorithms $PG$, $IS$, $KG$, $Hash$, $pHash$ are specified as same as in $\epsilon$-UPHDH's definition, i.e.,   Definition \ref{UPHDH}.
  \item $IS$ is a PPT algorithm that takes a security parameter $k$,
        a family parameter $\Lambda$ as input and outputs a tuple, i.e., $(\dot{w},\dot{x},\ddot{x})\leftarrow  IS(1^{k},\Lambda)$.
  \item $IT$ is a PPT algorithm that takes a security parameter $k$,
        a family parameter $\Lambda$, two instances as input and outputs a bit , i.e., $b\leftarrow  IT(1^{k},\Lambda, x_{1},x_{2})$.
\end{itemize}
and $\mathcal{H}$ has the following properties
\begin{enumerate}
  \item The properties projection,  $\epsilon$-universality are specified as same as that in $\epsilon$-UPHDH's definition, i.e.,   Definition \ref{UPHDH}.
  \item Verifiability. First, for any sufficiently large $k$, any $\Lambda \in Range(PG(1^{k}))$, any $(\dot{w},\dot{x},\ddot{x})\in Range (IS(1^{k},\Lambda))$, it holds that $IT(1^{k},\Lambda, \dot{x},\ddot{x})=IT(1^{k},\Lambda, \ddot{x}, \dot{x})=1$. Second,  for any sufficiently large $k$, any $(\Lambda, x_{1},x_{2})$ such that $IT(1^{k},\Lambda, x_{1},x_{2})=1$, at least one of
      $x_{1},x_{2}$ is $\epsilon$-universal.
\end{enumerate}
\end{definition}

It is easy to see that  verifiability guarantees any instance $x$ holds  at most one of the properties
projection and universality. Therefore, we have the following lemma.

\begin{lemma} \label{llempet}
Let $\mathcal{H}=(PG, IS, IT, KG, Hash, pHash)$ be a  $\epsilon$-universal projective hash family, then
\begin{equation*}
\dot{L} \cap  \ddot{L} = \emptyset
\end{equation*}
where $\dot{L} \stackrel {def}{=} \{\dot{x}| \Lambda \leftarrow PG(1^{k}), (\dot{w},\dot{x},\ddot{x})\leftarrow  IS(1^{k},\Lambda) \}$ and  $\ddot{L} \stackrel {def}{=} \{\ddot{x}| \Lambda \leftarrow PG(1^{k}), (\dot{w},\dot{x},\ddot{x})\leftarrow  IS(1^{k},\Lambda)\}$.
\end{lemma}

\subsubsection{Background} \label{forQR}
Let $Gen(1^{k})$ be an algorithm that operates as follows.
\begin{itemize}
  \item $Gen(1^{k})$: $(p,q)\in_{U}\{(p,q)|(p,q)\in (\mathds{P},\mathds{P}),p,q>2, |p|=|q|=k,\gcd (pq,(p-1)(q-1)) =1\} $, $N\leftarrow pq$, finally outputs
      $N$.
\end{itemize}
The problem decisional N-th residuosity (DNR), first presented by \cite{paillier1999cr}, is how
to construct an algorithm to distinguish  two probability ensembles $DNR_{1} \stackrel{def}{=}\{ DNR_{1}(1^{k}) \}_{k\in\mathds{N}}$ and $DNR_{2}\stackrel{def}{=}\{ DNR_{2}(1^{k}) \}_{k\in\mathds{N}}$ which are formulate  as follows.
\begin{itemize}
  \item $DNR_{1}(1^{k})$: $N \leftarrow Gen(1^{k})$, $a \in_{U} Z^{*}_{N^{2}}$, $b \leftarrow a^{N} \mod {N^{2}}$, finally outputs $(N,b)$.
  \item $DNR_{2}(1^{k})$: $N \leftarrow Gen(1^{k})$, $b \in_{U} Z^{*}_{N^{2}}$, finally outputs $(N,b)$.
\end{itemize}

The DNR assumption is that there is no efficient algorithm solving the problem. In other words, it is assumed that $DNR_{1}\stackrel{c}{=}DNR_{2}$.

Our  instantiation of  $\epsilon$-UPHDH is build from a DNR-based instantiation of $\varepsilon$-VUPH  ($\varepsilon < 1$) presented by \cite{kalai2005ot}.  The instantiation of  $\varepsilon$-VUPH is stated as follows.
\begin{itemize}
    \item $PG(1^{k})$: $N \leftarrow Gen(1^{k})$, $a \in_{U} Z^{*}_{N^{2}}$, $T \leftarrow N^{\ulcorner 2\log N \urcorner}$, $g \leftarrow a^{N \cdot T} \mod{N^{2}}$,
        $\Lambda\leftarrow (N, g)$,  finally outputs  $\Lambda$.
   \item $IS(1^{k},\Lambda)$: $(N, g) \leftarrow \Lambda$, $r,v \in _U Z_{N}^*$,
         $w\leftarrow r$,    $\dot{x} \leftarrow
        g^{r} \mod N^2$, $\ddot{x}\leftarrow \dot{x}(1 + vN) \mod N^2$, finally outputs $(w,\dot{x},\ddot{x})$.
  \item $IT(1^{k},\Lambda,\dot{x},\ddot{x})$: $(N, g) \leftarrow \Lambda$. Checks that $N >2^{2k}$,
        $g,\dot{x}\in Z^{*}_{N^{2}}$. $d\leftarrow \ddot{x}/\dot{x} \mod N^{2}$ and checks $N|(d-1)$.
        $v\leftarrow (d-1)/N$ and checks $\gcd (v,N)=1$. Outputs $1$ if all the test pass
        and $0$ otherwise.
  \item $KG(1^{k},\Lambda)$: $(N, g) \leftarrow \Lambda$, $hk \in_U Z_{N^2}$,
        $pk \leftarrow g^{hk} \mod N^2$, finally outputs $(hk,pk)$.
  \item $Hash(1^k,\Lambda,x,hk)$: $(N,g) \leftarrow \Lambda$, $y\leftarrow x^{hk}\mod N^2$,   finally outputs $y$.
  \item $pHash(1^k ,\Lambda,x,pk,w)$: $(N, g) \leftarrow \Lambda$, $y\leftarrow pk^w \mod
        {N^2}$, finally outputs $y$.
\end{itemize}

\subsubsection{Detailed Construction}
We now present our DNR-based instantiation  of  $\epsilon$-UPHDH ($\epsilon<1$)  as follows.
\begin{itemize}
  \item $PG(1^{k})$: $N \leftarrow Gen(1^{k})$, $a \in_{U} Z^{*}_{N^{2}}$, $T \leftarrow N^{\ulcorner 2\log N \urcorner}$, $g \leftarrow a^{N \cdot T} \mod{N^{2}}$,
        $\Lambda\leftarrow (N, g)$,  finally outputs  $\Lambda$.
 \item  $IS(1^{k},\Lambda, \delta)$: $(N, g) \leftarrow \Lambda$, $r \in _U Z_{N}^*$, $\dot{x} \leftarrow g^{r} \mod N^2$,
        $\dot{w}\leftarrow (r, 0)$, $v \in_U Z_{N}^*$, $\ddot{x}\leftarrow g^{r}(1 + v N)
        \mod N^2$, $\ddot{w} \leftarrow(r,v)$,
       finally outputs  $(\dot{x},\dot{w})$ if $\delta=0$,  $(\ddot{x},\ddot{w})$ if $\delta=1$.
  \item $DI(1^{k},\Lambda,x,w)$: $(N, g) \leftarrow \Lambda$, $(r, v) \leftarrow w$,
        \begin{enumerate}
          \item if $v=0 \mod N$, operates as follows: checks that $N >2^{2k}$, $g,x\in Z^{*}_{N^{2}}$, $r\in Z_{N}^*$, $x = g^{r} \mod N^2$. Outputs $0$ if all the test pass.
          \item if $v\neq 0 \mod N$, operates as follows: checks that $N >2^{2k}$, $g,x\in Z^{*}_{N^{2}}$, $r\in Z_{N}^*$, $x = g^{r}(1+vn) \mod N^2$. Outputs $1$ if all the test pass.
        \end{enumerate}
  \item $KG(1^{k},\Lambda,x)$: $(N, g) \leftarrow \Lambda$, $hk \in_U Z_{N^2}$,
        $pk \leftarrow g^{hk} \mod N^2$, finally outputs $(hk,pk)$.
  \item $Hash(1^k,\Lambda,x,hk)$: $(N,g) \leftarrow \Lambda$, $y\leftarrow x^{hk}\mod N^2$,
        finally outputs $y$.
  \item $pHash(1^k ,\Lambda,x,pk,w)$: $(N, g) \leftarrow \Lambda$, $y\leftarrow pk^w \mod
        {N^2}$, finally outputs $y$.
\end{itemize}

\begin{theorem}  \label{DNR_is_hash}
Assuming  DNR is a hard problem, the hash system is a $\epsilon$-UPHDH ($\epsilon <1$).
\end{theorem}
\begin{IEEEproof}
It is easy to see that the   hash system directly inherits properties $\varepsilon$-universality and projection from  the instantiation  of $\epsilon$-VUPH. Following Lemma \ref{llempet}, the hash system
holds property distinguishability. It remains to prove that the hash system holds the property hard subset membership.

For this system, the probability ensembles $Hm_{1}$, $Hm_{2}$  mentioned in the definition of $\epsilon$-UPHDH  can be described as follows.
\begin{itemize}
  \item $Hm_{1}(1^{k})$: $\Lambda\leftarrow PG(1^{k})$, $(N, g)
        \leftarrow \Lambda$, $r \in_{U} Z_{N}^{*}$, $\dot{x} \leftarrow
        g^{r} \mod N^{2}$. Finally outputs $(\Lambda,\dot{x})$.
  \item $Hm_{2}(1^{k})$: $\Lambda\leftarrow PG(1^{k})$, $(N, g)
        \leftarrow \Lambda$, $r, v \in_{U} Z_{N}^{*}$, $\ddot{x}
        \leftarrow  g^{r}(1+ v N) \mod N^{2}$.
        Finally outputs $(\Lambda,\ddot{x})$.
\end{itemize}
It is clear that $Hm_{1}\stackrel{c}{=} Hm_{2}$. Therefore, the hash system holds the property hard subset membership.
\end{IEEEproof}

\subsection{A Construction Under The Decisional Quadratic Residuosity Assumption} \label{hash_qr}
We reuse $Gen(1^{k})$  defined in section \ref{forQR}. Let $J_{N}$ be the subgroup of $Z^{*}_{N}$ of  elements with  Jacobi symbol  $1$. The problem decisional quadratic residuosity (DQR) is how to construct an algorithm to distinguish the two probability ensembles $DQR_{1} \stackrel{def}{=}\{ DQR_{1}(1^{k}) \}_{k\in\mathds{N}}$ and $DQR_{2}\stackrel{def}{=}\{ DQR_{2}(1^{k}) \}_{k\in\mathds{N}}$ which are formulated  as follows.
\begin{itemize}
 \item $DQR_{1}(1^{k})$: $N \leftarrow Gen(1^{k})$, $x \in_{U} J_{N}$, finally outputs $(N,x)$.
  \item $DQR_{2}(1^{k})$: $N \leftarrow Gen(1^{k})$, $r \in_{U} Z^{*}_{N}$,
        $x \leftarrow r^{2}\mod N$, finally outputs $(N,x)$.
\end{itemize}

The DQR assumption is that there is no efficient algorithm solving the problem. That is, it is assumed that $DQR_{1}\stackrel{c}{=}DQR_{2}$.

As in section \ref{hash-nr}, the hash system  we aim to achieve is an instantiation of  $\epsilon$-UPHDH. We will build it on an instantiation of $\varepsilon$-VUPH presented by \cite{kalai2005ot} which is constructed under DQR assumption. Considering the space, we do not iterate the instantiation of $\varepsilon$-VUPH here, and directly present our instantiation of  $\epsilon$-UPHDH  as follows.
\begin{itemize}
  \item $PG(1^{k})$: $(p,q)\in_{U}(\mathds{P},\mathds{P})$, where $|p|=|q|=k$,
        $p<q<2p-1$, $p=q=3 \mod 4$,  $a \in_{U} Z^{*}_{N}$, $T \leftarrow 2^{\ulcorner
        \log N \urcorner}$, $g \leftarrow a^{2 \cdot T} \mod N$,
        $\Lambda\leftarrow (N, g)$,  finally outputs  $\Lambda$.
 \item  $IS(1^{k},\Lambda, \delta)$: $(N, g) \leftarrow \Lambda$, $r \in _U Z_{N}$, $\dot{x} \leftarrow g^{r} \mod N$,  $\ddot{x}\leftarrow N- g^{r} \mod N$, $\ddot{w} \leftarrow r$ ,
        finally outputs  $(\dot{x},\dot{w})$ if $\delta=0$,  $(\ddot{x},\ddot{w})$ if $\delta=1$.
  \item $DI(1^{k},\Lambda,x,w)$: $(N, g) \leftarrow \Lambda$, $r \leftarrow w$;
        checks that $N >2^{2k}$, $g,x\in Z^{*}_{N}$. Outputs $0$, if  $x = g^{r} \mod N$
        and all the test pass. Outputs $1$, if  $x =N - g^{r} \mod N$ and all the test
        pass.
  \item $KG(1^{k},\Lambda,x)$: $(N, g) \leftarrow \Lambda$, $hk \in_U Z_{N}$,
        $pk \leftarrow g^{hk} \mod N$, finally outputs $(hk,pk)$.
  \item $Hash(1^k,\Lambda,x,hk)$: $(N,g) \leftarrow \Lambda$, $y\leftarrow x^{hk}\mod
        N$, finally outputs $y$.
  \item $pHash(1^k ,\Lambda,x,pk,w)$: $(N, g) \leftarrow \Lambda$, $y\leftarrow pk^w
        \mod {N}$, finally outputs $y$.
\end{itemize}

\begin{theorem}
Assuming  DQR is a hard problem, the hash system is a $\epsilon$-UPHDH, where $\epsilon <1$.
\end{theorem}

This theorem can be proven in a similar way in which Theorem \ref{DNR_is_hash} is proven.

\ifCLASSOPTIONcaptionsoff
  \newpage
\fi

\bibliographystyle{plain}
\bibliography{ot-against-malicious}

\begin{thebibliography}{10}

\bibitem{aiello2001priced}
B.~Aiello, Y.~Ishai, and O.~Reingold.
\newblock {Priced oblivious transfer: How to sell digital goods}.
\newblock In {\em Advances in Cryptology-Eurocrypt'2001}, pages 119--135.
  Springer, 2001.

\bibitem{ajtai1996generating}
M.~Ajtai.
\newblock {Generating hard instances of lattice problems (extended abstract)}.
\newblock In {\em Proceedings of the twenty-eighth annual ACM symposium on
  Theory of computing}, pages 99--108. ACM, 1996.

\bibitem{barak2004strict}
B.~Barak and Y.~Lindell.
\newblock {Strict Polynomial-time in Simulation and Extraction}.
\newblock {\em SIAM Journal on Computing}, 33(4):783--818, 2004.

\bibitem{bernstein2008proving}
D.~Bernstein.
\newblock {Proving tight security for Rabin-Williams signatures}.
\newblock pages 70--87. Springer, 2008.

\bibitem{boneh2007space}
D.~Boneh, C.~Gentry, and M.~Hamburg.
\newblock {Space-efficient identity based encryptionwithout pairings}.
\newblock In {\em Foundations of Computer Science, 2007. FOCS'07. 48th Annual
  IEEE Symposium on}, pages 647--657. IEEE, 2007.

\bibitem{camenisch2007simulatable}
J.~Camenisch, G.~Neven, and A.~Shelat.
\newblock {Simulatable adaptive oblivious transfer}.
\newblock In {\em Advances in Cryptology-Eurocrypt'2007}, page 590.
  Springer-Verlag, 2007.

\bibitem{canetti2000security}
R.~Canetti.
\newblock {Security and composition of multiparty cryptographic protocols}.
\newblock {\em Journal of Cryptology}, 13(1):143--202, 2000.

\bibitem{canetti2004adaptive}
R.~Canetti, I.~Damgard, S.~Dziembowski, Y.~Ishai, and T.~Malkin.
\newblock {Adaptive versus non-adaptive security of multi-party protocols}.
\newblock {\em Journal of Cryptology}, 17(3):153--207, 2004.

\bibitem{canetti2001uccom}
R.~Canetti and M.~Fischlin.
\newblock {Universally composable commitments}.
\newblock In {\em Advances in Cryptology¡ªCRYPTO 2001}, pages 19--40. Springer,
  2001.

\bibitem{Canetti2004ro}
R~Canetti, O~Goldreich, and S~Halevi.
\newblock The random oracle methodology, revisited.
\newblock {\em Journal of the ACM (JACM)}, 51(4):557--594, 2004.

\bibitem{canetti2006limitations}
R.~Canetti, E.~Kushilevitz, and Y.~Lindell.
\newblock {On the limitations of universally composable two-party computation
  without set-up assumptions}.
\newblock {\em Journal of Cryptology}, 19(2):135--167, 2006.

\bibitem{carter1979universal}
J.L. Carter and M.N. Wegman.
\newblock {Universal classes of hash functions}.
\newblock {\em Journal of computer and system sciences}, 18(2):143--154, 1979.

\bibitem{cheon2006security}
J.~Cheon.
\newblock {Security analysis of the strong Diffie-Hellman problem}.
\newblock {\em Advances in Cryptology-EUROCRYPT 2006}, pages 1--11, 2006.

\bibitem{cramer2000efficient}
R.~Cramer, I.~Damg{\aa}rd, and P.~MacKenzie.
\newblock {Efficient zero-knowledge proofs of knowledge without intractability
  assumptions}.
\newblock In {\em Public Key Cryptography}, pages 354--373. Springer, 2000.

\bibitem{cramer2002universal}
R.~Cramer and V.~Shoup.
\newblock Universal hash proofs and a paradigm for adaptive chosen ciphertext
  secure public-key encryption.
\newblock In L.~Knudsen, editor, {\em Advances in Cryptology - Eurocrypt'2002},
  pages 45--64, Amsterdam, NETHERLANDS, 2002. Springer-Verlag Berlin.

\bibitem{crpeau1987equivalence}
C.~Cr{\'e}peau.
\newblock {Equivalence Between Two Flavours of Oblivious Transfers}.
\newblock In {\em Advances in Cryptology-Crypto'87}, page 354. Springer-Verlag,
  1987.

\bibitem{elgamal1985public}
T.~ElGamal.
\newblock {A public key cryptosystem and a signature scheme based on discrete
  logarithms}.
\newblock {\em IEEE Transactions On Information Theory}, 31(4):469--472, 1985.

\bibitem{even1985randomized}
S.~Even, O.~Goldreich, and A.~Lempel.
\newblock {A randomized protocol for signing contracts}.
\newblock {\em Communications of the ACM}, 28(6):647, 1985.

\bibitem{galbraith2008pairings}
S.D. Galbraith, K.G. Paterson, and N.P. Smart.
\newblock {Pairings for cryptographers}.
\newblock {\em Discrete Applied Mathematics}, 156(16):3113--3121, 2008.

\bibitem{garay2009somewhat}
J.A. Garay, D.~Wichs, and H.S. Zhou.
\newblock {Somewhat non-committing encryption and efficient adaptively secure
  oblivious transfer}.
\newblock In {\em Advances in Cryptology-Crypto'2009}, page 523. Springer,
  2009.

\bibitem{gennaro2006framework}
R.~Gennaro and Y.~Lindell.
\newblock {A framework for password-based authenticated key exchange}.
\newblock {\em ACM Transactions on Information and System Security (TISSEC)},
  9(2):234, 2006.

\bibitem{Gentry2008Trapdoors}
C.~Gentry, C.~Peikert, and V.~Vaikuntanathan.
\newblock Trapdoors for hard lattices and new cryptographic constructions.
\newblock In {\em Stoc'08: Proceedings of the 2008 Acm International Symposium
  on Theory of Computing}, pages 197--206 798.
\newblock full paper available on http://eprint.iacr.org/2007/432.

\bibitem{gold2001found}
O.~Goldreich.
\newblock {\em {Foundations of cryptography,volume 1}}.
\newblock Cambridge university press, 2001.

\bibitem{gold2004found}
O.~Goldreich.
\newblock {\em {Foundations of cryptography, volume 2}}.
\newblock Cambridge university press, 2004.

\bibitem{goldreich1996collision}
O.~Goldreich, S.~Goldwasser, and S.~Halevi.
\newblock {Collision-free hashing from lattice problems}.
\newblock In {\em Electronic Colloquium on Computational Complexity (ECCC)},
  volume~3, 1996.

\bibitem{goldreich1996zk}
O.~Goldreich and A.~Kahan.
\newblock {How to construct constant-round zero-knowledge proof systems for
  NP}.
\newblock {\em Journal of Cryptology}, 9(3):167--189, 1996.

\bibitem{goldreich1987play}
O.~Goldreich, S.~Micali, and A.~Wigderson.
\newblock {How to play any mental game}.
\newblock In {\em Proceedings of the nineteenth annual ACM symposium on Theory
  of computing}, pages 218--229. ACM, 1987.

\bibitem{green2007blind}
M.~Green and S.~Hohenberger.
\newblock Blind identity-based encryption and simulatable oblivious transfer.
\newblock In K.~Kurosawa, editor, {\em Advances in Cryptology-Asiacrypt'2007},
  pages 265--282, Kuching, MALAYSIA, 2007. Springer-Verlag Berlin.

\bibitem{Halevi1996Commitment}
Shai Halevi and Silvio Micali.
\newblock Practical and provably-secure commitment schemes from collision-free
  hashing.
\newblock In Neal Koblitz, editor, {\em Advances in Cryptology ¡ª CRYPTO ¡¯96},
  volume 1109 of {\em Lecture Notes in Computer Science}, pages 201--215.
  Springer Berlin / Heidelberg, 1996.

\bibitem{ishai3extending}
Y.~Ishai, J.~Kilian, K.~Nissim, and E.~Petrank.
\newblock {Extending oblivious transfers efficiently}.
\newblock In {\em Advances in Cryptology-Crypto'03}, pages 145--161. Springer.

\bibitem{Ishai2008fcot}
Y.~Ishai, M.~Prabhakaran, and A.~Sahai.
\newblock Founding cryptography on oblivious transfer - efficiently.
\newblock In D.~Wagner, editor, {\em Advances in Cryptology-Crypto'2008}, pages
  572--591, Santa Barbara, CA, 2008. Springer-Verlag Berlin.

\bibitem{kalai2005ot}
Yael~Tauman Kalai.
\newblock Smooth projective hashing and two-message oblivious transfer.
\newblock In {\em Advances in Cryptology ¨C EUROCRYPT 2005}, volume 3494, pages
  78--95. Springer, 2005.
\newblock ´òÓ¡.

\bibitem{katz2008handling}
J.~Katz and Y.~Lindell.
\newblock {Handling expected polynomial-time strategies in simulation-based
  security proofs}.
\newblock {\em Journal of Cryptology}, 21(3):303--349, 2008.

\bibitem{Kilian1988fcot}
J~Kilian.
\newblock Founding crytpography on oblivious transfer.
\newblock In {\em Proceedings of the twentieth annual ACM symposium on Theory
  of computing}, pages 20--31, Inc, One Astor Plaza, 1515 Broadway, New York,
  NY,10036-5701, USA, 1988. ACM New York, NY, USA.
\newblock STOC.

\bibitem{Leurent2009Risky}
Gatan Leurent and Phong Nguyen.
\newblock How risky is the random-oracle model?
\newblock In Shai Halevi, editor, {\em Advances in Cryptology - Crypto 2009},
  volume 5677 of {\em Lecture Notes in Computer Science}, pages 445--464.
  Springer Berlin / Heidelberg, 2009.

\bibitem{lindell2008efficient}
A.Y. Lindell.
\newblock {Efficient Fully-Simulatable Oblivious Transfer}.
\newblock In {\em Topics in cryptology: CT-RSA 2008: the cryptographers' track
  at the RSA conference 2008, San Francisco, CA, USA, April 8-11, 2008:
  proceedings}, page~52. Springer-Verlag New York Inc, 2008.

\bibitem{lindell2007efficient}
Y.~Lindell and B.~Pinkas.
\newblock {An Efficient Protocol for Secure Two-Party Computation in the
  Presence of Malicious Adversaries}.
\newblock In {\em Advances in Cryptology-Eurocrypt'2007}, pages 52--78.
  Springer-Verlag, 2007.

\bibitem{lu2009secloss}
Chi-Jen Lu.
\newblock On the security loss in cryptographic reductions.
\newblock In {\em Advances in Cryptology-Eurocrypt'2009}, volume 5479, pages
  72--87. Springer, 2009.

\bibitem{luby1996pseudorandomness}
M.G. Luby and M.~Luby.
\newblock {\em {Pseudorandomness and cryptographic applications}}.
\newblock Princeton University Press, 1996.

\bibitem{micciancio2008worst}
D.~Micciancio and O.~Regev.
\newblock {Worst-case to average-case reductions based on Gaussian measures}.
\newblock {\em SIAM Journal on Computing}, 37(1):267--302, 2008.

\bibitem{Micciancio2009Lattice}
Daniele Micciancio and Oded Regev.
\newblock {\em Lattice-based Cryptography}, pages 147--191.
\newblock Springer Berlin Heidelberg, 2009.

\bibitem{naor1999oblivious}
M.~Naor and B.~Pinkas.
\newblock {Oblivious transfer and polynomial evaluation}.
\newblock In {\em Proceedings of the thirty-first annual ACM symposium on
  Theory of computing}, pages 245--254. ACM New York, NY, USA, 1999.

\bibitem{naor1999adaptive}
M.~Naor and B.~Pinkas.
\newblock {Oblivious transfer with adaptive queries}.
\newblock In {\em Advances in Cryptology-Crypto'99}, pages 573--590. Springer,
  1999.

\bibitem{naor2001efficient}
M.~Naor and B.~Pinkas.
\newblock {Efficient oblivious transfer protocols}.
\newblock In {\em Proceedings of the twelfth annual ACM-SIAM symposium on
  Discrete algorithms}, page 457. Society for Industrial and Applied
  Mathematics, 2001.

\bibitem{naor2005computationally}
M.~Naor and B.~Pinkas.
\newblock {Computationally secure oblivious transfer}.
\newblock {\em Journal of Cryptology}, 18(1):1--35, 2005.

\bibitem{paillier1999cr}
P.~Paillier.
\newblock Public-key cryptosystems based on composite degree residuosity
  classes.
\newblock In J.~Stern, editor, {\em Advances in Cryptology-Eurocrypt'99},
  volume 1592 of {\em Lecture Notes in Computer Science}, pages 223--238.

\bibitem{pedersen1991non}
T.P. Pedersen.
\newblock {Non-interactive and information-theoretic secure verifiable secret
  sharing}.
\newblock In {\em Advances in Cryptology-Crypto'1991}, volume~91, pages
  129--140. Springer, 1991.

\bibitem{peikert2008ot}
C.~Peikert, V.~Vaikuntanathan, and B.~Waters.
\newblock A framework for efficient and composable oblivious transfer.
\newblock In D.~Wagner, editor, {\em Advances in Cryptology-CRYPTO'2008}, pages
  554--571, Santa Barbara, CA, 2008. Springer-Verlag Berlin.

\bibitem{rabin1981exchange}
M.~Rabin.
\newblock {How to exchange secrets by oblivious transfer}.
\newblock Technical report, Technical Report TR-81, Harvard Aiken Computation
  Laboratory, 1981, 1981.

\bibitem{regev2009lattices}
O.~Regev.
\newblock {On lattices, learning with errors, random linear codes, and
  cryptography}.
\newblock {\em Journal of the ACM (JACM)}, 56(6):34, 2009.

\bibitem{Schnorr1991signature}
CP~Schnorr.
\newblock Efficient signature generation by smart cards.
\newblock {\em Journal of Cryptology}, 4(3):161--174, 1991.

\bibitem{shor1997polynomial}
P.~W. Shor.
\newblock Polynomial-time algorithms for prime factorization and discrete
  logarithms on a quantum computer.
\newblock {\em SIAM Journal on Computing}, 26(5):1484--1509, 1997.

\bibitem{shor1994algorithms}
P.W. Shor.
\newblock {Algorithms for quantum computation: Discrete logarithms and
  factoring}.
\newblock In {\em Annual Symposium On Foundations Of Computer Science},
  volume~35, pages 124--124. Citeseer, 1994.

\bibitem{shor1999polynomial}
P.W. Shor.
\newblock {Polynomial-time algorithms for prime factorization and discrete
  logarithms on a quantum computer}.
\newblock {\em SIAM review}, 41(2):303--332, 1999.

\bibitem{wegman1981new}
M.N. Wegman and L.~Carter.
\newblock {New hash functions and their use in authentication and set
  equality}.
\newblock {\em Journal of computer and system sciences}, 22(3):265--279, 1981.

\bibitem{yao1986generate}
A.C.C. Yao.
\newblock {How to generate and exchange secrets}.
\newblock In {\em Foundations of Computer Science, 1985., 27th Annual Symposium
  on}, pages 162--167, 1986.

\end{thebibliography}
%\bibliographystyle{IEEEtran}
%\bibliography{IEEEabrv, ot-against-malicious}
\end{document}